\documentclass[structabstract,longauth]{aa} 
\pdfoutput=1
\usepackage{natbib,amssymb,graphicx,graphics}
\usepackage{float}
\usepackage{graphicx}
\usepackage{txfonts}

\def\L'{m$_{L'}$\ }

\def\LSun{{\rm L_{\odot}}}

\def\Si{{S_{\rm init}}}
\def\Teff{{T_{\rm eff}}}
\newcommand{\kB}{k_{\rm B}}
\newcommand{\Sunits}{\kB/{\rm baryon}}

\def\kap{{\rm \kappa}}

\DeclareGraphicsExtensions{.jpg, .png, .bmp, .eps, .pdf, .tiff}

\begin{document}
   \title{Characterization of the gaseous companion $\kappa$ Andromedae b\thanks{The LBT is an international collaboration among institutions in the United States, Italy and Germany. LBT Corporation partners are: The University of Arizona on behalf of the Arizona university system; Instituto Nazionale di Astrofisica, Italy; LBT Beteiligungsgesellschaft, Germany, representing the Max-Planck Society, the Astrophysical Institute Potsdam, and Heidelberg University; The Ohio State University, and The Research Corporation, on behalf of The University of Notre Dame, University of Minnesota and University of Virginia.}}

   \subtitle{New Keck and LBTI  high-contrast observations}

\author{
 M. Bonnefoy \inst{1}
  \and
T. Currie \inst{2}
  \and
G.-D.  Marleau \inst{1}
  \and
 J. E. Schlieder \inst{1}
   \and
J. Wisniewski \inst{3}
\and 
 J.Carson \inst{4, 1}
  \and
K. R. Covey \inst{5}
 \and
T. Henning \inst{1}
 \and
 B. Biller \inst{1}
\and
P. Hinz \inst{6}
  \and  
H. Klahr \inst{1}
\and
A. N. Marsh Boyer \inst{7}
  \and
N. Zimmerman \inst{1}
  \and
M. Janson \inst{8}
  \and
M. McElwain \inst{9}
  \and
C. Mordasini \inst{1}
  \and
 A. Skemer \inst{6}
  \and
V. Bailey \inst{6}
\and
D. Defr\`{e}re \inst{6}
  \and
C. Thalmann \inst{10, 1}
  \and
M. Skrutskie  \inst{11}
\and
F. Allard \inst{12}
\and
D. Homeier \inst{12}
  \and
M. Tamura \inst{13, 14}
  \and
M. Feldt \inst{1}
\and
A. Cumming \inst{15}
\and
C. Grady \inst{9}
\and
W. Brandner \inst{1}
\and
R. Kandori \inst{13}
\and
M. Kuzuhara \inst{13, 16}
\and
M. Fukagawa \inst{17}
\and
J. Kwon \inst{13}
\and
T. Kudo \inst{18}
\and
J. Hashimoto \inst{13}
\and
N. Kusakabe \inst{13}
\and
L. Abe \inst{19}
\and
T. Brandt \inst{8}
\and
S. Egner \inst{18}
\and
O. Guyon \inst{18}
\and
Y. Hayano \inst{18}
\and
M. Hayashi \inst{13}
\and
S. Hayashi \inst{18}
\and
K. Hodapp \inst{16}
\and
M. Ishii \inst{18}
\and
M. Iye \inst{13}
\and
G. Knapp \inst{8}
\and
T. Matsuo \inst{20}
\and
K. Mede \inst{14}
\and
M. Miyama \inst{21}
\and
J.-I. Morino \inst{13}
\and
A. Moro-Martin \inst{22}
\and
T. Nishimura \inst{18}
\and
T. Pyo \inst{18}
\and
E. Serabyn \inst{23}
\and
T. Suenaga \inst{24}
\and
H. Suto \inst{13}
\and
R. Suzuki \inst{13}
\and
Takahashi \inst{14}
\and
M. Takami \inst{25}
\and	
N. Takato \inst{18}
\and
H. Terada \inst{18}
\and
D. Tomono \inst{18}
\and
E. Turner \inst{8, 26}
\and
M. Watanabe \inst{27}
\and
T. Yamada \inst{28}
\and
H. Takami \inst{13}
\and
T. Usuda \inst{18}}
\offprints{bonnefoy@mpia-hd.mpg.de}
\institute{
  Max Planck Institute for Astronomy, K\"{o}nigstuhl 17, D-69117 Heidelberg, Germany
\and
Department of Astronomy \& Astrophysics, University of Toronto. 50 St. George Street, Room 101, Toronto Ontario Canada, M5S 3H4
\and
HL Dodge Department of Physics \& Astronomy, University of Oklahoma, 440 W Brooks St, Norman, OK 73019, USA
\and
Department of Physics \& Astronomy, College of Charleston, 58 Coming St., Charleston, SC 29424, USA
\and
Lowell Observatory, 1400 W. Mars Hill Road, Flagstaff AZ 86001 USA
\and
Steward Observatory, Department of Astronomy, University of Arizona, 933 N. Cherry Ave, Tucson, AZ 85721, USA
\and
Lehigh University, College of Art and Science, Department of Physics, 27 Memorial Dr W  Bethlehem, PA 18015, USA
\and
Department of Astrophysical Sciences, Princeton University, NJ 08544, USA
\and
Exoplanets and Stellar Astrophysics Laboratory, Code 667, Goddard Space Flight Center, Greenbelt, MD 20771, USA
\and
Astronomical Institute “Anton Pannekoek”, University of Amsterdam, Science Park 904, 1098 XH Amsterdam, The Netherlands
\and
Department of Astronomy, University of Virginia, Charlottesville, VA 22904, USA
\and
CRAL, UMR 5574, CNRS, Universit\'{e} de Lyon, \'{E}cole Normale Sup\'{e}rieure de Lyon, 46 All\'{e}e d'Italie, F-69364 Lyon Cedex 07, France
\and
National Astronomical Observatory of Japan, 2-21-1 Osawa, Mitaka, Tokyo 181-8588, Japan
\and
Department of Earth and Planetary Science, Graduate School of Science, 7-3-1 Hongo, Bunkyo-ku, Tokyo 113-0033, Japan
\and
Department of Physics, McGill University, 3600 rue University, Montr\'{e}al, Qu\'{e}bec H3A 2T8, Canada
\and
Institute for Astronomy, University of Hawaii, 640 N. Aohoku Place, Hilo, HI 96720, USA
\and
Department of Earth and Space Science, Graduate School of Science, Osaka University, 1-1 Machikaneyama, Toyonaka, Osaka 560-0043, Japan
\and
Subaru Telescope, 650 North Aohoku Place, Hilo, HI 96720, USA
\and
Laboratoire Lagrange (UMR 7293), Universite de Nice-Sophia Antipolis, CNRS, Observatoire de la Cote d'Azur, 28 avenue Valrose, F-06108 Nice Cedex 2, France
\and
Department of Astronomy, Kyoto University, Kitashirakawa-Oiwake-cho, Sakyo-ku, Kyoto 606-8502, Japan
\and
Hiroshima University, 1-3-2, Kagamiyama, Higashihiroshima, Hiroshima 739-8511, Japan
\and
Departamento de Astrof\'{i}sica, CAB (INTA-CSIC), Instituto Nacional Tecnica Aeroespacial, Torrej\'{o}n de Ardoz, E-28850 Madrid, Spain
\and
Jet Propulsion Laboratory, California Institute of Technology, Pasadena, CA 171-113, USA
\and
Department of Astronomical Sciences, Graduate University for Advanced Studies (Sokendai), Mitaka, Tokyo 181-8858, Japan
\and
Institute of Astronomy and Astrophysics, Academia Sinica, P.O. Box 23-141, Taipei 10617, Taiwan
\and
Kavli Institute for Physics and Mathematics of the Universe, The University of Tokyo, 5-1-5, Kashiwanoha, Kashiwa, Chiba 277-8568, Japan
\and
Department of Cosmosciences, Hokkaido University, Kita-ku, Sapporo, Hokkaido 060-0810, Japan
\and
Astronomical Institute, Tohoku University, Aoba-ku, Sendai, Miyagi 980-8578, Japan
}

   \date{Received 21/06/2013; accepted 06/08/2013}

 
  \abstract
   {We previously reported the direct detection of a low mass companion at a projected separation of 55$\pm$2 AU around the B9 type star $\kappa$ Andromedae. The properties of the system (mass ratio, separation) make it a benchmark for the understanding of the formation and evolution of gas giant planets and brown dwarfs on wide-orbits.}
   {We present new angular differential imaging (ADI)  images of the system  at 2.146 ($\mathrm{K_{s}}$), 3.776 (L'), 4.052 ($\mathrm{NB\_{4.05}}$) and 4.78 $\mu$m (M') obtained with Keck/NIRC2 and LBTI/LMIRCam, as well as more accurate near-infrared photometry of the star with the MIMIR instrument.  We aim to determine the near-infrared spectral energy distribution (SED) of the companion and use it to characterize the object.}
   {We used analysis methods adapted to ADI to extract the companion flux. We compared the photometry of the object to reference young/old objects and  to a set of seven PHOENIX-based atmospheric models of cool objects accounting for the formation of dust. We used evolutionary models to derive mass estimates considering a wide range of plausible initial conditions. Finally, we used dedicated formation models to discuss the possible origin of the companion.}
   {We derive a more accurate $\mathrm{J=15.86\pm0.21}$, $\mathrm{H=14.95\pm0.13}$, $\mathrm{K_{s}=14.32\pm0.09}$ mag for  $\kappa$ And b. We redetect the companion in all our high contrast observations.  We confirm previous contrasts obtained at $\mathrm{K_{s}}$ and L' band. We derive $\mathrm{NB\_{4.05}=13.0 \pm 0.2}$ and $\mathrm{M'=13.3 \pm 0.3}$ mag and estimate $\mathrm{Log_{10}(L/L_{\odot})=-3.76\pm0.06}$. Atmospheric models yield $\mathrm{T_{eff}=1900^{+100}_{-200}\:K}$. They do not set constrains on  the surface gravity.  ``Hot-start" evolutionary models predict masses of  $\mathrm{14^{+25}_{-2}\: M_{Jup}}$ based on the luminosity and  temperature estimates, and considering a conservative age range for the system ($30^{+120}_{-10}$ Myr).  ``warm-start" evolutionary tracks constrain the mass to $\mathrm{M \geq 11M_{Jup}}$.}
   {The mass of $\kappa$ Andromedae b mostly falls in the brown-dwarf regime, due to remaining uncertainties in age and
mass-luminosity models. According to the formation models,  disk instability in a primordial disk could account for the position and a wide range of plausible masses of $\kappa$ And b.}


   \maketitle
%

\section{Introduction}

\begin{table*}
\begin{minipage}[!H]{\linewidth}
\caption{Log for the new high-contrast observations of $\kappa$ Andromedae.}
\label{table:logobs}
\centering
\renewcommand{\footnoterule}{}  
\begin{tabular}{lllllllll}
\hline \hline
Date & Instrument    &   Band   &DIT & NDIT & $\mathrm{N_\mathrm{exp}} $ & Parallactic angle / start  & Parralactic angle / end   &  Remarks   \\
    &      &      &     (s)  &      &                   & ($\mathrm{\degr}$)     & ($\mathrm{\degr}$)   &       \\
\hline
2012/10/06    &   LMIRCam  &   M   &  0.757   &   30   &   87   &   177.49   &   123.54  &   Saturated exposures\\ 
2012/10/06    &  LMIRCam   &   M    &  0.029  &   4  &  36    &   165.97  &  124.98  &   Unsaturated exposures \\
2012/10/30    & {NIRC2} &$\mathrm{L^\prime}$&0.3&70&30&160.92&153.58&\\
2012/10/30    & NIRC2 & $\mathrm{L^\prime}$ & 0.3 & 70 & 30 & 140.35 & 134.46\\
2012/11/03    & NIRC2 & $\mathrm{NB\_{4.05}}$ & 0.3 & 100 & 10 & 163.14 & 159.94\\
2012/11/03    & NIRC2 & $\mathrm{K_{s}}$& 5 & 3 & 15 & 155.27 & 151.92\\
2012/11/03    & NIRC2 & $\mathrm{NB\_{4.05}}$ & 0.3 & 100 & 30 & 144.95 & 137.40\\
2012/11/03    & NIRC2 & $\mathrm{K_{s}}$& 5 & 3 & 15 & 133.43 & 131.58\\
\hline
\end{tabular}
\end{minipage}
\end{table*}

During the last 15 years, radial velocity and transit surveys have provided a detailed inventory of  the population of giant planets within $\sim$3 AU around stars of different masses, ages, and metallicities \citep[e.g][and ref therein]{2009A&A...495..335L, 2011AJ....141...16J, 2011A&A...533A.141S, 2012A&A...543A..45M,2013A&A...549A.109B,2013arXiv1304.4328S, 2013arXiv1304.6755N}. Correlations between planet frequencies and the host-star metallicity \citep{1997MNRAS.285..403G, 2001A&A...373.1019S, 2005ApJ...622.1102F, 2011A&A...533A.141S, 2012A&A...543A..45M}, the host-star mass \citep{2007A&A...472..657L, 2010ApJ...709..396B}, and between the heavy-element content of gas giants  with host star metallicity \citep{2006A&A...453L..21G, 2011ApJ...736L..29M} favour the hypothesis of a formation by core-accretion \citep[hereafter CA;][]{1996Icar..124...62P, 2009A&A...501.1139M, 2009A&A...501.1161M, 2011A&A...526A..63A, 2012A&A...541A..97M}. Core-accretion considers that a core of solids (ice, rock) forms through collisions of planetesimals in the protoplanetary disk at a distance of a few AU  from the central star.  Once the core  has reached a critical mass \citep{1980PThPh..64..544M, 1986Icar...67..391B},  its gravitational potential causes a rapid capture of the surrounding gas which ultimately forms a massive gas envelope. Additional migration mechanisms have been proposed \citep[e.g][and ref. therein]{1986ApJ...309..846L, 2004A&A...417L..25A} to explain the population of giant planets orbiting very close to their parent stars.

Conversely, high-contrast and high-angular resolution imaging is probing the population of wide-orbit ($>5 $AU) gaseous companions around a variety of young (age $\leq$300 Myr) and nearby (d $\leq$ 150 pc) stars, ranging from M dwarfs to early-type (F to B) stars \citep{2002A&A...384..999N, 2003A&A...404..157C, 2005AJ....130.1845L, 2005ApJ...625.1004M, 2007ApJS..173..143B, 2007A&A...472..321K, 2007ApJ...670.1367L, 2010A&A...509A..52C, 2011ApJ...736...89J, 2012A&A...539A..72D, 2012ApJ...753..142B, 2012A&A...544A...9V, 2013A&A...553A..60R, 2013arXiv1306.1233N}.  The majority of planetary mass companions have been discovered  along an extended range of wide orbits (15 AU to several hundreds of AU)  around low mass  (MGK) stars \citep[e.g ][]{2004A&A...425L..29C, 2010ApJ...714L..84T, 2010ApJ...719..497L}.  The high-mass ratio with their host and the high separations makes the fragmentation of pre-stellar cores during collapse \citep[e.g.][]{2012MNRAS.419.3115B} a candidate for the formation of these wide systems.

Low mass ($\leq$ 15 $\mathrm{M_{Jup}}$) gaseous companions discovered more recently at moderate separations ($\leq$  100 AU) around the massive stars HR 8799 \citep{2008Sci...322.1348M, 2010Natur.468.1080M} and $\beta$ Pictoris \citep{2009A&A...493L..21L, 2010Sci...329...57L} might represent a previously unexplored population of gaseous companions \citep[][]{2012A&A...544A...9V, 2013A&A...553A..60R, 2013arXiv1306.1233N}.  The extended debris disks identified around these stars, shaped by the companions \citep[][]{
2009ApJ...705..314S, 2012A&A...542A..40L}, suggest these systems emerged from a primordial gaseous disk. This picture is reinforced by recent resolved images of transition disks around Herbig stars \citep{2011ApJ...732...42A, 2012A&A...546A..24R} with cavities which extend beyond the separations of the aforementioned companions, and might have been carved by planets  \citep{2013Natur.493..191C}.  Nevertheless, the subsequent formation pathway remains unclear. Core accretion (CA) could eventually explain the properties of directly imaged planets with the narrowest orbits \citep[HR 8799 e and d, $\beta$ Pictoris b; ][]{2008ApJ...673..502K, 2009A&A...501.1139M, 2011ApJ...727...86R}. But  associated CA formation timescales become too long compared to the mean lifetime of primordial disks and require higher disk surface density for an in-situ formation at more than $\sim$15 AU \citep[][]{2009ApJ...695L..53B, 2009ApJ...707...79D}. A revision of the way solids are accreted \citep{2010A&A...520A..43O}, or additional outward migration mechanisms  must be considered \citep[][and ref therein]{2009A&A...502..679C, 2012ARA&A..50..211K} to explain planets found at larger radii if formed initially by CA. Gravitational instability within disks \citep[G.I.; ][]{1978M&P....18....5C} has been considered as an alternative mechanism for these objects \citep{2011ApJ...731...74B} and can also be associated with migration \citep[e.g.][]{2012ApJ...746..110Z} and ejection \citep[e.g.][]{2013A&A...552A.129V}. Here, protostellar disks develop global instabilities (if cool enough) and fragment into bound clumps that contract to form giant planets. This mechanism operates on much shorter timescales than CA (a few orbital periods). However, recent surveys suggest it might not dominate at wide ($> 30$~AU) separations \citep[][]{2011ApJ...736...89J, 2012ApJ...745....4J, 2013A&A...553A..60R}. To conclude, it is not clear how these 8-78 AU companions relate to  the more distant low-mass brown-dwarfs companions found around other massive (1.35-2.5) young (age $\leq$ 150 Myr) stars \citep[\object{HR 7329 B}, \object{HIP 78530 B}, \object{HD 1160 B}, \object{HR 6037 BaBb}, \object{HD 23514 B};][]{2000ApJ...541..390L, 2011ApJ...730...42L, 2012ApJ...750...53N, 2010A&A...521L..54H, 2013arXiv1306.1233N, 2012ApJ...748...30R}.\\

Direct imaging offers the possibility to collect multiple-band photometry and spectra emitted by the photospheric layers of the companions in the near-infrared \citep[1-5 $\mu$m; ][]{2007ApJ...656..505M, 2011ApJ...729..128C, 2010A&A...512A..52B, 2013arXiv1306.3709B}. These data can  provide a stringent characterization of the chemical (composition) and physical properties (mass, radius, effective temperature) of the sources, which are at the basis of our understanding of their formation processes \citep[][]{2013arXiv1302.1160B, 2013Sci...339.1398K}.  They can also give glimpses of the physics and chemistry at play in the cool and complex atmospheres of the sources \citep[e.g.][]{2011ApJ...729..128C, 2012ApJ...753...14S, 2013arXiv1306.3709B}.  The peculiar near-infrared spectro-photometric properties  of the companions are now better understood as a consequence of the low  temperature and surface gravity atmosphere, which in some cases can lead to the formation of thick layers of dust and/or trigger non-equilibrium chemistry \citep{2010ApJ...710L..35J, 2011ApJ...732..107S, 2011ApJ...733...65B, 2011ApJ...735L..39B,  2011ApJ...737...34M, 2011ApJ...729..128C,2012ApJ...753...14S, 2012ApJ...754..135M,2013AJ....145....2F}. 

In the course of the Strategic Explorations of Exoplanets and Disks with Subaru \citep[SEEDS,][]{2009AIPC.1158...11T}, we identified in early-2012 a low-mass companion at a projected separation of 1" around the massive ($\mathrm{2.5\pm0.1\:M_{\odot}}$) B9 IVn \citep{1994AJ....107.1556G} and  nearby \citep[$51.6\pm0.5$ pc, ][]{2007A&A...474..653V} star $\kappa$ Andromedae \citep[hereafter $\kappa$ And,][]{2013ApJ...763L..32C}. The host star kinematics make it a high probability member \citep{2011ApJ...732...61Z, 2013ApJ...762...88M, 2013ApJ...763L..32C} of the $\sim$30 Myr old  Columba moving group \citep{2008hsf2.book..757T}. At these ages, we estimated $\kap$ And b has a mass of $\mathrm{12.8_{-1.0}^{+2.0}\:M_{Jup}}$, making it the first planetary/brown-dwarf companion directly imaged around such a massive star.  

However, the mass estimate of $\kappa$ And b relied on the  near-infrared photometry of the star, derived from 2MASS images \citep{2003tmc..book.....C} where $\kappa$ And A is saturated. More importantly, it was derived from predictions of the so-called ``hot-start" evolutionary models, which assume that the object starts its evolution following a spherical collapse from an arbitrary large initial radius \citep{1997ApJ...491..856B, 2000ApJ...542..464C, 2003A&A...402..701B}. The alternative ``cold-start" models \citep{marl07, sb12, 2012A&A...547A.111M} hypothetize that the gaseous material accreted onto planet embryo passes through a super-critical accretion shock, and looses all its gravitational energy, therefore leading to objects with low initial entropies ($S_{init}$). As a consequence, these models predict lower temperatures and lower luminosities at early ages than ``hot-start" evolutionary models for a given objet. The more recent ``warm-start" models \citep{sb12, mc13} generalise the previous cases by exploring the impact of initial conditions on the cooling curves through the choice of the initial entropy ($\mathrm{S_{init}}$) of the object. With these models,  joint constraints on the mass and initial entropy of companions can be derived from a brightness and temperature measurement.

We present new high-contrast near-infrared images of  $\kappa$ And b obtained using NIRC2 at the W.M. Keck Observatory and LMIRCam on the Large Binocular Telescope Interferometer (LBTI) at 2.146 ($\mathrm{K_{s}}$), 3.776 (L'), 4.052 ($\mathrm{NB\_{4.05}}$) and 4.78 $\mu$m (M'). They confirm and complement the current set of photometric data of the companion. We also present additional unresolved observations of the system in the near-infrared. Unresolved observations provide accurate photometry of the primary star and, as a consequence, of the companion from 1 to 2.5 $\mu$m, at wavelengths where the effect of atmospheric dust can be studied. Altogether, they enable us to refine the companion properties and discuss its formation mechanism. 

This paper is organized as follows: we describe in Section \ref{section:obs} the observations and the related analysis of the data; we present in section \ref{section:analresults} our main results. Section \ref{section:analresults} is split into four subsections. We first rederive a more conservative age estimate for the system in Subsection \ref{section:systage}. We compare the photometry of $\kappa$ And b to empirical reference objects  in Subsection \ref{subsec:compemp}, and to atmospheric models in Subsection \ref{subsec:atmo}. We give new mass estimates based on ``hot", ``cold", and ``warm-start" evolutionary models in Subsection \ref{subsec:evolm}. We discuss the properties of $\kappa$ And b and review the possible formation scenarios  in section \ref{section:discussion}.


\section{Observation and data reduction}
\label{section:obs}
\subsection{Seeing-limited observations}

\begin{figure*}
\centering
\includegraphics[scale=0.285,trim=20mm 0mm 20mm 0mm,clip]{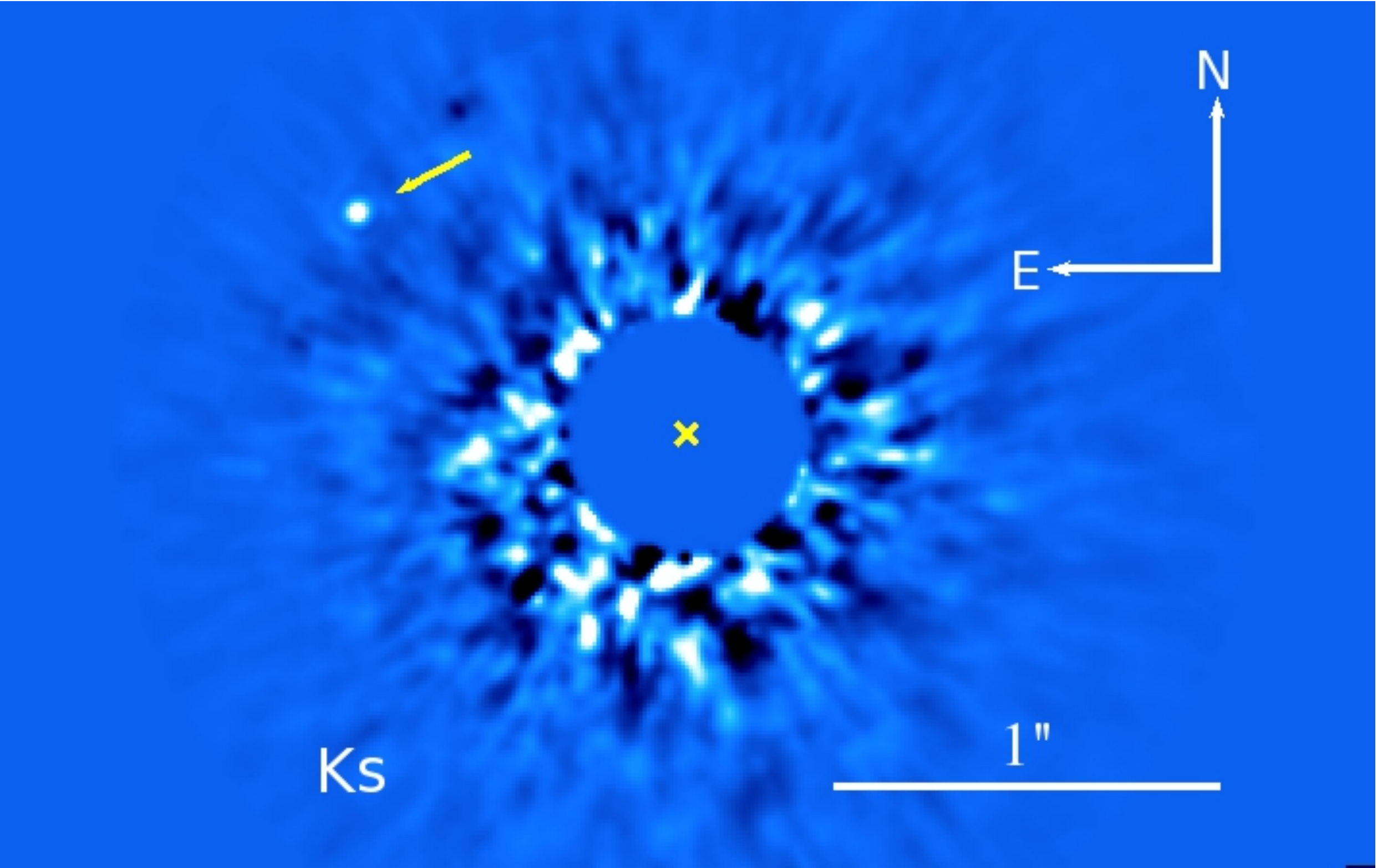}
\includegraphics[scale=0.285,trim=20mm 0mm 20mm 0mm,clip]{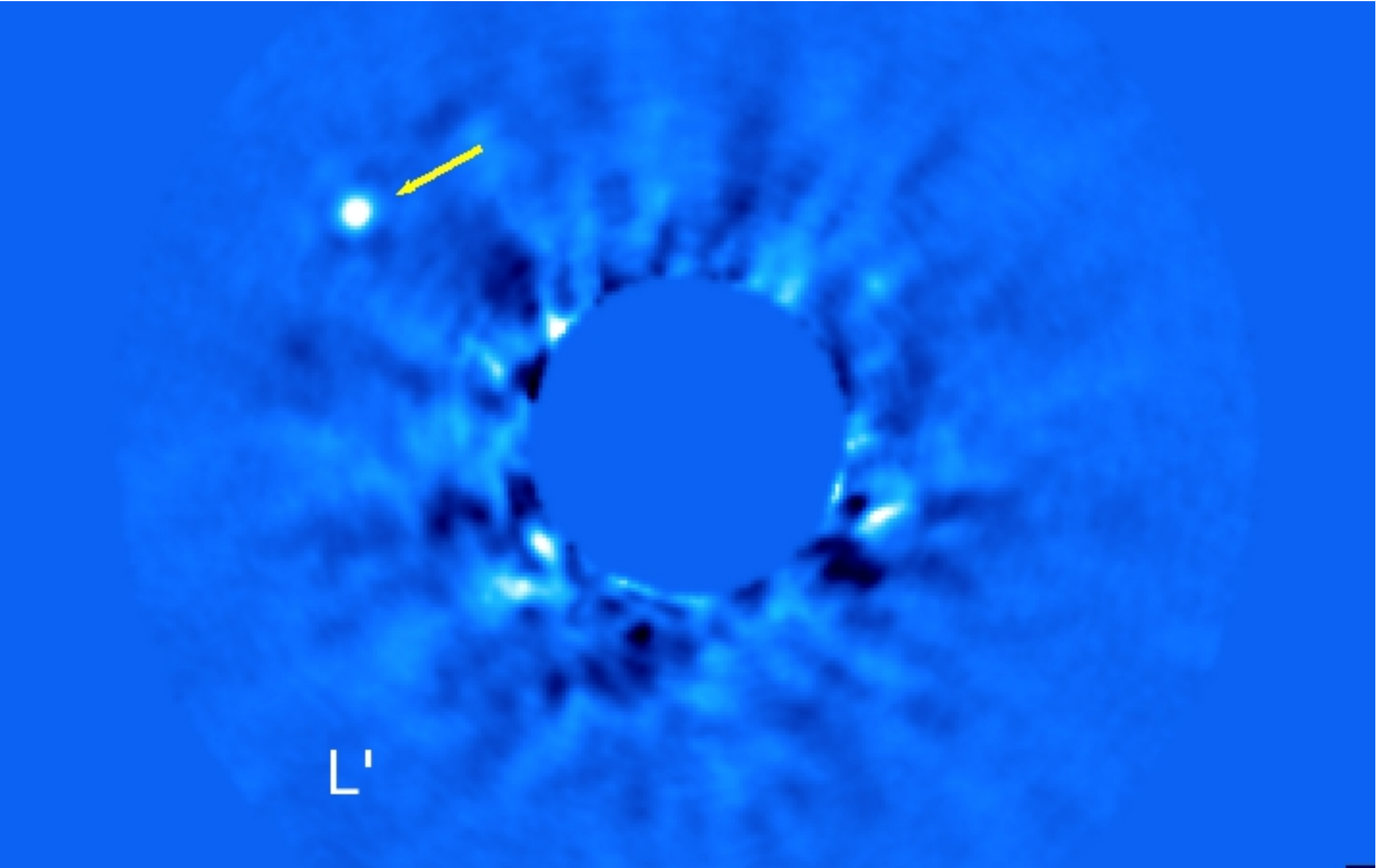}
\includegraphics[scale=0.285,trim=20mm 0mm 20mm 0mm,clip]{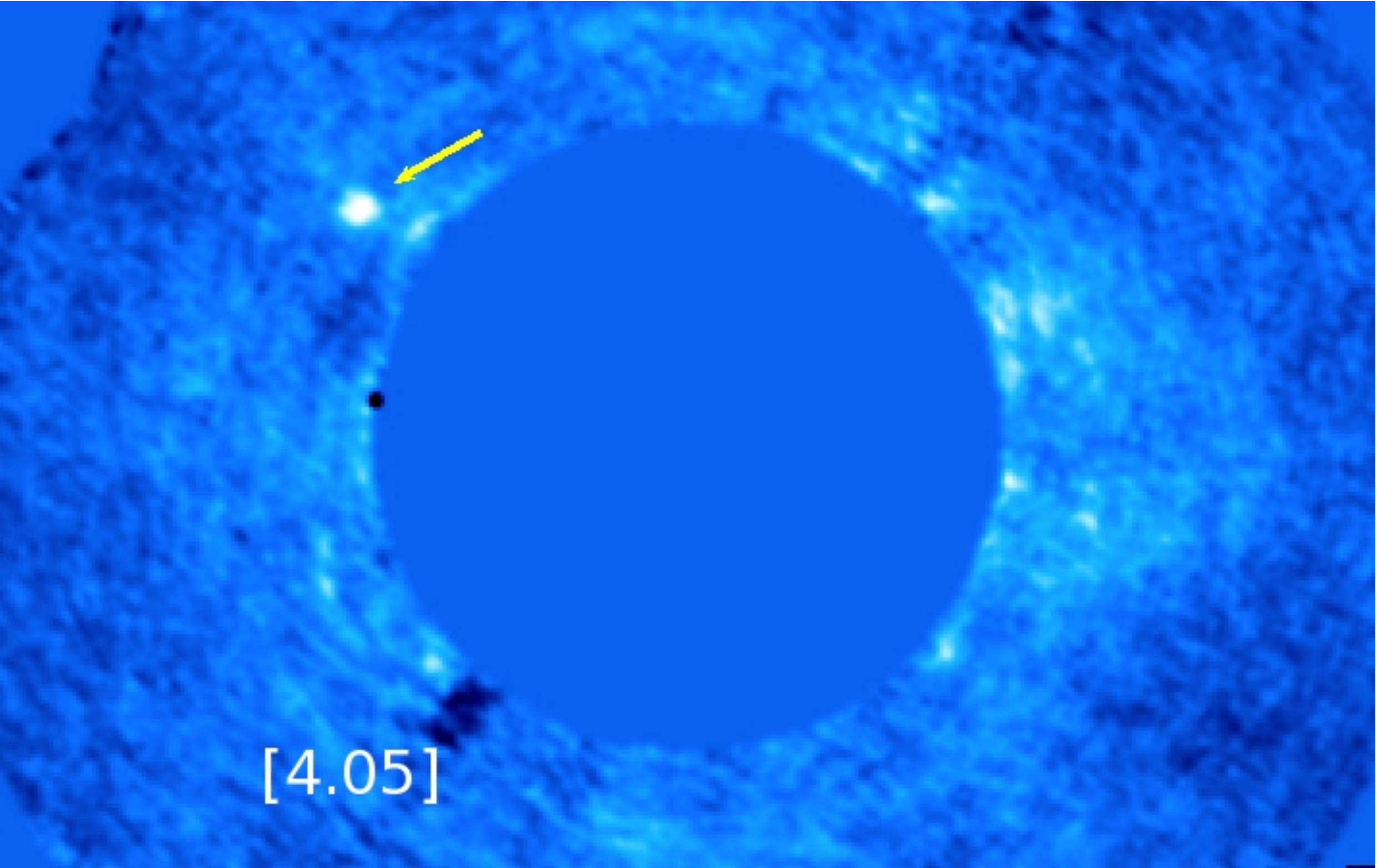}
\caption{Reduced Keck/NIRC2 images at $\mathrm{K_{s}}$/2.16 $\mu m$ (left), $\mathrm{L^\prime}$/3.78 $\mu m$ (middle), and NB$\_{4.05}$ (right) of 
$\kappa$ And showing the detection of the companion $\kappa$ And b. We adjust the image color scale such that the 
white region identifying $\kappa$ And b (pointed to by an arrow) roughly covers an area equal to the image FWHM.  The 
cross identifies the stellar centroid position in the $\mathrm{K_{s}}$ data and is at the same point in all three images.} 
\label{keckimages}
\end{figure*}

The accuracy of $\kappa$ And b near infrared photometry  (J=$16.3\pm0.3$, H=$15.2\pm0.2$, $\mathrm{K_{s}=14.6\pm0.4}$) reported in \cite{2013ApJ...763L..32C}  was predominantly limited by the accuracy of the 2MASS photometry of the star (the only near-infrared photometry publicly avaliable at that time; Table~\ref{Table:phot}). The low accuracy arises from the strong saturation of the star in the  2MASS images \footnote{The saturation can be seen directly on the 2MASS images: http://irsa.ipac.caltech.edu/cgi-bin/2MASS/IM/nph-im\_pos}.

We obtained infrared photometry of $\kappa$ And A in the J, H, and $\mathrm{K_{s}}$ filters using the MIMIR instrument \citep{2007PASP..119.1385C},
mounted on the 1.8m Perkins telescope at Lowell Observatory, on 2012 October 30.  At the f/5 focus,
the 1024 x 1024 Aladdin array covered a field of view (FOV) of 10 $\times$ 10 arcmin with a pixel
scale of 0$\farcs$579 pixel$^{-1}$.  Standard bias, dark, flat field, and sky corrections were applied
to all data using calibration data obtained during the night. Five 0.1s exposures were obtained
at each of six dither positions separated by $\sim$60 arcsec in every filter.  Due to
the brightness of our science target, we defocussed the telescope until stars took the appearance
of $\sim$60 arcsec wide donuts.  Our observations of $\kappa$ And A were bracketed by observations
of HR 8962, using the same exposure time and focus settings, to enable us to determine the photometric zeropoint.

Photometry was determined using apertures with a radius of 70 pixels and a 10 pixel wide background
sky annulus starting at a radius of 80 pixels from each centroid position.  Our final photometry was computed
from 10 data frames (2 dither positions) in which our large apertures did not intersect with known regions of bad
pixel clusters in the array.  We determined the zeropoint for our observations by comparing our measured
photometry for HR 8962 against published 2MASS values \citep{2003tmc..book.....C}.  HR 8962 is known to be a visual
binary (see e.g. \citealt{2002A&A...396..933M, 2009AJ....138..813H}) whose separation of 0$\farcs$5 is within both our aperture definition
and the aperture definition of 2MASS, and thus should not affect our zeropoint determination.  The
photometry we determined for 2MASS J23373296+4423122, another bright object in the FOV of our
HR 8962 data, was within 2-$\sigma$ of published 2MASS photometry for the source, thereby confirming
our zeropoint determination.

Aperture photometry on the source and of the photometric reference yields $\mathrm{J=4.26\pm0.04}$, $\mathrm{H=4.31\pm0.05}$, and $\mathrm{K_{s}=4.32\pm0.05}$ mag for $\kappa$ And A (see also Table~\ref{Table:phot}). We find negligible photometric shifts ($\leq0.004$ mag) between MIMIR and HiCIAO photometric systems using the corresponding filter pass-bands, a flux-calibrated spectrum of Vega \citep{2007ASPC..364..315B}, and of a B9 star from the  \cite{1998PASP..110..863P}  library. This enables us to revise the original photometry of $\kappa$ And b \citep{2013ApJ...763L..32C} to $\mathrm{J=15.86\pm0.21}$, $\mathrm{H=14.95\pm0.13}$, $\mathrm{K_{s}=14.32\pm0.09}$ mag. 
 
 \begin{table}
\begin{minipage}{\columnwidth}
\caption{Near-infrared photometry of $\kappa$ And A}
\label{Table:phot}
\centering
\renewcommand{\footnoterule}{}  
\begin{tabular}{lllll}
\hline \hline 
Epoch		 		&   J		&	H		& $\mathrm{K_{s}}$  &  Ref.	\\   			
\hline
29/10/1998    &  $4.62\pm0.27$   & $4.60\pm0.22$   & $4.57\pm0.36$  &  1  \\   
30/10/2012   &   $\mathrm{4.26\pm0.04}$  &   $\mathrm{4.31\pm0.05}$  &   $\mathrm{4.32\pm0.05}$  &  2 \\
\hline
\end{tabular}
\end{minipage}
\tablefoot{References: [1] - 2MASS / \cite{2003tmc..book.....C}, [2] - this work
}
\end{table}

 \subsection{High-resolution spectroscopy}

We obtained a R $\sim$31500 optical ($\sim$3600--10000\AA) spectrum of $\kappa$ And A on UT 2012 October 24 with the ARC Echelle Spectrograph \citep[ARCES,][]{2003SPIE.4841.1145W} mounted on the Apache Point Observatory (APO) 3.5 m telescope. The spectrum was obtained using the default $1\farcs6 \times 3\farcs2$ slit and an exposure time of 45 seconds.  A ThAr lamp exposure was obtained after the integration to facilitate accurate wavelength calibration. The data were reduced using standard \textit{IRAF} techniques.  We used this spectrum to derive new estimates of the surface gravity and effective temperature of the star.  Our results and method are detailed in Section~\ref{section:analresults} and Appendix~\ref{AppendixB}.

\subsection{High-contrast observations}
	\subsubsection{Keck/NIRC2}
We observed $\kappa$ Andromedae on October 30, 2012 and November 3, 2012 with the NIRC2 camera 
  fed by the adaptive optics system of Keck II \citep{2004ApOpt..43.5458V}. We used the $\mathrm{K_{s}}$ ($\mathrm{\lambda_{c}=2.16 \mu m}$) and $\mathrm{L^\prime}$ ($\mathrm{\lambda_{c}=3.78 \mu m}$)  broadband filters and the Br$\alpha$ ($\mathrm{\lambda_{c}=4.05 \mu m}$) narrow band filter (hereafter $\mathrm{NB\_{4.05}}$, Table \ref{table:logobs}).  All data were 
taken in the narrow camera mode \citep[9.952 mas pixel$^{-1}$,][]{2010ApJ...725..331Y} in either multi-correlated 
double sampling ($\mathrm{K_{s}}$) or correlated double sampling ($\mathrm{L^\prime}$, $\mathrm{NB\_{4.05}}$).  
Observing conditions both nights were photometric with above average seeing (FWHM$_{natural}$ 
$\sim$ 0\farcs{}4--0\farcs{}5) and average AO performance.
In all cases, we used the ``large hex" pupil plane mask.  

The $\mathrm{K_{s}}$ data were taken through the partially transmissive 0\farcs{}6 diameter coronagraphic mask 
in coadded 15 second exposures.  The $\mathrm{L^\prime}$ data were taken in coadded 20 second exposures 
with the primary star's point-spread function (PSF) core saturated. We dithered the star off of the detector to obtain sky frames in the middle of our observing sequence.  
For the NB$\_{4.05}$ data, the PSF core was unsaturated, and we took three sequences of 10 science frames followed 
by 5 sky frames ($t_{exp}$ = 30 s).

All data were taken in ``vertical angle" or \textit{angular differential imaging} mode \citep{2006ApJ...641..556M}.  
On both nights, we observed $\kappa$ And immediately after unsaturated observations of HR 8799, which is at a similar 
right ascension, beginning at an hour angle of $\sim$ 0.5. These unsaturated observations were later used to derive the  photometry of $\kappa$ And b (see below). Still, over the course of our sequence, 
we achieved $\sim$ 24--26$^{o}$ of field rotation, or $\approx$ 5--10 $\lambda$/D at the angular separation 
of $\kappa$ And b.

Basic image processing followed standard steps previously used to process NIRC2 data \citep{2012ApJ...755L..34C,2012ApJ...757...28C}.
Briefly, for the $\mathrm{K_{s}}$ data, we employ standard dark subtraction and flat-fielding corrections, identify and 
interpolate over hot/cold pixels.  For the thermal IR data, we subtracted off a median-combined sky image comprised 
of sky frames taken closest in time to the science frames of interest.  For all data sets, we applied the distortion 
correction from \cite{2010ApJ...725..331Y}.  After copying each image into a larger blank image, we performed image registration 
by finding directly the stellar centroid position ($\mathrm{K_{s}}$ and $\mathrm{NB\_{4.05}}$) or estimating it by cross-correlating 
the first image in our sequence by a 180$^{o}$ rotation of itself ($\mathrm{L^\prime}$).  We then determined the relative 
offsets for other images in the sequence by solving for the peak in the cross-correlation function between the 
first (reference) image and all subsequent images.

To extract a detection of $\kappa$ And b, we used the \textit{A-LOCI} pipeline 
described in \citet{2012ApJ...760L..32C}.  Because $\kappa$ And b is at a wide separation ($r$ 1\farcs{}05) and previous data obtained 
at similar wavelengths yielded very high SNR detections, we adopt conservative algorithm settings that 
minimally bias the planet flux. This included a large rotation gap 
($\delta$ = 0.8--1.5), a high cross-correlation cutoff for the contrast-limited $\mathrm{K_{s}}$ and $\mathrm{L^\prime}$ data 
($corr_{lim}$ =0.9--0.95) and, a lower cutoff for the background-limited NB$\_{4.05}$ data ($corr_{lim}$=0.2), 
and a moving pixel mask \citep{2012ApJ...760L..32C}, yielding planet throughputs (estimated by implanting synthetic 
point sources into registered images) ranging between 0.92 and 1.  

Figure \ref{keckimages} shows the final Keck/NIRC2 images, where we detect $\kappa$ And b at a SNR of 30, 20, and 7 at $\mathrm{K_{s}}$, $\mathrm{L^\prime}$, 
and $\mathrm{NB\_4.05}$.   Exterior to our inner working angle of 0\farcs{}3, we do not detect any other point sources, as was the case 
with previous  $\mathrm{H}$, $\mathrm{K_{s}}$, and $\mathrm{L^\prime}$ data \citep{2013ApJ...763L..32C}.   To measure the brightness of $\kappa$ And b, 
we perform aperture photometry with a diameter roughly equal to the image FWHM ($\sim$ 5--10 pixels), correcting 
for the very minor throughput loss induced by our processing.  To flux calibrate $\kappa$ And b in $\mathrm{m_{K_{s}}}$, we use the 
derived coronagraph spot extinction from \citet{2012ApJ...755L..34C} of 6.91 $\pm$ 0.15 mag.  This yielded a parent/companion contrast of $\Delta$m = 10.04 $\pm$ 0.15 mags.  Using our MIMIR photometry, this translates into a companion 
brightness of $\mathrm{m_{K_{s}}}$ = 14.36 $\pm$ 0.15.  For the $\mathrm{L^\prime}$, we used the unsaturated 
images of HR 8799 A \citep[$\mathrm{m_{NB\_{4.05}}}$ $\simeq$ $\mathrm{m_{L^\prime}}$ = 5.220 $\pm$ 0.018;][]{2008Sci...322.1348M}
to flux-calibrate $\kappa$ And b, deriving $\mathrm{m_{L^\prime}}$ = 13.13 $\pm$ 0.07.  Finally, for the NB$\_{4.05}$ data, both HR 8799 A 
and $\kappa$ And A were unsaturated.  Assuming $\mathrm{m_{L^\prime}}$ $\simeq$ $\mathrm{m_{NB\_{4.05}}}$, we then derive $\mathrm{m_{NB\_{4.05}}}=\mathrm{m_{L^\prime}}$ = 4.32 $\pm$ 0.05 
 for $\kappa$ And A and $\mathrm{m_{NB\_{4.05}}}$ = 13.0 $\pm$ 0.2 for $\kappa$ And b.

We measure a separation of $\rho=1.029\pm0.005$" and a position angle of $\theta=55.3\pm0.3^{\circ}$ in the $\mathrm{K_{s}}$ band images. This value is in agreement with the astrometry derived from IRCS  ($\rho=1.044"$, $\theta=55.2$)  and HiCIAO data presented by \cite{2013ApJ...763L..32C}. We nevertheless refrained from making an updated proper motion analysis, due to possible systematic offsets on $\rho$ and $\theta$ introduced by the instrument change.
    \subsubsection{LBTI/LMIRCam}
    
   We observed $\kappa$ And with the LMIRCam near-infrared camera \citep{2008SPIE.7013E.100H, 2010SPIE.7735E.118S}  at the LBTI on October 10, 2012. The LBTI was operated in single-aperture mode in order to avoid extra-overheads associated with the alignment of the telescope beams.  The telescope+instrument do not have a derotator. Therefore, it automatically operates in a mode that enables passive \textit{angular differential imaging} \citep{2006ApJ...641..556M}. We obtained 151 frames consisting of 30$\times$0.758s coadded exposures each with a M'-band filter ($\lambda_{c}=4.78 \mu m$, $\mathrm{FWHM=0.37 \mu m}$). The integration time was chosen in order to saturate the core of the stellar point spread function (PSF) to a radius of 110 mas. The 1h45 min spent on the target produced a field rotation of   53.9$^{\circ}$. The source was nodded in the instrument field of view (1.8" nod)  every 3 mins in order to  properly remove the background contribution. We recorded four frames corresponding to eight 0.029s coadded exposures following each telescope nod. These unsaturated exposures monitored the evolution of the PSF during the observing sequence, and were used to derive the contrast ratio of the system components. 
   
   Data were reduced with the MPIA-LBTI angular differential imaging pipeline. The pipeline carried out all the basic cosmetics steps (removal of detector stripes, sky subtraction, bad pixels interpolation) following the selection of pairs of nodded exposures. The  position of the source in the resulting images was found using a bidimensional Moffat fitting function. Images were recentered and placed in a master cube of frames using subpixel shifts.  We selected 87 frames with high-Strehl ratio or high-throughput to build a final master cube.  We applied six algorithms  to estimate and remove the flux distribution of the star in each input frame of the master cube. We removed a radial profile in each image, derotated them in order to re-align them with the North using the parallactic angle at the time of the exposure (non-ADI, or NADI), and median-combined these frames to produce a final residual image. Alternatively, we  built and removed the PSF of the star taking the median of all input frames contained in the cube  (classical ADI, or CADI). We also applied the RADI \citep{2006ApJ...641..556M, 2012A&A...542A..41C}, LOCI \citep{2007ApJ...660..770L}, and a custom algorithm based on Principal-Component Analysis \citep[Zimmerman et al., in prep; see also][]{2012ApJ...755L..28S, 2012MNRAS.427..948A}.

    \begin{figure}
   \centering
   \includegraphics[width=\columnwidth]{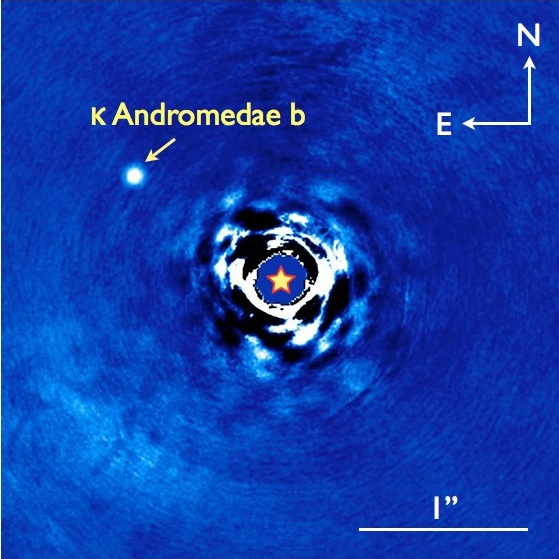}
      \caption{LBTI/LMIRCam high-contrast image of the $\kappa$ And system at M band. The companion is recovered.}
         \label{Fig:imdet}
   \end{figure}

 We also developped and applied an extension of the RADI algorithm, called ``correlated radial ADI" (or CODI), in order to optimise the reduction of these data.  Each cube image $i$ was first divided into concentric annuli. For each annuli $AN_{i}$, we selected $N$ reference annuli $AN_{ref_{k}=1-N}$ taken from other images contained in the cube for which the companion has moved by $n \times$ FWHM.  We computed and sorted the correlation factors between the $N$ reference annuli and the processed  annuli. We selected the  $\alpha$ ($\in  \mathbb N$) most correlated reference annuli $AN_{{ref}_{k=1-\alpha}}$ and retrieved the  coefficients of the linear relation between pixels of the reference annuli  and of the processed annuli $AN_{i} = a_{k,i} \times  AN_{{ref}_{k}} + b_{k,i} $ (with  $a_{k,i}$ and $b_{k,i}  \in  \mathbb R$ derived from a linear regression and a matrix inversion). This way, we were able to account \textit{simultaneously} for any variation of the  atmospheric transmission and of a piston flux offset (e.g. residuals from the background subtraction, etc) between the reference and processed annuli. The $\alpha$ model annuli were median-combined to create a final model $AN_{model}$ of $AN_{i}$. The operation was repeated for each concentric annuli and each cube image.  We considered a separation criterium of 1.5 FWHM at the separation of the companion for the CADI, RADI, CODI, LOCI, and PCA-based analyses. 

$\kappa$ And b is detected in each individual input frame. It is then naturally retrieved in all post-processed frames with a signal to noise from 20 to 38 reached with the RADI and CADI algorithm, respectively (see Figure \ref{Fig:imdet}).  We integrated the  flux of $\kappa$ And b over an aperture of 16 pixels in radius (1.5 FWHM) and used the non-saturated exposures to derive the contrast ratio between the star and its companion.  Values were corrected from the inevitable flux losses associated to angular differential imaging using 3 fake planets injected at position angles of -67$^{\circ}$, 115$^{\circ}$,  and 233$^{\circ}$.

All algorithms converge to a contrast of $\mathrm{\Delta M'=8.9 \pm 0.3}$ mag for $\kappa$ And b. The variation of the PSF in unsaturated exposures dominates the final error budget. We estimate that $\kappa$ And A has a M-band magnitude of 4.4 mag considering the mean $\mathrm{K_{s}-M}$ colors of B9 (IV-V) stars of the \cite{1996yCat..41190547V} and IRTF \citep{2003MNRAS.345..144L} catalogues ($\mathrm{K_{s}-M=-0.023\pm 0.030}$).  We also find a similar  color ($\mathrm{K_{s}-M=-0.023\pm0.006}$) using ATLAS9  models \citep{1997A&A...318..841C} taken at the temperature of the star (see section \ref{section:systage} and Appendix \ref{AppendixB}), the $\mathrm{K_{s}}$, and M' band filter transmission curves. These colors can be used as a reasonable guess for $\kappa$ And A because of the lack of  excess emission for this star at these wavelengths (see Section \ref{section:systage}). Therefore, we derive M'=$13.3 \pm 0.3$ mag $\kappa$ And b. 

We did not get observations of an astrometric field necessary to derive a reliable astrometry from these data.

\section{Analysis and Results}
\label{section:analresults}

\subsection{Re-evaluation of the system age}
\label{section:systage}
The mass of $\kappa$ And A and its companion, and insights into the companion's origin, are tied to the correct determination of the system age. For this reason, we re-investigated the various age indicators available for $\kappa$ And A based on our new measurements of the star and material found in the literature.\\

   \begin{figure}
   \centering
   \includegraphics[width=\columnwidth]{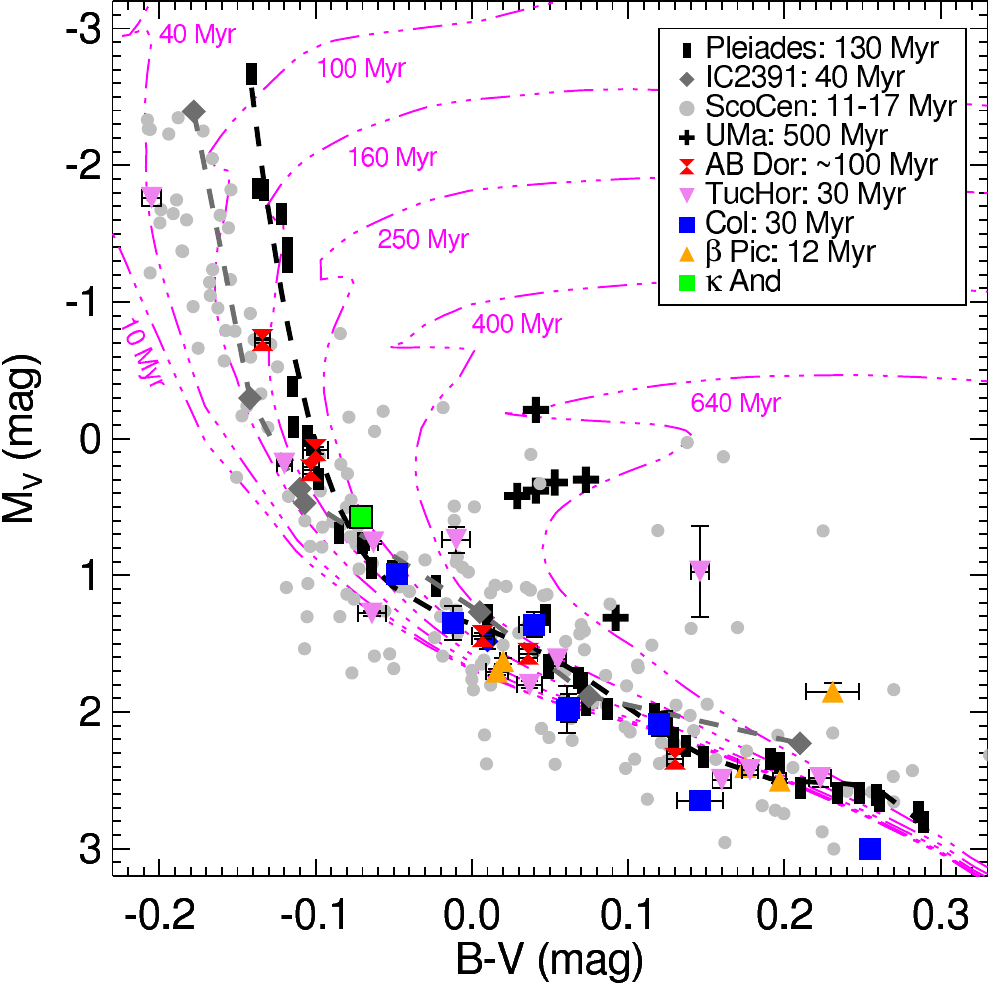} 
     \caption{Position of $\kappa$ And A in a color-magnitude diagram (CMD) compared to other A and B-type member of the Pleiades, the IC 2391 cluster, Scorpius-Centaurus (ScoCen), Ursa Majoris group, and other young moving groups in the solar neighborhood (see legend).  The star's position in the diagram is consistent with some members of all the plotted samples except the Ursa Majoris moving group (UMa), which is clearly older and more evolved. The placement in the CMD suggests that $\kappa$ And is not much older than the upper limit of the Pleiades, 150 Myr. The dashed-dotted lines are \cite{2012A&A...537A.146E} tracks that include rotation.  These models predict an age of $\lesssim$250 Myr for $\kappa$ And, but they overestimate the ages of B-type cluster members.  The dotted lines are 5th order polynomial fits to the Pleiades and IC 2391 sequences and are shown to highlight them for clarity.} 
         \label{Fig:CMDJosh}
   \end{figure}

   \begin{figure}
   \centering
   \includegraphics[width=\columnwidth]{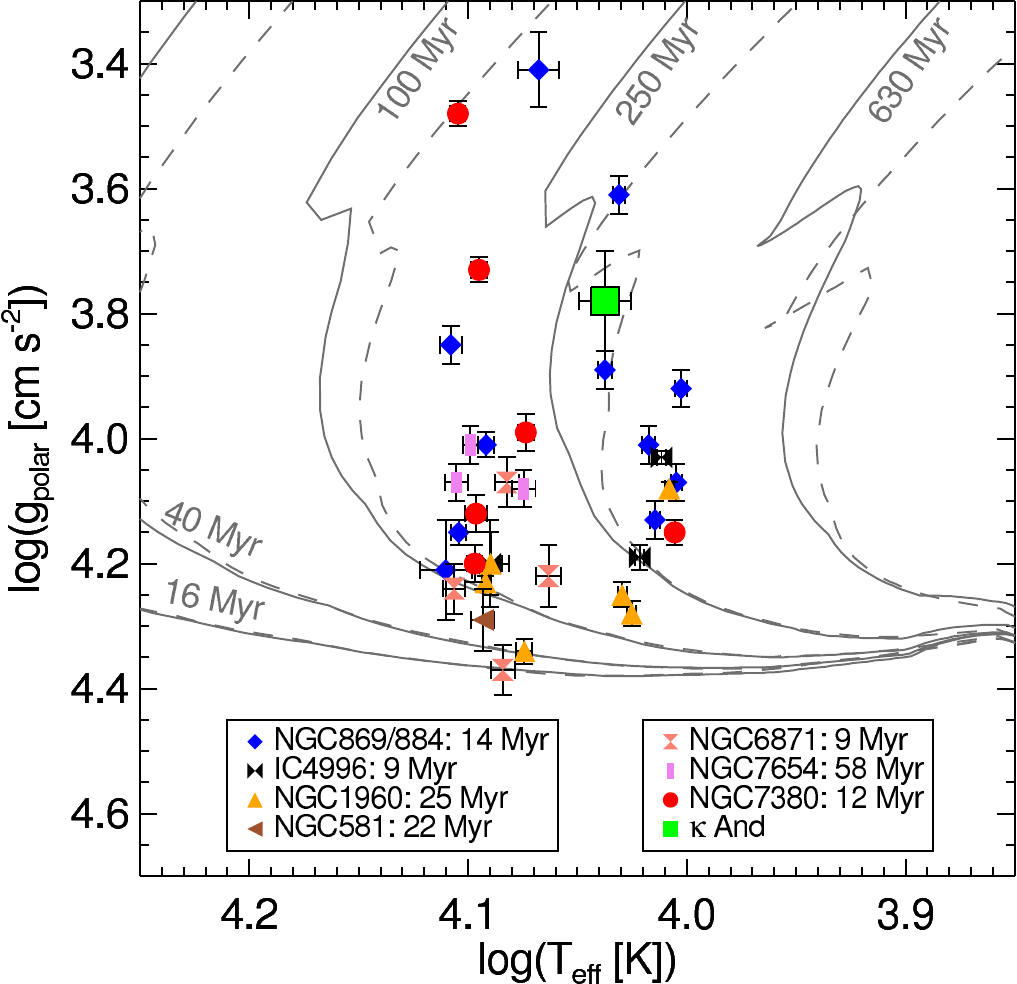} 
     \caption{Position of $\kappa$ And A in a temperature-gravity diagram compared to the models of \cite{2012A&A...537A.146E} for rotating (solid lines;  V/V$_{crit}$ =0.4 when stars reach the zero-age main-sequence) and non-rotating (dashed limes). We also show the same parameters for similarly typed members of young clusters (see legend).  The models do not reproduce the ages of the cluster members and there is significant scatter in the polar gravity estimates within a given cluster.  We interpret this as indication that these comparisons do no place a meaningful constraint on the age of $\kappa$ And.} 
         \label{Fig:rotnorot}
   \end{figure}

\cite{2013ApJ...763L..32C} suggest a possible age range of 20 - 120 Myr, and a nominal age of $\sim$30 Myr, for the system based on  the kinematics and color-magnitude diagram (CMD) position of $\kappa$ And A. Their kinematic study was based on the \cite{2013ApJ...762...88M} online tool\footnote{http://www.astro.umontreal.ca/$\sim$malo/banyan.php}, which computes probabilities of membership in the TW Hydrae, Tucana-Horologium, Columba, Carina, and Argus associations, the $\beta$ Pictoris and AB Doradus moving groups, and the field population using Bayesian methods. We found a  95.6\%, 0.7\%, and 3.8\% chance that $\kappa$ And belongs to Columba, $\beta$ Pictoris, and the field, respectively. The star has a 0\% probability to belong to the remaining groups. The same tool applied to HR8799 yields 98.1\%, 0.7\%, and 1.1\%  probability to Columba, $\beta$ Pictoris, and the field. Recently, \cite{2012ApJ...761...57B} put independent constraints on the age of HR8799 which are consistent with the high probability of Columba membership derived using this tool. One might argue that the \cite{2013ApJ...762...88M} Bayesian analysis tool assumes $\kappa$ And is a member of Columba in the priors of the calculation.  This will artificially inflate the probability of group membership since the star's kinematics are partly being used to define those of the group.  An independent analysis of the $\kappa$ And's UVW space velocities shows that they are consistent with other proposed group members with measured parallaxes at the $<$2$\sigma$ level.  $\kappa$ And's X and Z Galactic distances are also completely consistent with those of the previously mentioned members with well constrained distances.   The star's Galactic Y distance (46.5 pc) however, falls above the mean value for Columba's bona fide members \citep[-26.3 pc; using Table 3 of ][and removing $\kappa$ And from the sample]{2013ApJ...762...88M}.  While discrepant from the group members originally proposed in \cite{2008hsf2.book..757T}, this is consistent with other proposed Columba members from \cite{2011ApJ...732...61Z} that lie at northern declinations. Additionally, the dispersion in Y values for proposed members is large (26.5 pc) and might be underestimated since few surveys have searched for new members in the north.  Thus, there are several lines of evidence that support $\kappa$ And's kinematic membership in the proposed Columba association. A full kinematic traceback study of proposed Columba group members may shed more light on $\kappa$ And's reliability as a member and the past history of the association as a whole \citep[e.g.][]{2007MNRAS.377..441O, 2007ApJS..169..105M, 2013ApJ...762..118W}. Since consistent kinematics are a necessary but not sufficient criteria for moving group membership, we investigate the age of $\kappa$ And using several different methods that are independent of its kinematics.  

The existence of excess IR flux in the spectral energy distribution (SED) of $\kappa$ And A is indication of remnant, circumstellar material (a debris disc) and may place constraints on its age \citep[see discussions in][]{2008ARA&A..46..339W}.  To construct $\kappa$ And A's SED, we used the Virtual Observatory (VO) SED Analyzer \citep[VOSA, ][]{2008A&A...492..277B}\footnote{http://svo2.cab.inta-csic.es/theory/vosa/}.  The tool allows the user to use both publicly available and user provided data to construct the SED of a source and fit it with their choice of model.  We queried photometric catalogs available through the VO to compile a complete SED of $\kappa$ And A.  We recovered data from $\sim$0.13 to 100 $\mu$m from the following sources:  the International Ultraviolet Explorer \citep[IUE,][]{1978Natur.275..372B}, A catalogue of compiled UBV photometry \citep{1994cmud.book.....M}, the Tycho-2 catalogue \citep{2000A&A...355L..27H}, the 2MASS All-Sky Point Source Catalog \citep{2003tmc..book.....C}, the WISE All-Sky Data Release \citep{2012yCat.2311....0C, 2010AJ....140.1868W}, the AKARI/IRC mid-IR all-sky Survey \citep{2010A&A...514A...1I}, and the IRAS Catalog of Point Sources, Version 2.0 \citep{1988iras....7.....H}. Not all of the recovered data were useful for constructing the SED. As mentioned in Section 2.1, the 2MASS photometry of $\kappa$ And A is saturated, we therefore replaced it with our new determinations of the $JHK_s$ photometry.  The WISE W1 and W2 data were also beyond the saturation limit listed in the \emph{Explanatory Supplement}\footnote{http://wise2.ipac.caltech.edu/docs/release/allsky/expsup/}, so we did not include these points in the construction of the SED.  The IRAS 25, 60, and 100 $\mu$m data are listed as upper limits, thus, they were also not used.  To supplement the SED, we also checked for additional mid-IR data in the $Spitzer$ and $Herschel$ archives (not queried by the VO).  The star was not observed by either telescope. We therefore used the VOSA to perform a $\chi^2$ fit of an ATLAS9 model \citep{1997A&A...318..841C} to the remaining reliable photometric data.  The SED fit reveals no significant excess above the expected photospheric flux in an 11000 K model out to 22 $\mu$m.  Unfortunately, this imposes no constraint on the age of $\kappa$ And A.  \citet{2008ARA&A..46..339W} shows the evolution of 24 and 70 $\mu$m excess in A-type stars (a reasonable proxy for $\kappa$ And's B9 type) with ages up to 800 Myr (his Figure 6).  The fraction of stars with measured excess is a function of age, however more than half of the stars observed at 24 $\mu$m have no detectable excess.  Even at 70 $\mu$m, not all of the young targets presented in \cite{2008ARA&A..46..339W} exhibit excesses.  Thus, longer wavelength observations ($>$22 $\mu$m) may yet reveal IR excess in the SED of $\kappa$ And A, but the absence of excess in the current data does not provide useful information regarding its age.

$\kappa$ And is the earliest-type (B9) proposed member of the Columba association.  The ages of later-type stars in the association are well constrained by a combination of CMD positions and lithium depletion. These diagnostics indicate an association age of $\sim$30 Myr \cite{2008hsf2.book..757T}. Age determinations like lithium depletion are not applicable to young, early-type stars. Therefore, to constrain $\kappa$ And's age independently of its kinematics, we rely on HR diagrams and model comparisons.  We first place $\kappa$ And A in a $\mathrm{M_{v}}$ vs. B-V CMD to critically compare its position to those of early-type members of several young, open clusters with well defined ages (see Figure \ref{Fig:CMDJosh} and Appendix \ref{AppendixA}). $\kappa$ And's CMD position is consistent with similarly typed members of the Scorpius-Centaurus subgroups \citep[ScoCen, 11-17 Myr, ][]{2012ApJ...756..133C} and the Pleiades \citep[$130\pm20$ Myr,][]{2004ApJ...614..386B}.  There are no members of the IC 2391 cluster \citep[30-50 Myr,][]{2004ApJ...614..386B, 2013MNRAS.431.1005D} with CMD placement in the immediate vicinity of $\kappa$ And, however the approximate shape of the cluster sequence would place them close.  $\kappa$ And A is clearly younger than the $\sim$500 Myr Ursa Majoris moving group \citep{2003AJ....125.1980K}, where the earliest type stars are much more evolved.  The other proposed members of young moving groups \citep{2013ApJ...762...88M} occupy small regions of color space and exhibit significant scatter in the CMD, therefore comparison to them cannot effectively constrain the star's age.  The sample of ScoCen stars is also scattered. This may be partly attributed to large distance uncertainties. This aspect of the comparison sample may skew the interpretation of $\kappa$ And's age to younger values. Thus, while the CMD position of $\kappa$ And is consistent with populations of stars at several different ages, we infer that the star's placement is most consistent with the Pleiades.  We also compare the CMD positions' of the stars to the rotating models (V/V$_{crit}$=0.4 at the zero-age main-sequence) of \citet{2012A&A...537A.146E}.  The models reproduce the distribution of A-type stars on the main-sequence, but, they deviate from the known ages of stars that are beginning to evolve to redder colors.  For example, B-type IC 2391 and Pleiades members are approximately coincident with the 100-160 Myr and 160-250 Myr isochrones, respectively.   The age of $\kappa$ And is predicted to be $\lesssim$250 Myr. However, since member ages of well defined clusters are overestimated by the models, we do not use them to place limits on the age of $\kappa$ And.  Rather, we use the comparison to the empirical Pleiades sequence to suggest an upper age limit of $\sim$150 Myr for the system.


We also made new estimates of the atmospheric parameters of $\kappa$ And A (see the details in Appendix~\ref{AppendixB}) and compared these values to the tracks of \cite{2012A&A...537A.146E} in both the rotating and non-rotating cases. Figure~\ref{Fig:rotnorot} shows our measured polar surface gravity and effective temperature for the star (see Table~\ref{tab:sum}).  The polar surface gravity is a correction to the measured surface gravity that accounts for the rapid rotation of $\kappa$ And \citep[$v$sin$i$=130-190 km s$^{-1}$,][]{2006ApJ...648..591H,2000AcA....50..509G, 2002ApJ...573..359A,  2005AJ....129.1642F}.  We also plot the same parameters for samples of mid to late-B type stars in other young clusters with ages $\sim$10-60 Myr \citep{2010ApJ...722..605H, 2012AJ....144..158M}.  The models predict an age $\gtrsim$250 Myr for $\kappa$ And. However, they systematically overestimate the ages of all but a few of the cluster members.  There is also large scatter in the estimated polar gravity of stars within the same cluster.  We interpret these features as indication that placement in this kind of diagram does not provide a meaningful constraint on the age of $\kappa$ And.  The figure also illustrates the inherent difficulty in accurately measuring atmospheric parameters of young, early type stars. 

 \begin{figure*}
   \centering
   \includegraphics[width=\textwidth]{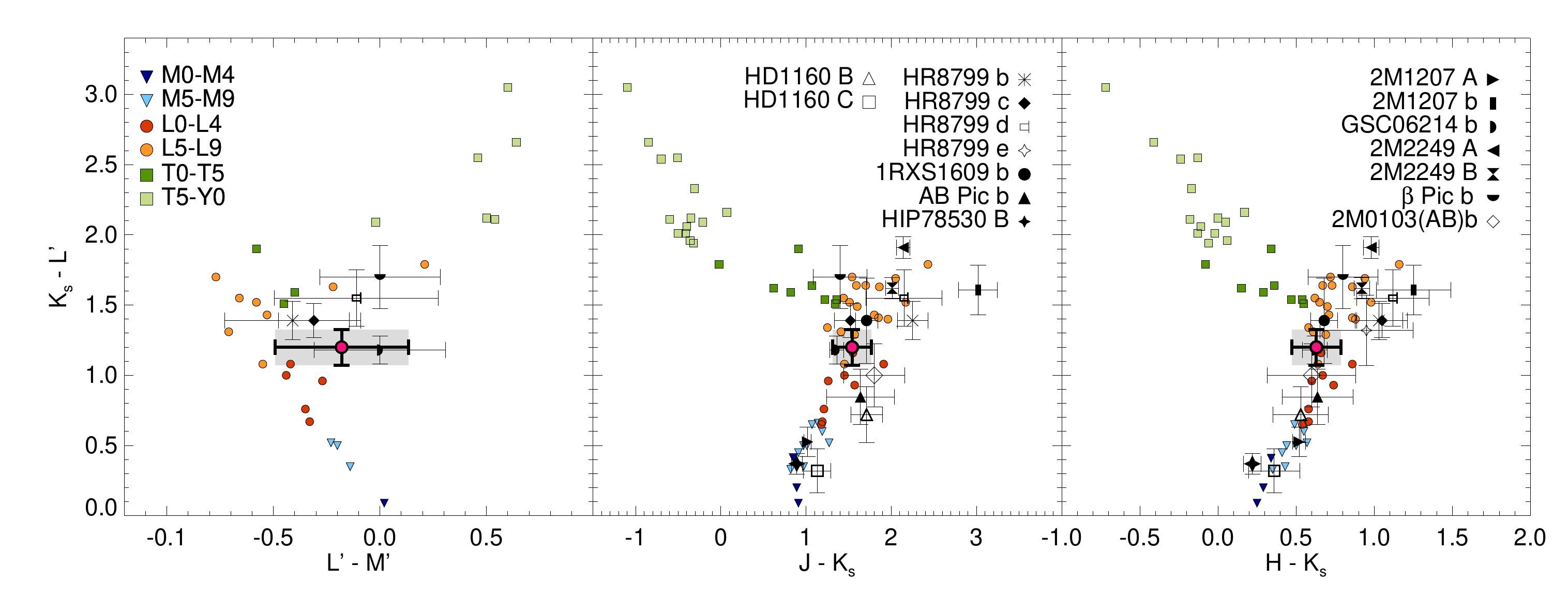}
      \caption{$\kappa$ And b (magenta dot) in color-color diagrams compared to known MLTY field dwarfs and brown-dwarfs, and young directly imaged  companions. The photometry of young companions is referenced in \cite{2013arXiv1302.1160B, 2013arXiv1306.3709B}  (and ref. therein), \cite{2012ApJ...750...53N}, and \cite{2013arXiv1303.4525D}. The photometry of MLTY field dwarfs is taken from \cite{2010ApJ...710.1627L, 2013ApJ...763..130L}.}
         \label{Fig:ccdiag}
   \end{figure*}

The age suggested by the empirical CMD is moderately older than that suggested by Columba moving group membership \citep[$30^{+20}_{-10}$ Myr,][]{2010Natur.468.1080M} and used by \cite{2013ApJ...763L..32C}.  A possible explanation that could reconcile the age estimated following these two approaches is unresolved binarity of the star.  Future monitoring of the radial velocity or observations at higher angular resolution (i.e. sparse aperture masking) may be able to constrain this hypothesis. Our high resolution spectrum does not resolve $\kappa$ And as a tight binary. However, if it were a binary, and the period were long enough, the radial velocity amplitude between resolved component lines could be overshadowed by rotational broadening in the spectrum.  Another possible explanation is large intrinsic scatter in the observed parameters of early-type stars due to the effects of inclination and rapid rotation.  When a star rotates rapidly, centrifugal force leads to a deformation of the photosphere, and thus a surface gravity gradient, between the equator and poles.  This gives rise to a temperature gradient across the observed stellar surface where the equator is cooler than the pole.  This is known as ``gravity darkening" \citep[][and references therein]{1924MNRAS..84..684V, 2006ApJ...648..591H}. Consequently, the measured temperature, surface gravity, and other observables are dependent on the viewing angle of the rotation axis with respect to the observer.  The end result of these intrinsic effects in early-type stars is that positions in diagrams comparing observed parameters (e.g. temperature-gravity, temperature-luminosity, color-magnitude) are degenerate in mass, age, and rotation.

Evidence of these effects has been directly observed in interferometric and spectroscopic measurements of the B7V/B8IV star Regulus \citep[$\alpha$ Leo,][]{2005ApJ...628..439M, 2011ApJ...732...68C}. The authors find the star is observed edge on and has $v$sin$i\approx$320 km s$^{-1}$.  This combination of rotation and inclination angle lead to an observed $>$3000 K temperature gradient between the pole and equator.  As a result, the true luminosity of the star is larger than that estimated from photometry.  Once the effects of rotation and inclination are taken into account, the HR diagram age of Regulus is reduced by nearly 100 Myr \citep{2011ApJ...732...68C}.  These results may help to reconcile the previously estimated age difference between Regulus and its $\sim$176\farcs~K2V companion.  \cite{2001A&A...379..162G} estimate ages for the primary and secondary of 150 Myr and 30-50 Myr respectively from model luminosity-temperature diagrams. \cite{2008ApJ...682L.117G} also discovered a low-mass, short period ($\sim$40 days) companion to Regulus using long term radial velocity monitoring.  They propose the companion is either a white-dwarf or an M dwarf.  If the companion is a white dwarf, correct age determination of the system becomes more difficult because the evolution of the progenitor must be considered \citep{2009ApJ...698..666R}.  These deep investigations of the Regulus system may shed light on the observed scatter in samples of known age in Figure~\ref{Fig:CMDJosh} and Figure~\ref{Fig:rotnorot}.  The estimated $v$sin$i$ of $\kappa$ And is not as large as that of Regulus, but the previous example highlights that rotation/inclination induced effects on the measured physical parameters may be significant and change the age interpretation.  Further examples of interferometric studies of rotating, early-type stars and discussions of how new data led to revised understandings of long studied stars can be found in \citet[][and references therein]{2012A&ARv..20...51V}.

Future observations of $\kappa$ And and its companion will provide the means to refine the system age. The proximity and luminosity of $\kappa$ And make it amenable to a full interferometric analysis where the measured radius, oblateness, and inclination can place strong constraints on its evolutionary status \citep{2012A&ARv..20...51V}. Additionally, direct spectroscopic observations of $\kappa$ And b should allow for independent age constraints via gravity sensitive features.  In the absence of further observations, we use in the subsequent analyses: 1/ An age for the $\kappa$ And system based on the proposed kinematic membership to the Columba association ($30^{+20}_{-10}$ Myr) 2/ A more conservative, age range -- $30^{+120}_{-10}$ Myr -- defined by the lower age limit from kinematics and the upper age limit from the empirical CMD.



\subsection{Comparison of $\kappa$ And b to reference objects}
\label{subsec:compemp}

  \begin{figure}
   \centering
   \includegraphics[width=\columnwidth]{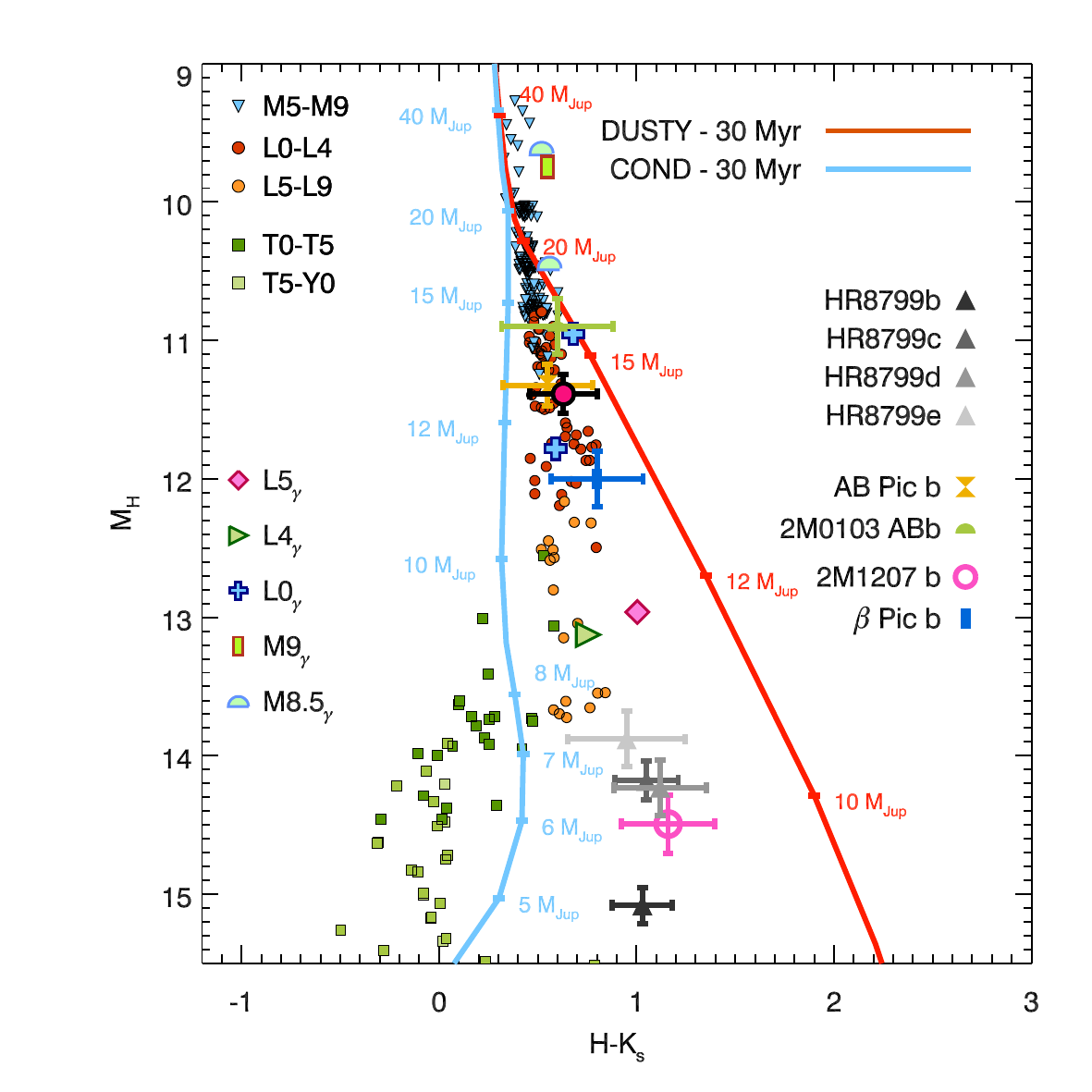}
      \caption{Position of $\kappa$ And b (magenta dot) in a color-magnitude diagram based on an improved near-infrared photometry.}
         \label{Fig:cmdiag}
   \end{figure}

  \begin{figure}
   \centering
   \includegraphics[width=\columnwidth]{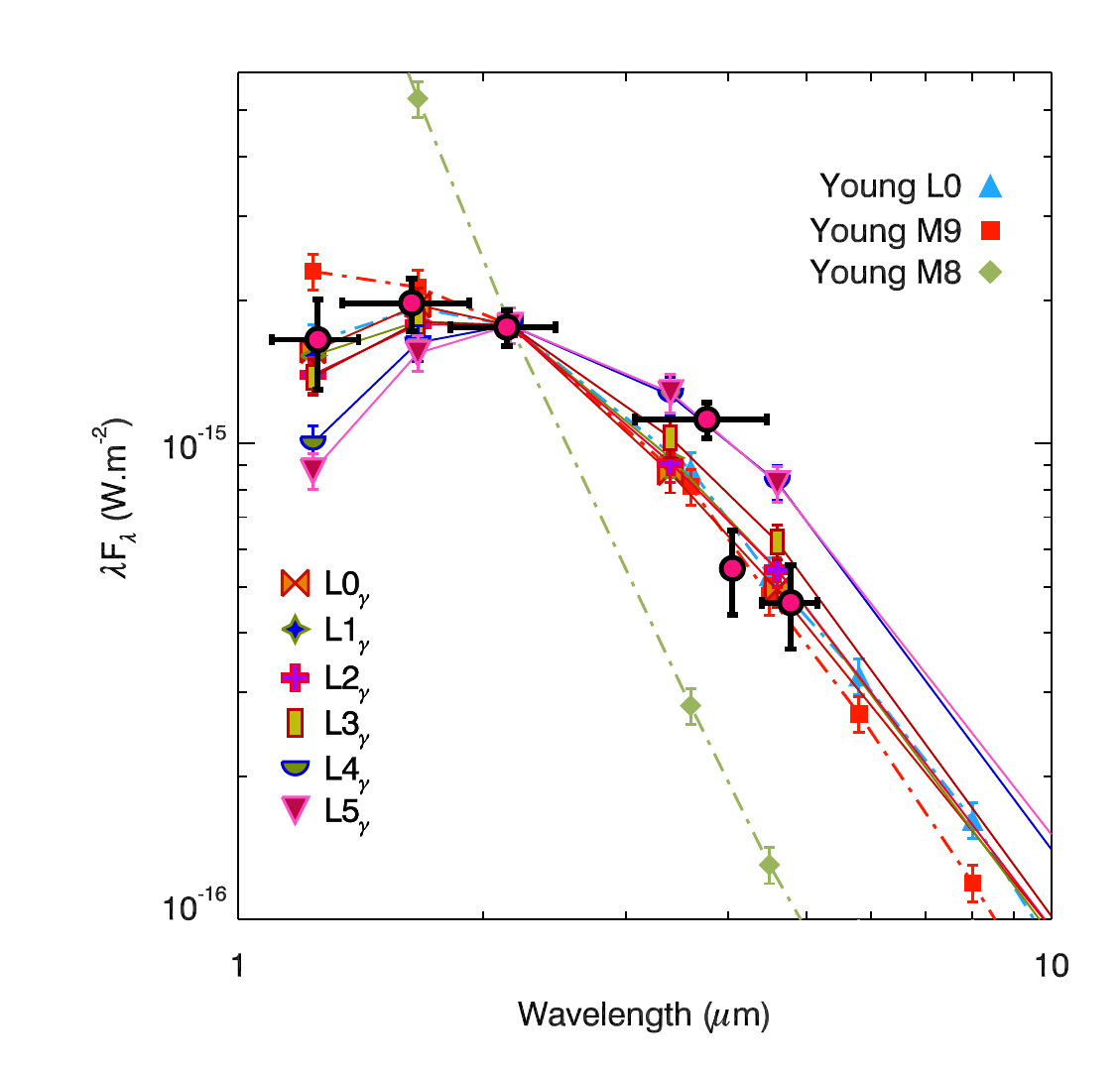}
      \caption{$\kappa$ And b  spectral energy distribution (magenta dots) compared to the mean spectral energy distribution of young M8-L0 dwarfs \citep{2010ApJS..186..111L} and of L0-L5$_{\gamma}$ dwarfs \citep{2013AJ....145....2F}.}
         \label{Fig:SEDemp}
   \end{figure}

We first used the photometry of $\kappa$ And b derived in Section \ref{section:obs} to study the companion location in color-color (Figure \ref{Fig:ccdiag}) and color-magnitude diagrams (Figure \ref{Fig:cmdiag}). The new $\mathrm{J,\:H,\:K_{s}}$-band based colors of $\kappa$ Andromeda b are close to those of other late-M to mid-L class companions (USCO CTIO 108B, M9.5, \cite{2008ApJ...673L.185B}; 2MASS J01033563-5515561(AB)b \citep{2013arXiv1303.4525D}; GSC 06214-00210 b, L0, \cite{2011ApJ...726..113I}; AB Pic b, L0, Bonnefoy et al. 2013b; $\beta$ Pictoris b, $\sim$L2, \cite{2013arXiv1302.1160B}; 1RXS J160929.1-210524b, L4, \cite{2008ApJ...689L.153L}; CD-35 2722B, L4, \cite{2011ApJ...729..139W}). These colors are also similar to those of young L0-L3  dwarfs \citep[L$_{\gamma}$ dwarfs; ][]{2005ARA&A..43..195K} listed by \cite{2013AJ....145....2F}, and of field L dwarfs with availiable 2MASS photometry listed in \cite{2006ApJ...637.1067B}, \cite{2008AJ....136.1290R}, and \cite{2012ApJS..201...19D}.  The companion  still lies close to GSC 06214-00210 b  and to L4-L6 dwarfs in $\mathrm{K_{s}-L'}$ vs $\mathrm{J-K_{s}}$ and  $\mathrm{K_{s}-L'}$ vs $\mathrm{H-K_{s}}$ diagrams.  The companion colors are intermediate between those of 1RXS J160929.1-210524b and AB Pic b in these diagrams. $\kappa$ And falls intermediate between the sequence of field dwarfs and of the 1-3 Myr old companion GSC 06214-00210 b in a $\mathrm{K_{s}-L'}$ vs  $\mathrm{L'-M'}$ diagram. GSC 06214-00210 b is surrounded by circum(sub)stellar material, and shows signs of ongoing accretion \citep{2011ApJ...743..148B, 2013ApJ...767...31B}. Therefore, the location of $\kappa$ And b in this diagram suggests that it has an intermediate age between GSC 06214-00210 b and the population of field objects, or that it does not exhibit excess emission as strong as for GSC 06214-00210 b.

 $\kappa$ And b falls on the sequence of L0-L4 field dwarfs in color-magnitude diagrams.  The two $\mathrm{L0_{\gamma}}$ objects of \cite{2013AJ....145....2F} with trigonometric parallaxes bracket the companion photometry. This is also self-consistent with the good match to AB Pic b \citep[classified as L0 following the good match with the young L0$\gamma$ dwarf 2MASS J01415823-4633574; ][]{2006ApJ...639.1120K, 2010A&A...512A..52B}.  The companion falls above the 12$^{+8}_{-4}$ Myr old exoplanet $\beta$ Pictoris b, and at an intermediate location between those of other known 30 Myr old  companions to 2MASS J01033563-5515561AB and HR 8799. This is consistent with $\kappa$ And being more massive than the exoplanets HR 8799bcde \citep[$\leq$7-10-10-10 $\mathrm{M_{Jup}}$;][]{2010Natur.468.1080M, 2011ApJ...729..128C, 2012ApJ...755...38S} and less massive than 2MASS J01033563-5515561(AB)b \citep[12-14 $\mathrm{M_{Jup}}$;][]{2013arXiv1303.4525D},  if we assume the system is a member of the Columba moving group (see Section \ref{section:systage}). 

   \begin{figure*}
   \centering
   \includegraphics[width=\textwidth]{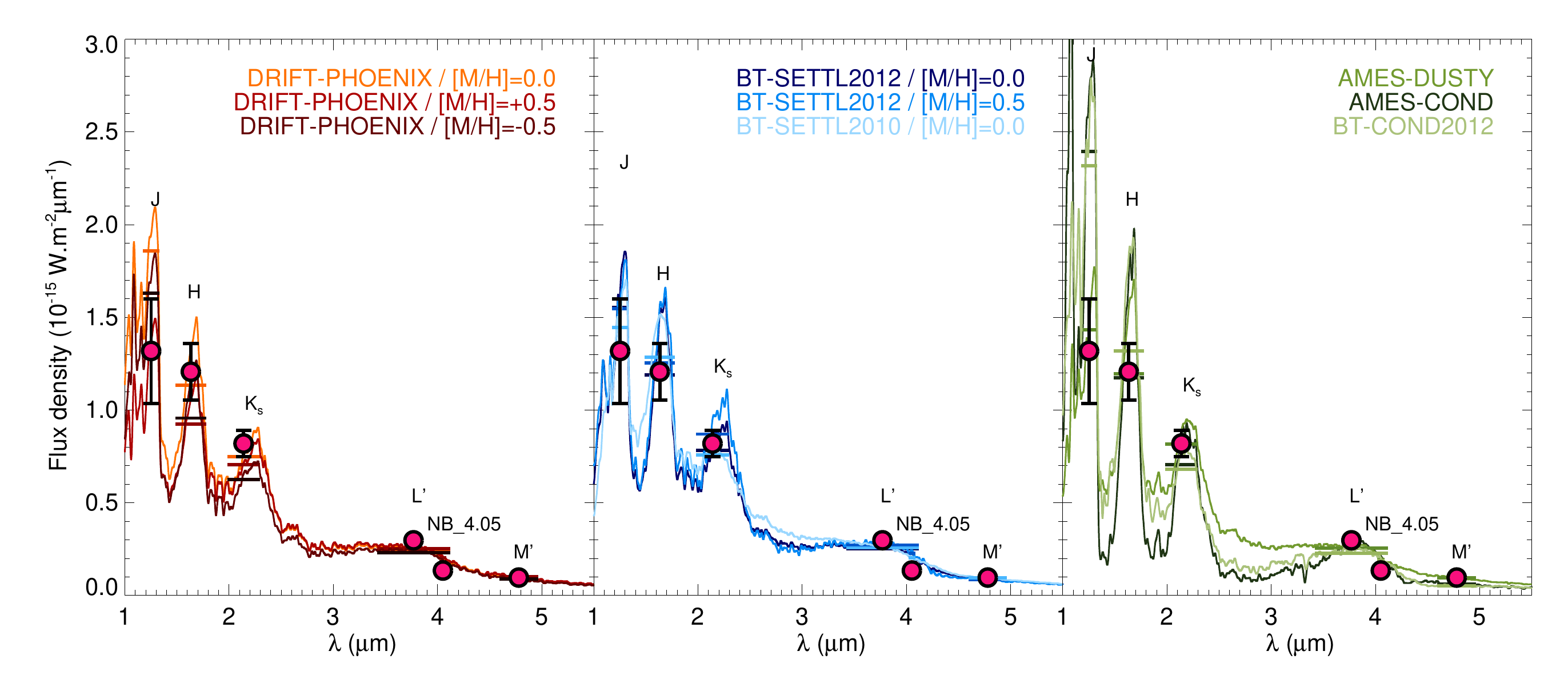}
      \caption{Best fitted synthetic flux (horizontal bars) to the spectral-energy distribution of $\kappa$ And b (magenta dots) considering the DRIFT-PHOENIX models with three different metallicities (left panel), the BT-SETTL models (middle panel), the AMES-DUSTY, AMES-COND, and BT-COND models (right panel). The corresponding synthetic spectra are overlaid in each panel.} 
         \label{Fig:synthspec}
   \end{figure*}

We  finally analyze the companion SED  from the available photometry, the corresponding filter pass bands, and a flux calibrated spectrum of Vega \citep{2007ASPC..364..315B}, as shown in Figure \ref{Fig:SEDemp}. We also built, and report on the figure, the SED of typical young M8-L0 dwarfs using the 2MASS and Spitzer colors\footnote{We used the Gemini magnitude-flux converter for that purpose (http://www.gemini.edu/sciops/instruments/midir-resources/imaging-calibrations/fluxmagnitude-conversion).} reported in Table 13 of \cite{2010ApJS..186..111L} and normalized to the $\mathrm{K_{s}}$ band flux of $\kappa$ And b (the error on the normalization factor were  propagated). We reconstructed the SED of L$_{\gamma}$ dwarfs taking the mean of the photometry reported in Table 3 of \cite{2013AJ....145....2F} for a given spectra type, and normalizing the flux to the one of $\kappa$ And b in the $\mathrm{K_{s}}$ band . The SED of  $\kappa$ And b falls between those of typical M9 and L0 dwarfs. It is incompatible with SEDs of  L0  and L3$_{\gamma}$ dwarfs. 

In summary, our empirical analysis suggests that $\kappa$ And b  is a M9 to L3 dwarf.  We use the bolometric correction $BC_{K}$ of 3.39$\mathrm{\pm0.03}$ mag \citep{2010ApJ...714L..84T} derived for the prototype  $\mathrm{L0_{\gamma}}$ dwarf 2MASS J01415823-4633574 \citep[][and considering the $BC_{K}$ of the Taurus M9.5 member KPNO-Tau 4 and of another young L0 field dwarf 2MASS J02411151−0326587 for the error bar]{2006ApJ...639.1120K}  in order to estimate $\mathrm{Log_{10}(L/L_{\odot})=-3.76\pm0.06}$ for $\kappa$ And b. This bolometric correction remains close to the one valid for M9-L8 field dwarfs \citep[$BC_{K}=3.19-3.33$ mag;][]{2010ApJ...722..311L}. 


\subsection{Atmospheric models}
\label{subsec:atmo}

\begin{figure*}
   \centering
   \begin{tabular}{cccc}
   \includegraphics[width=5.8cm]{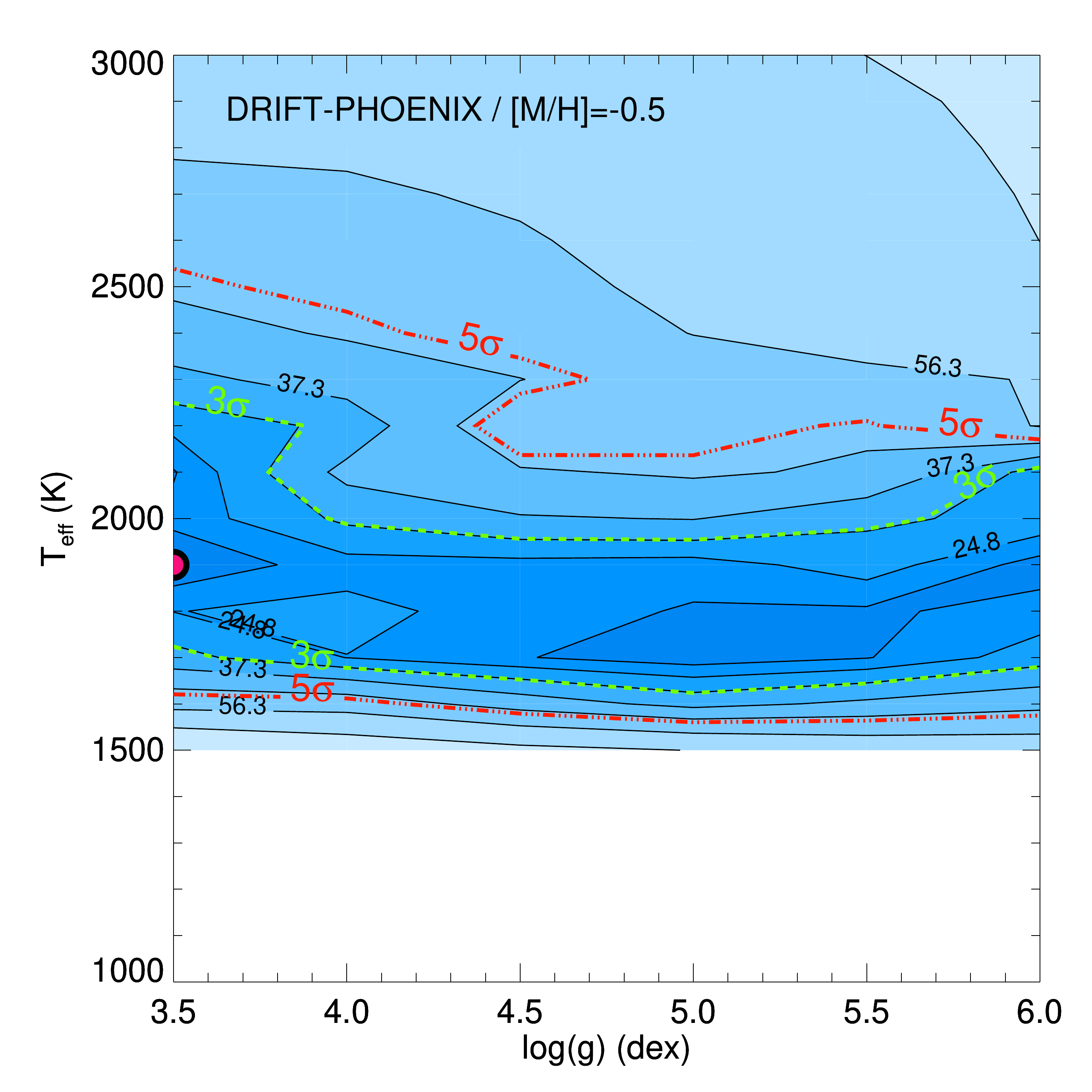} &
   \includegraphics[width=5.8cm]{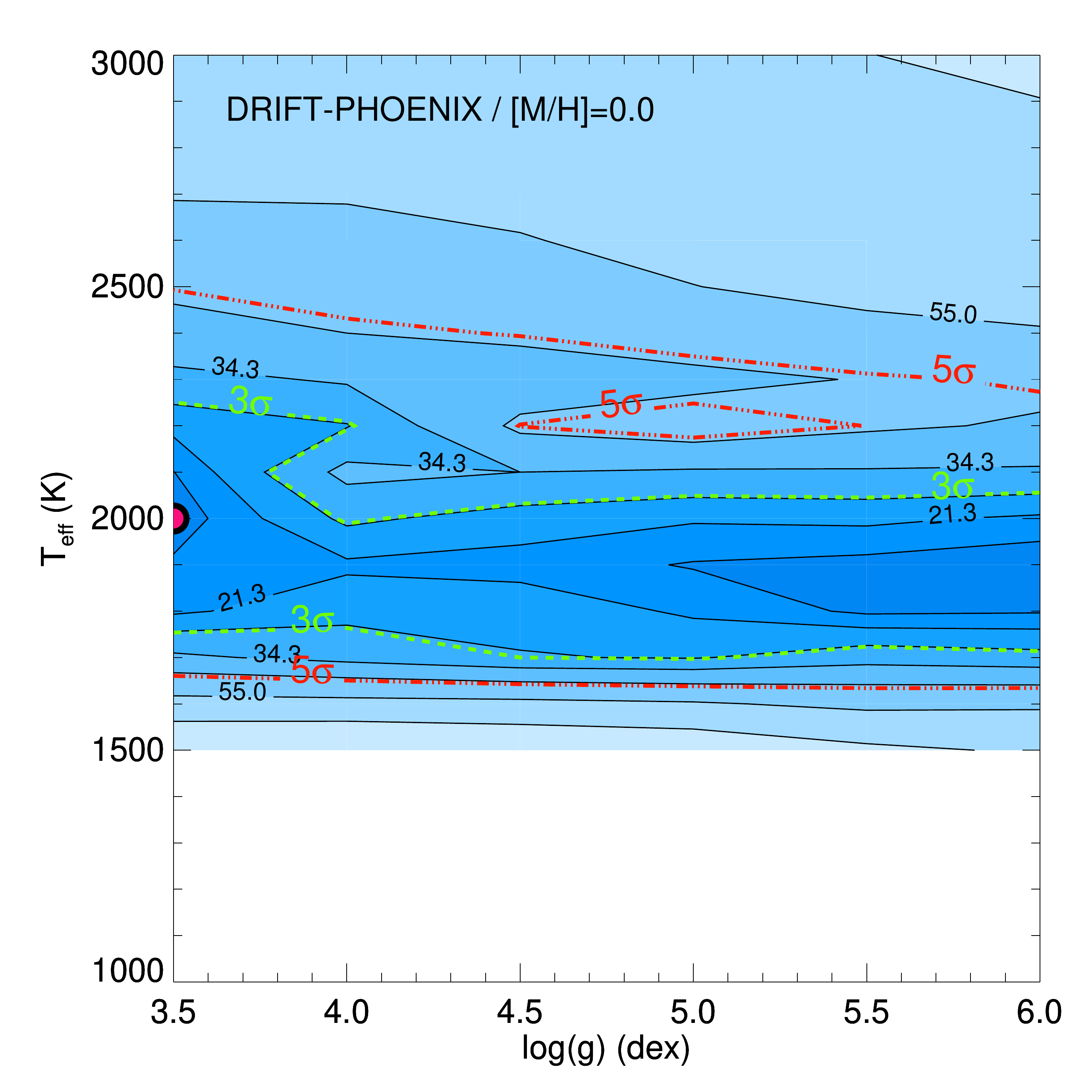} &
   \includegraphics[width=5.8cm]{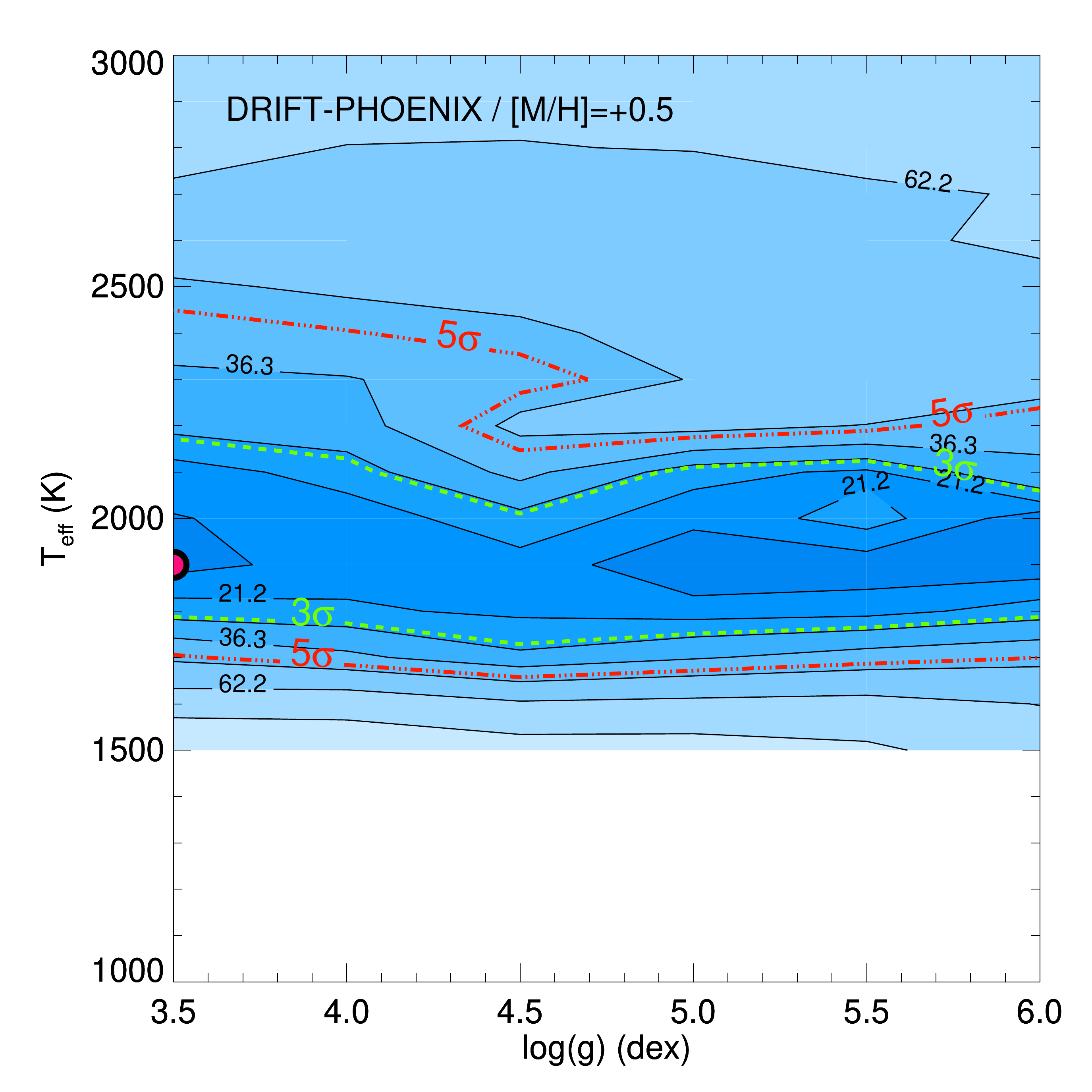} \\
   \includegraphics[width=5.8cm]{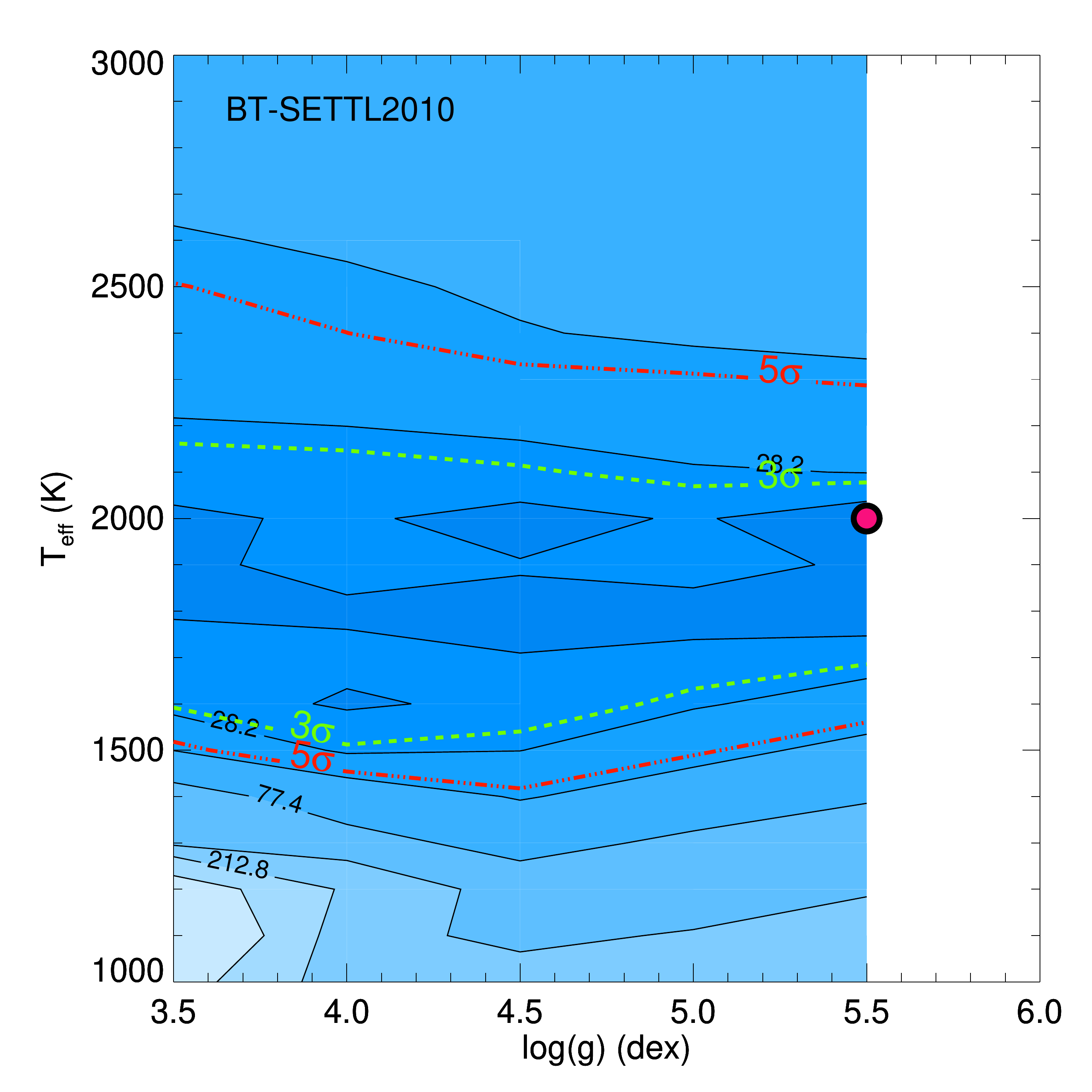} &
   \includegraphics[width=5.8cm]{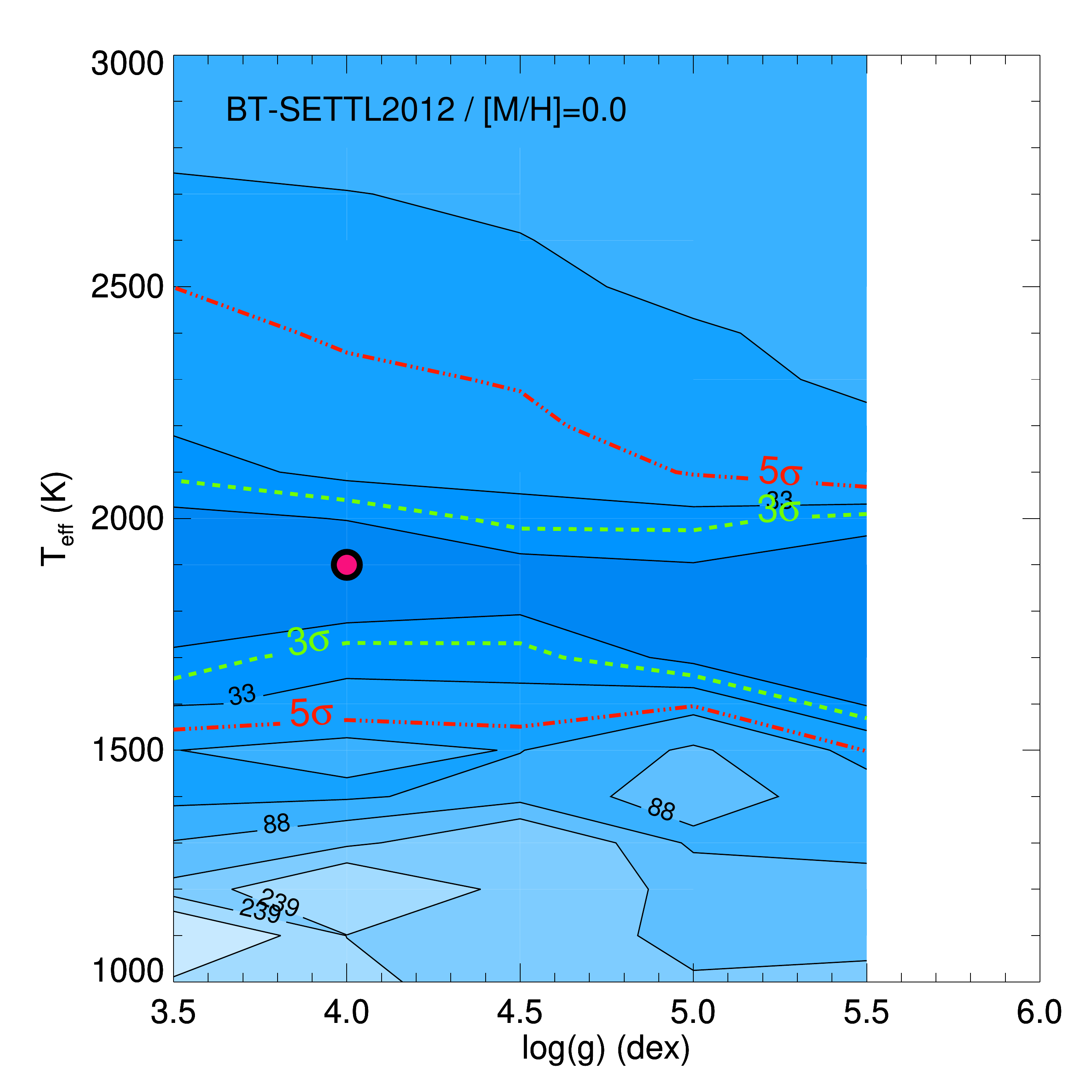} &
   \includegraphics[width=5.8cm]{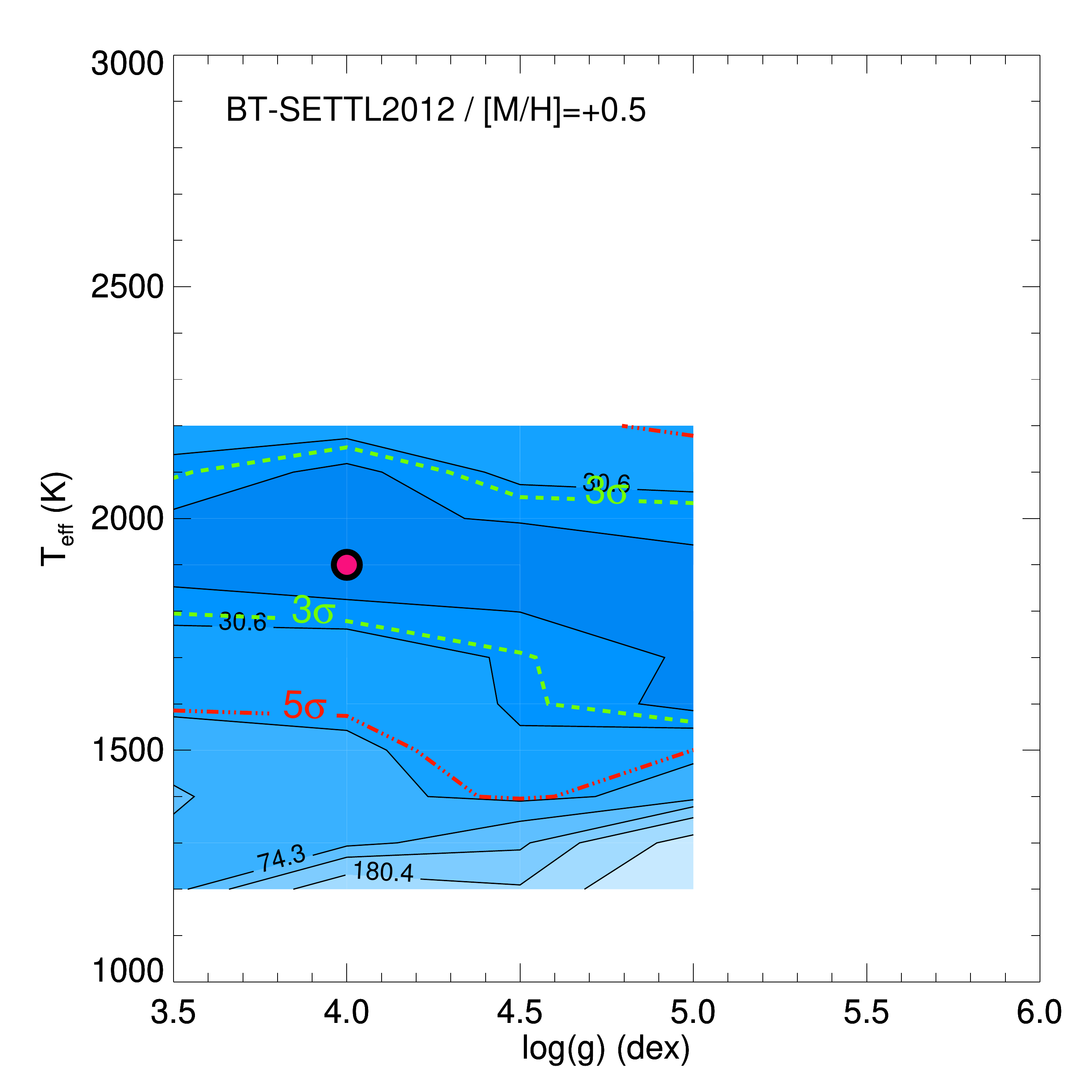} \\
   \includegraphics[width=5.8cm]{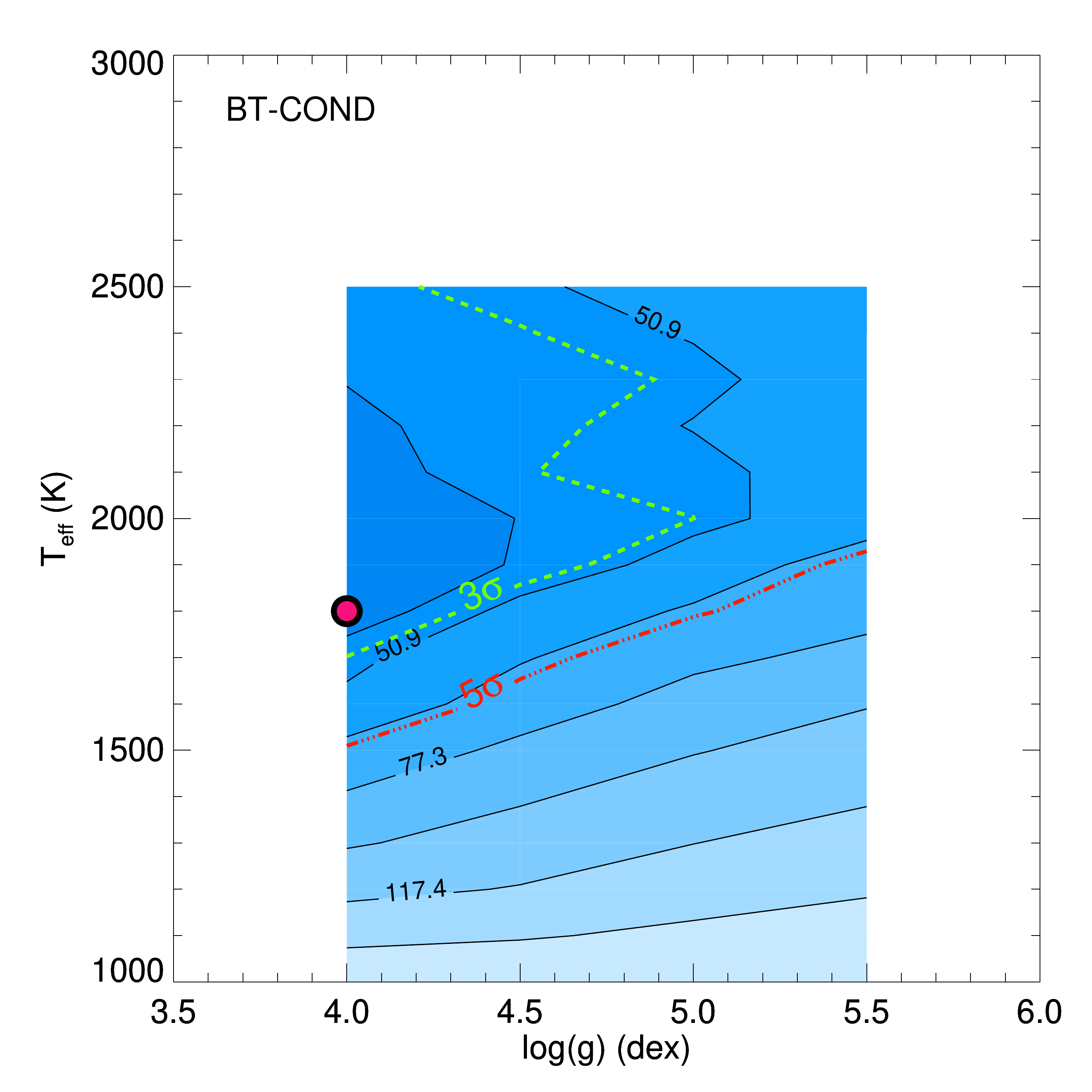} &
   \includegraphics[width=5.8cm]{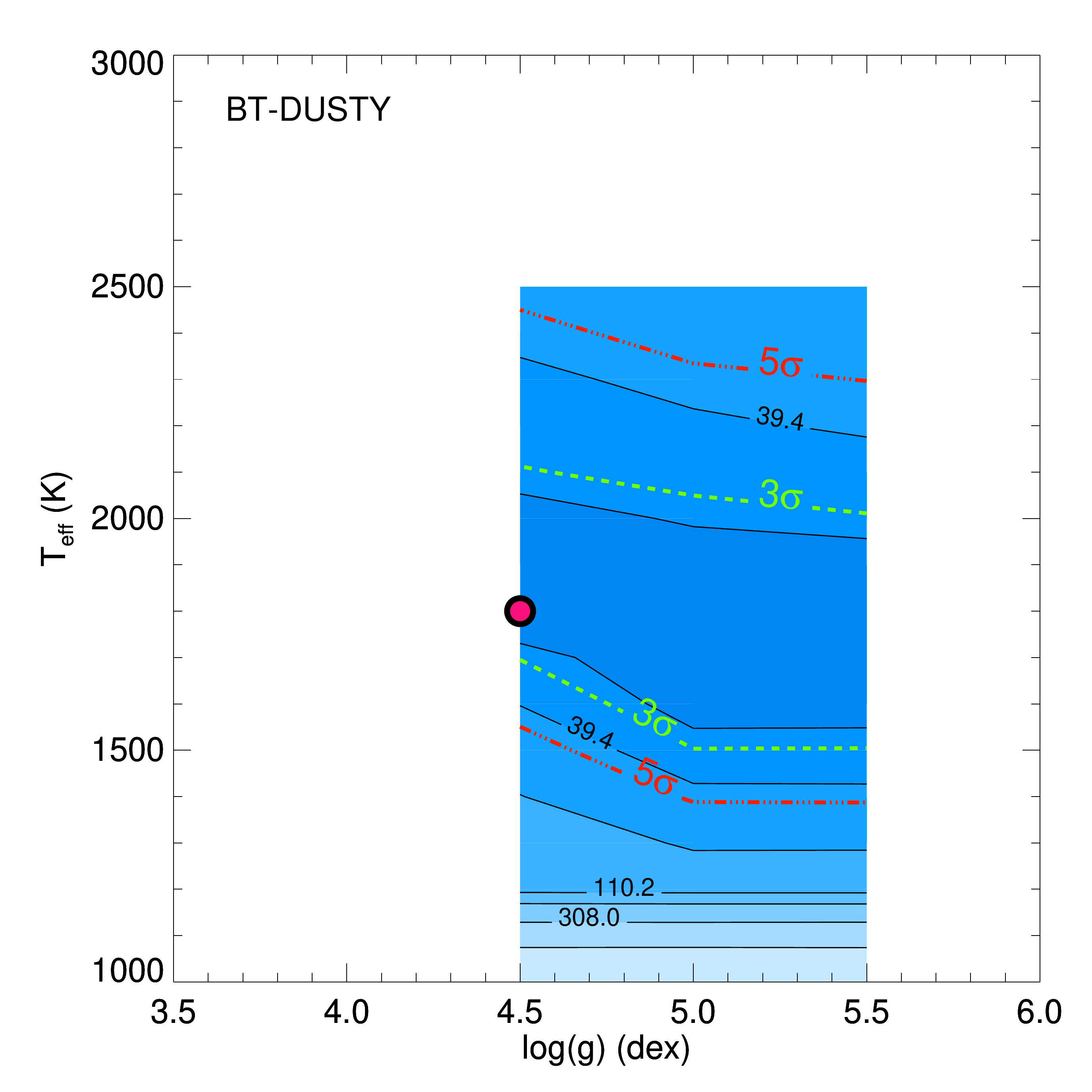} &
   \includegraphics[width=5.8cm]{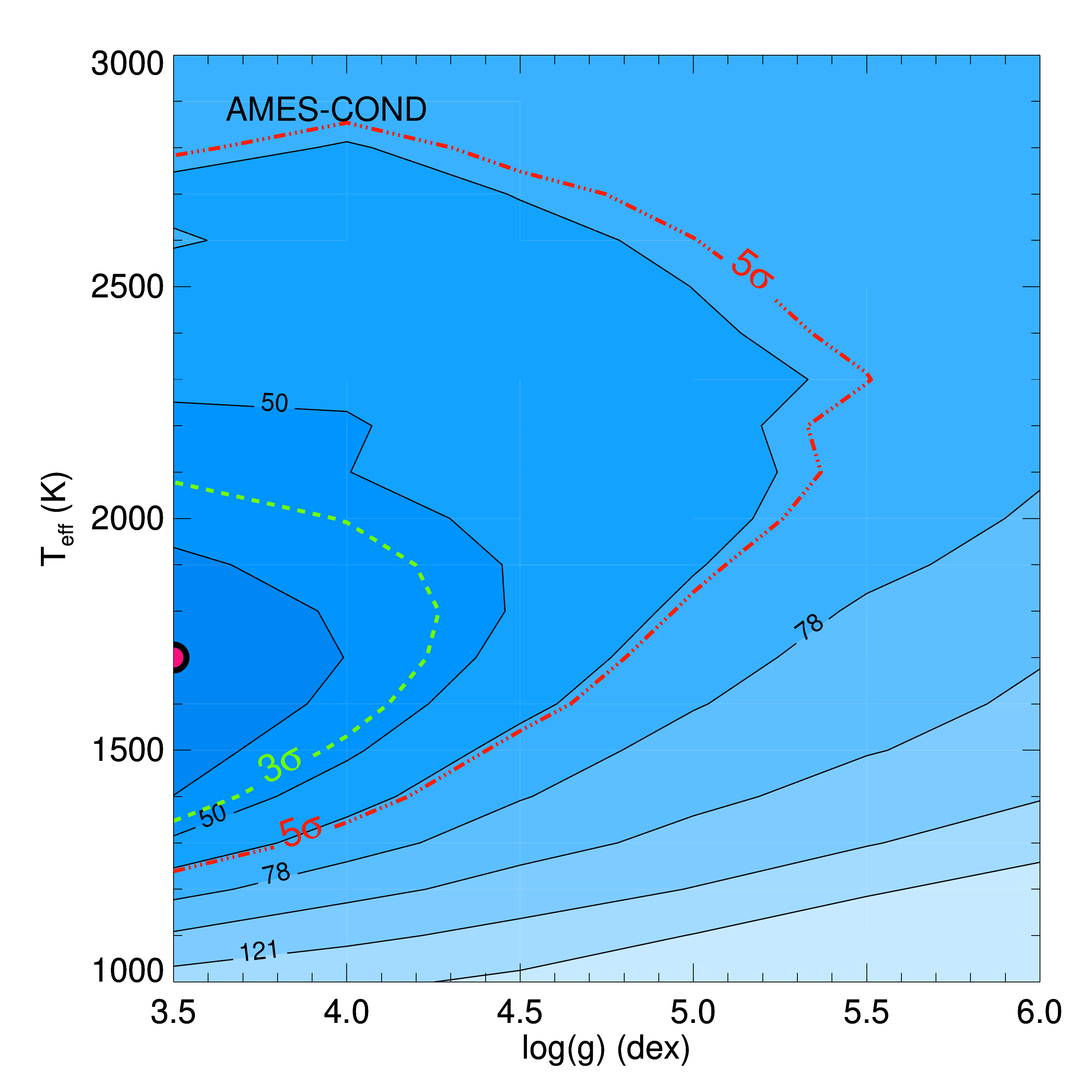} \\
  &  &  \includegraphics[width=5.8cm]{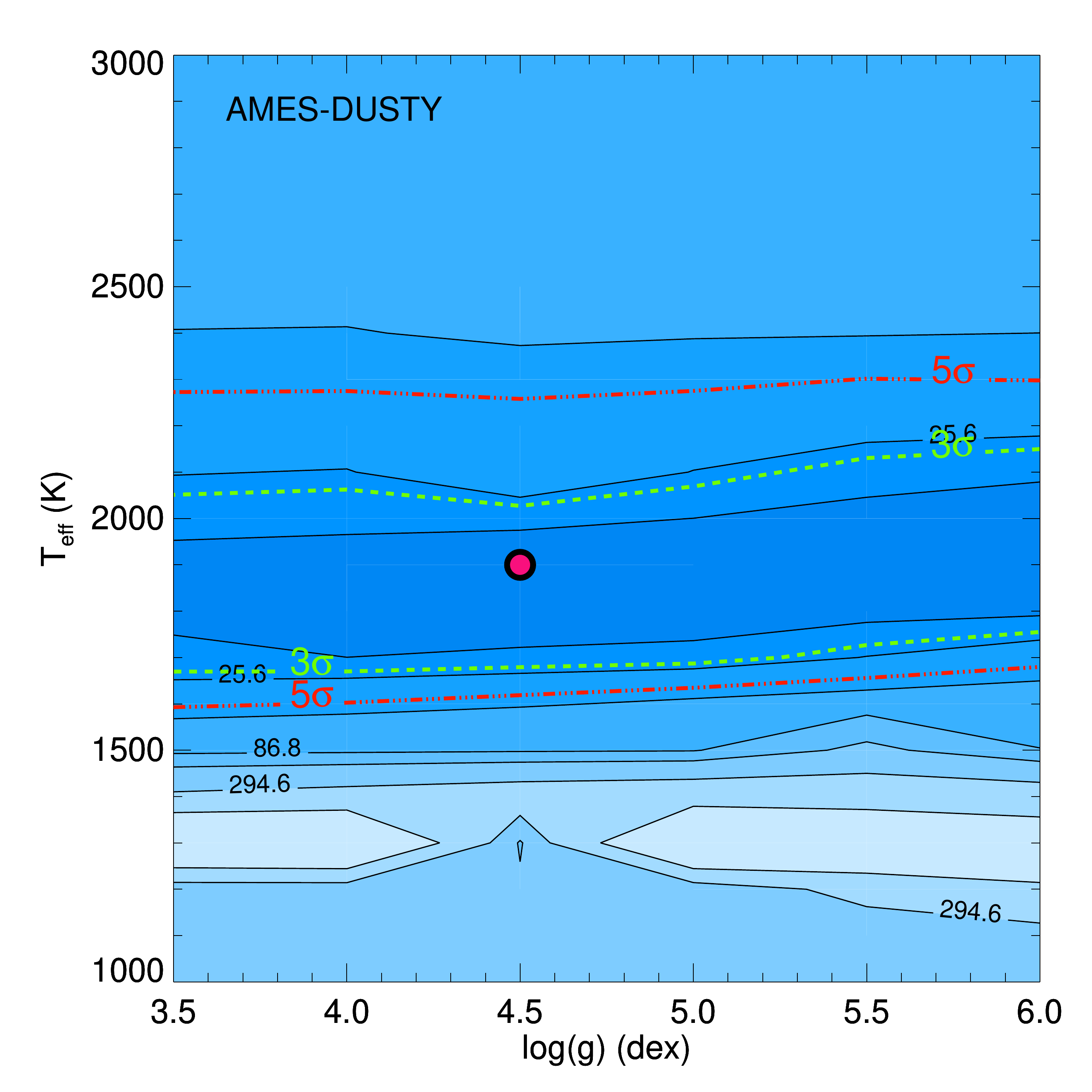}   \\
   
   \end{tabular}
      \caption{$\chi^{2}$ maps corresponding to the comparison of  the  spectral energy distribution of $\kappa$ And b to synthetic fluxes derived from atmospheric models for given log g and $\mathrm{T_{eff}}$. Minima are indicated by magenta dots. We overlay contours corresponding to 3$\sigma$ (green) and 5$\sigma$ (red) confidence levels.}
         \label{Fig:Figrefchi2}
   \end{figure*}

We compared the SED of $\kappa$ And b to synthetic fluxes generated from seven atmospheric grids (AMES-DUSTY, AMES-COND, BT-DUSTY, BT-COND, BT-Settl 2010, BT-Settl 2012, DRIFT-PHOENIX) in order to determine the atmospheric parameters of the companion ($\mathrm{T_{eff}}$, log g) and evaluate systematic errors on these parameters introduced by the models (Bonnefoy et al. 2013b, submitted). The models and the fitting procedure are described in \cite{2013arXiv1302.1160B}.  We used the 2010 release of the BT-Settl models (BT-Settl 2010), despite the models do not incorporate the up-to-date physics. Indeed, \cite{2013arXiv1306.3709B} showed that BT-Settl 2010 synthetic spectra tend to better reproduce the near-infrared (1.1-2.5 $\mu$m) spectra of young objects at the M-L transition, such as $\kappa$ And b. Results are reported in Table \ref{atmoparPHOENIX}. The best fitted synthetic fluxes are displayed in Figure \ref{Fig:synthspec}. $\chi^{2}$ maps of the fit (with 3 and 5 $\sigma$ confidence levels overlaid) are shown in Figure \ref{Fig:Figrefchi2}. 

The models successfully reproduce the companion SED for $\mathrm{T_{eff}=1900_{-200}^{+100}}$ K. This corresponds roughly to the extreme values determined for a 3$\sigma$ confidence level using the most advanced models BT-SETTL12 (Figure \ref{Fig:Figrefchi2}).  The temperature coincides with the latest range derived for the young L0 objects AB Pic b and 2MASS J01415823-4633574 from near-infrared spectra and SED fits \citep[$\mathrm{T_{eff}=1800_{-200}^{+100}}$ K; see][]{2013arXiv1306.3709B}. We also find that models with photospheric dust produce better fits to the data.  These models provide a better constraint on $\mathrm{T_{eff}}$, but not on the surface gravity  (log g=$4.5 \pm 1.0$). This was  already the case for  $\beta$ Pictoris b \citep{2013arXiv1302.1160B}. The high content in photospheric dust of $\kappa$ And b is also consistent with constraints on the spectral type derived in Section \ref{subsec:compemp}. To conclude, we note that the degenerate effects of metallicity, and surface gravity do not affect the temperature determination of the companion.

The effective temperature  and bolometric luminosity derived in Section \ref{subsec:compemp} give a semi-empirical radius estimate of $\mathrm{1.2_{-0.1}^{+0.2}\:R_{Jup}}$ for $\kappa$ And b. This radius is consistent with the value derived by adjusting synthetic fluxes to the observed companion SED exepted for the BT-DUSTY models, which temperature determination is likely biased by the limited coverage of the grid (Table \ref{atmoparPHOENIX}). The radius matches  predictions from ``hot-start" evolutionary tracks corresponding to the companion luminosity and temperatures for ages between 30 and 250 Myr. 

\begin{table}
\begin{minipage}{\columnwidth}
\caption{Best-fit atmospheric parameters for $\kappa$ And b}
\label{atmoparPHOENIX}
\centering
\renewcommand{\footnoterule}{}  
\begin{tabular}{lllll}
\hline \hline 
Atmospheric model		 		&   $\mathrm{T_{eff}}$		&	 $\mathrm{log\:g}$	& R		& $\mathrm{\chi^{2}}$	\\   			
									&				(K)					&		$\mathrm{(cm.s^{-2}}$)	&	($\mathrm{R_{Jup}}$)	\\	
\hline
AMES-Dusty				&		1900						&	4.5			&	1.25		&	7.53   \\
AMES-Cond		 		&		1700				&	3.5			&	1.47		&	32.02	\\
BT-Settl 2010				&		2000						&	5.5			&	1.18		&	10.25	\\
BT-Cond 2012				&		1800						&	4.0			&	1.34		&	33.55	\\
BT-Dusty 2012\tablefootmark{a}					&		1800		&	4.5			&	1.65		&	14.10	\\
BT-Settl 2012	 [M/H]=0.0		&		1900						&	4.0			&	1.26		&	12.05	\\
BT-Settl 2012	 [M/H]=+0.5		&		1900						&	4.0		&	1.24		&	12.59	\\
DRIFT-P. [M/H]=0.0			&		2000						&	3.5			&	1.13		&	13.29    \\
DRIFT-P. [M/H]=+0.5			&		1900			&	3.5			&	1.16		&	12.35    \\
DRIFT-P. [M/H]=-0.5			&		1900			&	3.5			&	1.16		&	16.42    \\
\hline
\end{tabular}
\end{minipage}
\tablefoot{ 
\tablefoottext{a}{Analysis limited to log g $\geq$ 4.5.} 
}
\end{table}

\subsection{The mass of $\kappa$ And b}
\label{subsec:evolm}
   We compared $\mathrm{T_{eff}}$ and the luminosity estimates derived in previous sections to predictions of evolutionary models in order to re-estimate the companion mass.  We considered the  two distinct age ranges determined in Section\ref{section:systage}.  

\subsubsection{Classical hot- and cold-start models}
\begin{figure}
   \centering
   \includegraphics[width=\columnwidth]{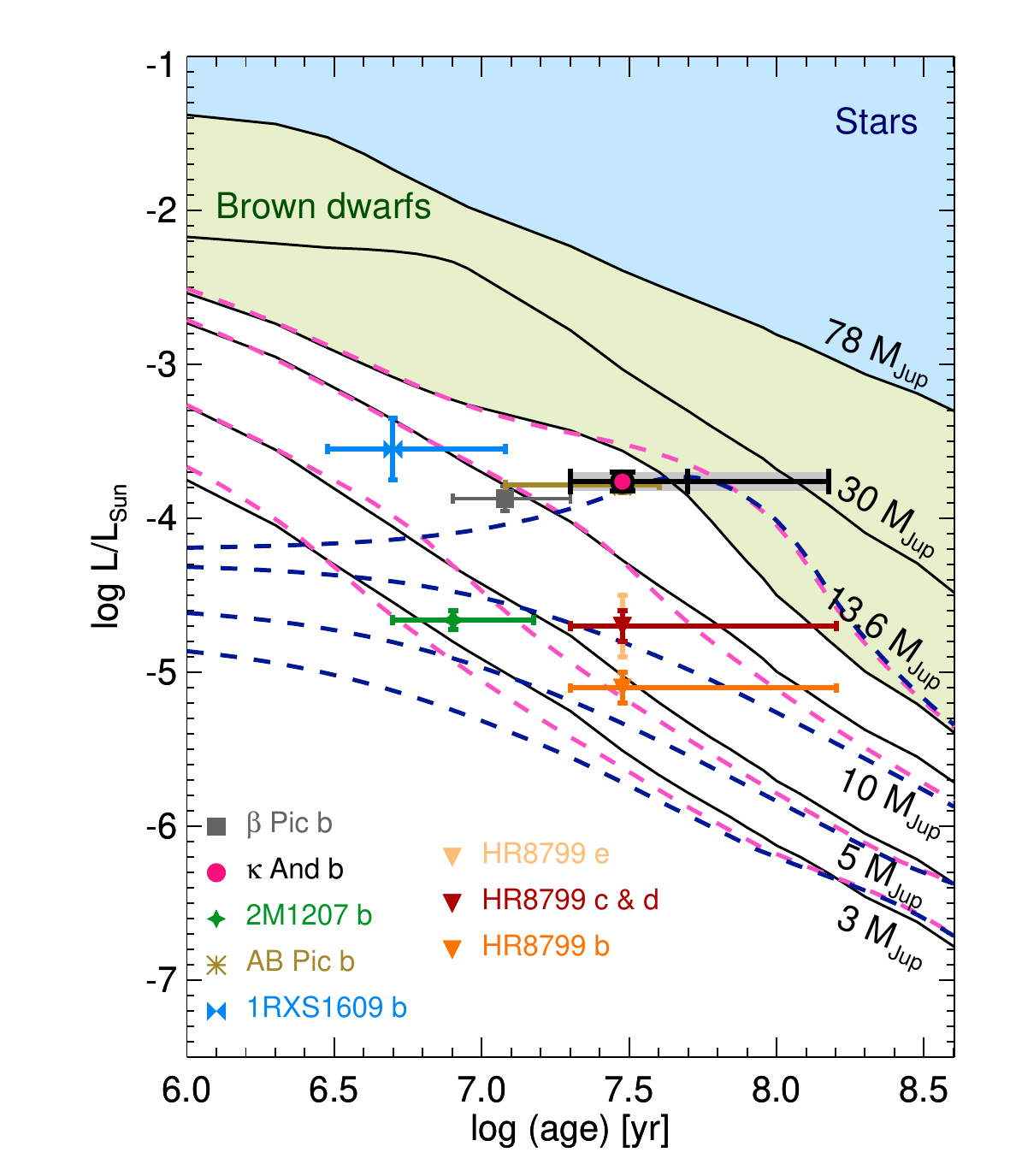}
      \caption{Evolution of the luminosity of gaseous objects predicted by the COND models (black solid line), and \cite{mc13} models with typical ``hot-start" (light pink dashed curve; 3, 5, 10, 13.6 $\mathrm{M_{Jup}}$), and ``cold-start"  initial conditions (dark blue dashed curve; 3, 5, 10, 13.6 $\mathrm{M_{Jup}}$). We overlay measured luminosity of young low mass companions. A more complete version of this figure can be found in \cite{mc13}.}
         \label{Fig:lumCOND}
   \end{figure}

        We first used the ``hot-start" models of \cite{2003A&A...402..701B} (hereafter COND03). We compare in Figure \ref{Fig:lumCOND}  masses predicted  by these models for different companions. $\kappa$ And b mass estimates are in the brown-dwarf regime if the system is 150 Myr old.
        
         We used alternatively  the ``hot-start" models of  \cite{2008ApJ...683.1104F} (FM08), and \cite{sb12} (SB12). FM08 and SB12 models explore the impact of chemical enrichment related to the formation process (1x and 5x solar for the FM08 models, 1x and 3x solar for the SB12 models) on the object emergent flux and evolution.   Results are reported in Table \ref{massHS}.  These models predict masses above 10 $\mathrm{M_{Jup}}$ (maximum mass covered by these models). $\kappa$ And b's temperature and bolometric luminosity give the same estimates. 
    
   We also considered the ``cold-start"  version of FM08 and SB12 models. Model predictions do not extend to sufficient high masses (M $\geq$ 10 $\mathrm{M_{Jup}}$) to reproduce the luminosity and temperature of $\kappa$ And b.

\begin{table}
\begin{minipage}{\columnwidth}
\caption{Mass of $\kappa$ And b prediced by ``hot-start" evolutionary models}
\label{massHS}
\centering
\renewcommand{\footnoterule}{}  
\begin{tabular}{llll}
\hline \hline 
Model		 		    &    age   &       Mass from $\mathrm{T_{eff}}$		&	 Mass from $\mathrm{L/L_{\odot}}$		\\   			
						     				    &    (Myr) &		($\mathrm{M_{Jup}}$)	 	&	($\mathrm{M_{Jup}}$)	                        \\	
\hline
COND03                   &   $30^{+20}_{-10}$      &     $14^{+11}_{-2}$      &         $13\pm1$                            \\
COND03                   &   $30^{+120}_{-10}$      &     $14^{+25}_{-2}$  &         $13^{+22}_{-1}$                    \\
FM08              &   $30^{+20}_{-10}$          &       $> 10$      &         $> 10$                                         \\
SB12               &     $30^{+20}_{-10}$        &      $> 10$                 &        $> 10$                                   \\
\hline
\end{tabular}
\end{minipage}
\end{table}

\subsubsection{Warm-start models}
\label{subsub:warmstart}
The formation mechanism of $\kappa$ And b  is not known (see section \ref{subsec:form}) and, more importantly, the outcome, in terms of initial brightness, of the different formation scenarios cannot yet be predicted.   We show in Figure \ref{Fig:lumCOND}  the impact of initial conditions with two cooling curves of \cite{mc13}. The ``cold-start" curves correspond to cases with initial entropie $\mathrm{S_{init}}$ of 9.5  Boltzmann units per baryon ($\mathrm{k_{B}}$/baryon). The ``hot-start" cooling curves correspond to $\mathrm{S_{init}}$=13 $\mathrm{k_{B}}$/baryon for a mass of 3 $\mathrm{M_{Jup}}$, and 14 $\mathrm{k_{B}}$/baryon for masses from 5 to 13.6 $\mathrm{M_{Jup}}$.  It is then essential to take into consideration models with a wide range of possible entropies, including those of the ``hot-start" and ``cold-start" models. 
We examined the predictions of two sets of  ``warm-start" models \citep{sb12, mc13} for $\kappa$ And b for that reason. \\

\begin{figure}
   \centering
   \includegraphics[width=\columnwidth]{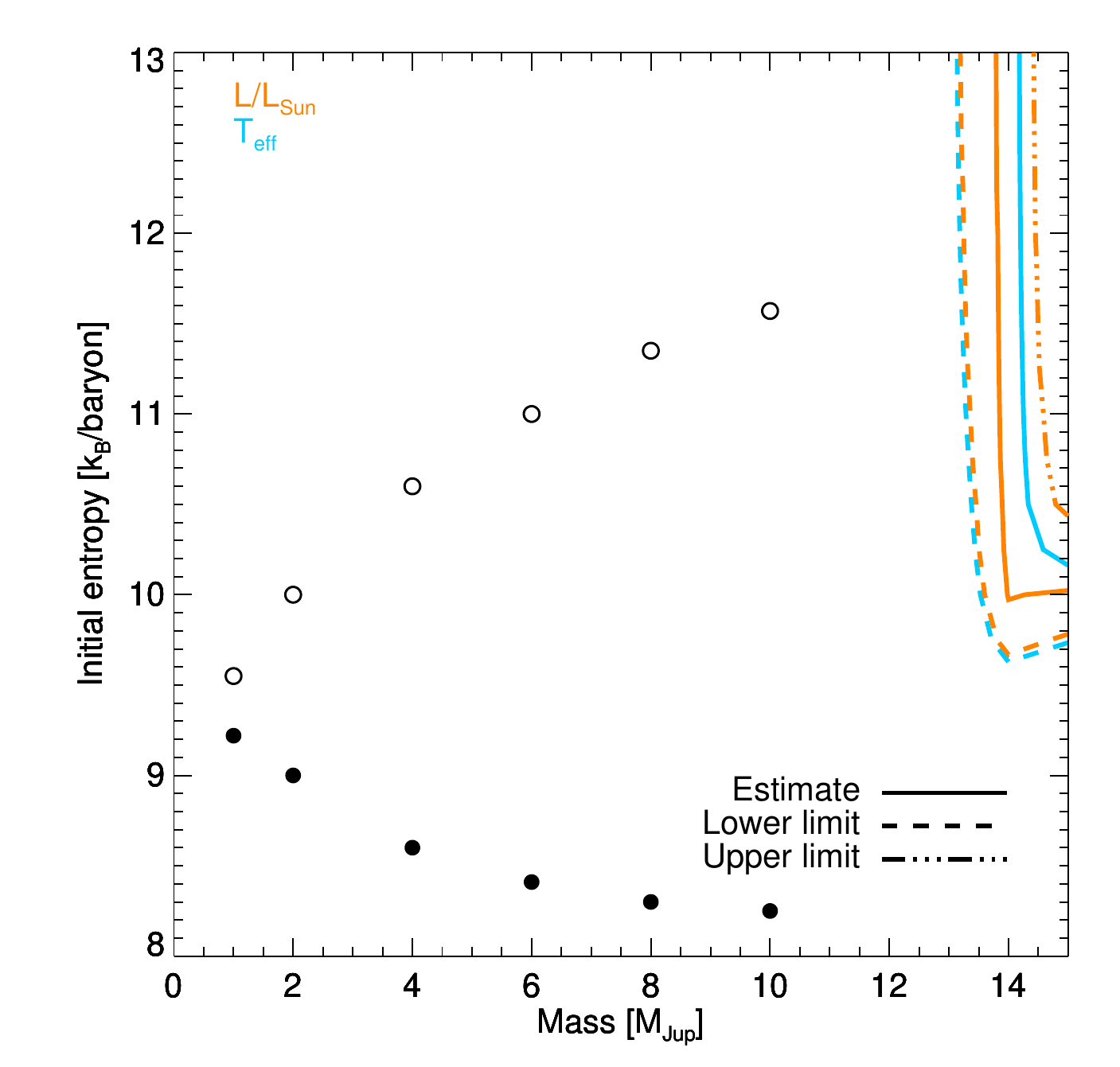}
      \caption{Predictions of the ``warm-start" evolutionary models of SB12 for $\kappa$ And b for a system at $t\approx 30^{+20}_{-10}$ Myr. The extreme values of the companion age, $\mathrm{T_{eff}}$, and luminosities define a range of masses and initial entropies lying between the dashed and dotted-dashed curves. We also overlay the initial entropies considered in ``hot-start" (open circles) and ``cold-start" (dots) models of FM08 \citep[based on][]{marl07}. }
         \label{Fig:wsSB12lumteff}
   \end{figure}

The ``warm-start"  models of SB12 consider  $\mathrm{S_{init}}$ from 8 to 13 $\mathrm{k_{B}}$/baryon, in 0.25 $\mathrm{k_{B}}$/baryon increments, and masses from 1 to 15 $\mathrm{M_{Jup}}$. The models incorporate a deuterium-burning phase \citep[][A. Burrows, priv. com.]{2011ApJ...727...57S}. Temperature and radii predictions were provided by the authors (D. Spiegel priv. com.) for this set of input parameters.  We combined them to create bolometric luminosity predictions. The luminosity and effective temperature predicted by the models tend to increase with the object mass and initial entropy, and decrease with ages. Therefore, we determined the combination of initial entropies and masses corresponding to $\kap$ And b's measured luminosity and temperature assuming a system age of $t=30$ Myr and propagating the associated uncertainties. We show the results in Figure \ref{Fig:wsSB12lumteff} for an age of $t=30^{+20}_{-10}$ Myr. The mass of $\kappa$ And b is greater or equal to 13 $\mathrm{M_{Jup}}$ according to these models. Predicted masses from the luminosity agree with the ones derived from the temperature, althougth predictions do not extent to sufficient high masses to reproduce the upper limit on the estimated $T_{eff}$ of $\kappa$ And b. The companion properties can also not be reproduced for ages of 150 Myr for the same reason.

We also derived absolute flux predictions of SB12 models for the given filter passbands  and the four sets of boundary conditions (cloud-free models at solar metallicity - cf1s, cloud-free models with three times the solar metallicity - cf3s, hybrid clouds at solar metallicity - hy1s, hybrid clouds  with three times the solar metallicity - hy3s) used for $\kappa$ And b following the same method as in \cite{2013arXiv1302.1160B}. These synthetic fluxes were compared to the observed SED. The results are reported in Table \ref{tab:warm-start}. The comparison is biased by the limited mass coverage of the models. We note however, that solutions found within the models boundaries correspond to initial entropies intermediate between those of hot and cold-start models, placing the mass at the typical planet/brown-dwarf boundary \citep[$\sim$13.6 $M_{Jup}$][]{spiegel11, 2012A&A...547A.105M, boden13}. These solutions correspond to $\mathrm{T_{eff}}$ values that are in good agreement with those determined from the companion SED.\\

\begin{table}[t]
\begin{minipage}[ht]{\columnwidth}
\caption{Best fit photometric predictions of the ``warm-start" evolutionary models. Solutions found at the edges of the parameter space (mass, $\mathrm{S_{init}}$) covered by the models are highlighted in italic.}
\label{tab:warm-start}
\centering
\renewcommand{\footnoterule}{}  
\begin{tabular}{lllll}
\hline \hline 
Atmospheric model			 &  	 Age     &			Mass								& 		$\mathrm{S_{init}}$				&	$\mathrm{\chi^{2}}$	\\
										 &		(Myr)		&			($\mathrm{M_{Jup}}$)	&	($\mathrm{k_{B}}$/baryon)   	&									\\		
\hline
Cloud free - 1x solar	 &			20		&			\textit{15}								&		\textit{9.75}									&			83.19					\\							
Cloud free - 3x solar	 &			20		&			\textit{15}								&		\textit{9.75}									&			57.77					\\							
Hybrid cloud - 1x solar		 &			20		&			14								&		9.75									&			14.56					\\							
Hybrid cloud - 3x solar		 &			20		&			14								&		9.75									&			11.92					\\							
Cloud free - 1x solar	 &			30	&			14								&		9.75									&			83.39					\\							
Cloud free - 3x solar	 &			30	&			14								&		9.75									&			58.32					\\							
Hybrid cloud - 1x solar		 &			30	&			14								&		10.25									&			13.73					\\							
Hybrid cloud - 3x solar		 &			30	&			14								&		10.00									&			11.29					\\							
Cloud free - 1x solar	 &			50	&			14								&		10.25									&			82.45					\\							
Cloud free - 3x solar	 &			50	&			14								&		10.50									&			57.66					\\							
Hybrid cloud - 1x solar		 &			50	&			\textit{14}								&		\textit{13.00}									&			19.30					\\							
Hybrid cloud - 3x solar		 &			50	&			\textit{14}								&		\textit{13.00}								&			16.27					\\							
Cloud free - 1x solar	 &			150	&			\textit{14}								&		\textit{13.00}								&			185.43					\\							
Cloud free - 3x solar	 &			150	&			14								&		12.75									&			175.71					\\							
Hybrid cloud - 1x solar		 &			150	&			\textit{14}								&		\textit{13.00}									&			173.50					\\							
Hybrid cloud - 3x solar		 &			150	&			\textit{14}								&		\textit{13.00}									&			168.67				\\							
\hline
\end{tabular}
\end{minipage}
\end{table}

In comparison, the models of  \citet{mc13}  have a much simpler outer boundary condition (hereafter MC13),
using a grey, solar-metallicity atmosphere.  We used them as they 
can be used to evaluate the impact of underlying hypotheses made in the models (e.g. atmosphere treatment, equation of state)
on the derived joint mass and $\mathrm{S_{init}}$ values. We  ran Markov Chain Monte Carlo simulations (MCMCs)
in mass and initial entropy as in \citet{mc13} with the related models to account for the uncertainties on the age,  $\mathrm{T_{eff}}$, and luminosity of $\kap$ And b.  
We assumed Gaussian distributions on $L$ and  $\mathrm{T_{eff}}$. We took 
normal or lognormal errorbars for the two considered age ranges  
($t= 30_{-10}^{+20}$~Myr and $t= 30_{-10}^{+120}$~Myr), and chose flat priors in $\Si$ and $M$.

\begin{figure}
\includegraphics[width=84mm]{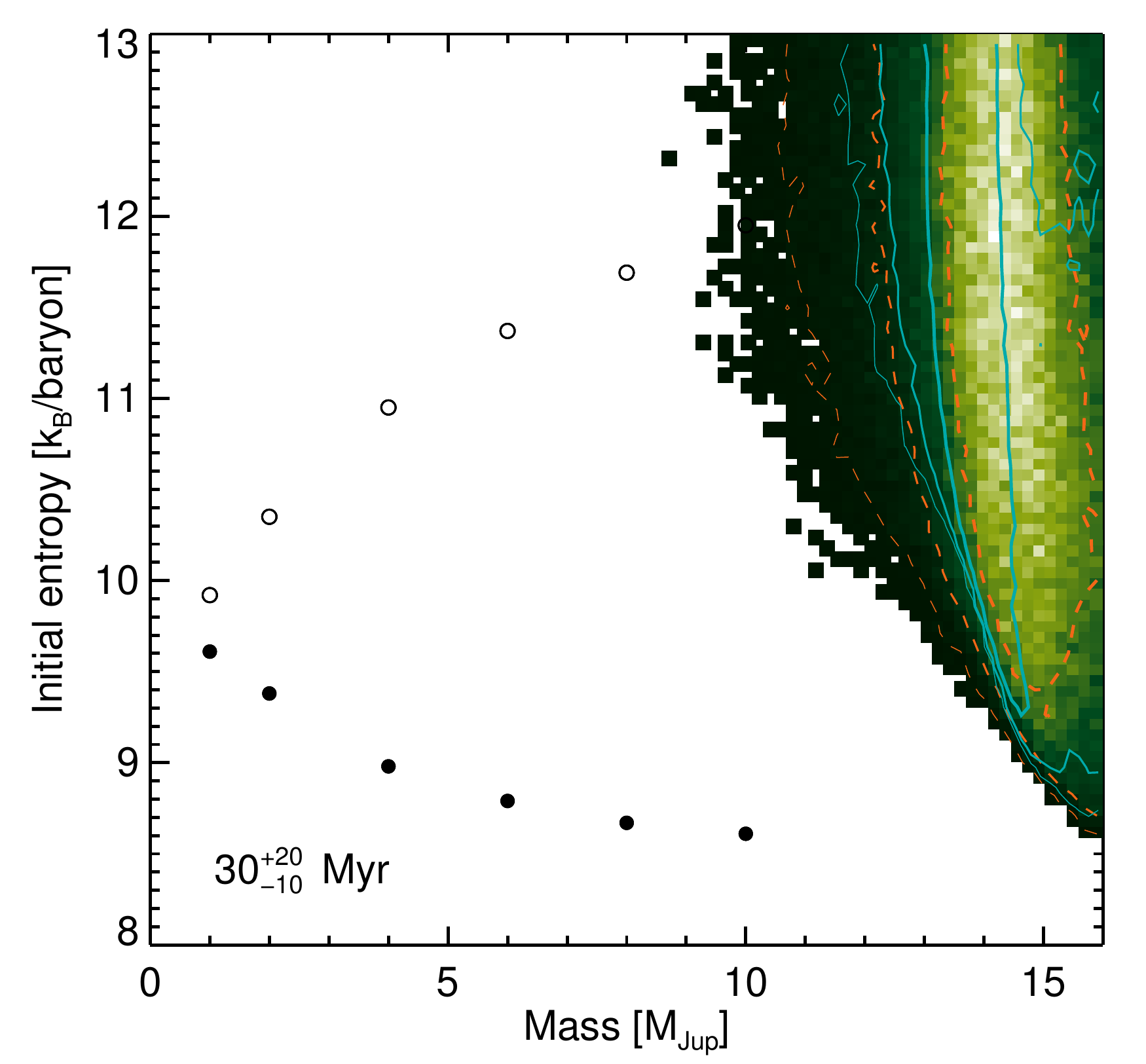}\\
\includegraphics[width=84mm]{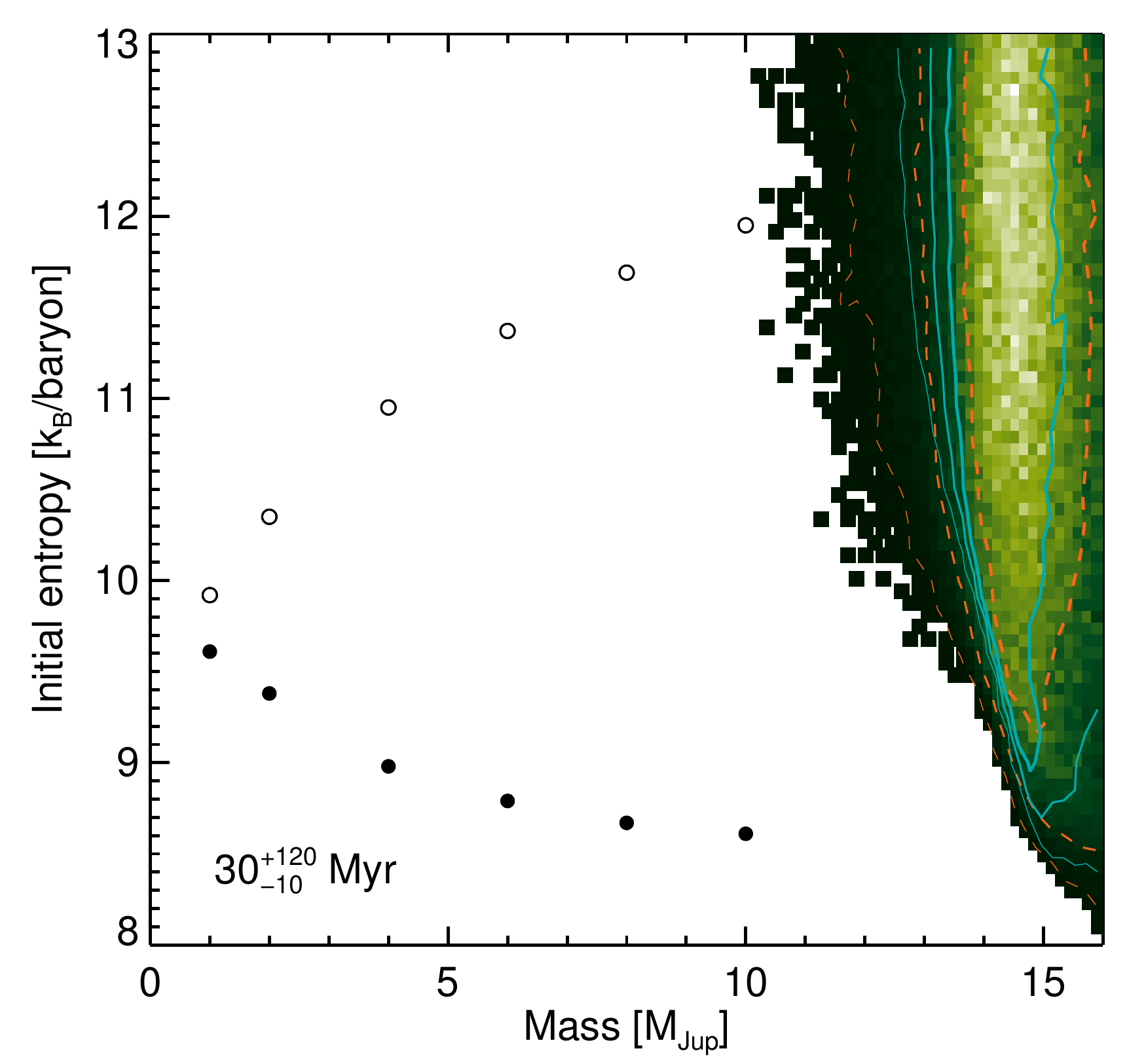}
\caption{Joint posteriors on the mass and intial entropy of $\kappa$~And~b obtained through MCMC runs with
the models of \citet{mc13}, which include deuterium burning,
from $\Teff$ (solid lines) and the bolometric luminosity (dashed lines) of $\kappa$ And b,
assuming Gaussian errorbars. Entropy values need to be decreased by 0.45~$\Sunits$ when
comparing with SB12. The contours indicate the 68-, 95- and 99~\%\ confidence
levels (thick to thin lines). The colour (from dark to light) is proportional to the joint posterior probability at each $M$ and $\mathrm{S_{init}}$ location in the $\Teff$-based run.
The age was taken as $t= 30_{-10}^{+20}$~Myr (top panel)
and $t= 30_{-10}^{+120}$~Myr (bottom panel).}
\label{fig:MCMC}
\end{figure}

Fig.~\ref{fig:MCMC} displays the 68-, 95- and 99 \% joint confidence regions
from the MCMC runs for both age groups. Open and closed circles are as in \cite{marl07}, show the approximate range of entropies
spanned by hot and coldest starts, respectively, but shifted upwards by $+0.38~\Sunits$ to match the luminosity in the models of MC13 (see therein).
 The results are  consistent with those of Fig.~\ref{Fig:wsSB12lumteff}. 

\begin{figure}
\includegraphics[width=84mm]{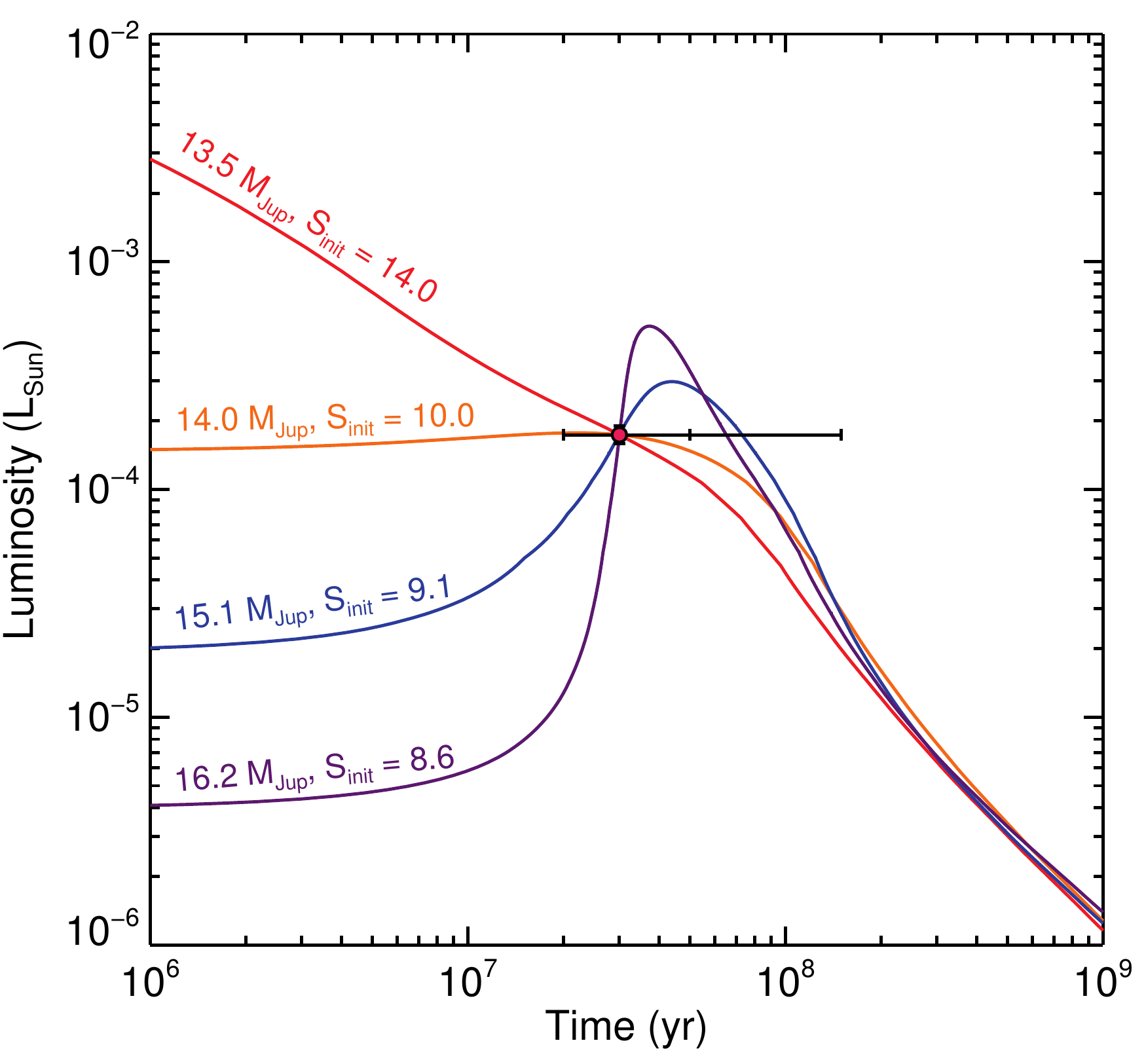}
\caption{
Examples of possible cooling curves for $\kappa$~And~b using
$\mathrm{log\:L/L_{\odot}=-3.76}$ at 30~Myr,
with the corresponding mass and initial entropy (to be decreased by
0.45~$\Sunits$
when comparing with SB12) of each curve indicated in the figure.
For the lower entropies, we find that $\kap$~And~b could be presently
undergoing a deuterium
`flash' \citep[][Marleau \& Cumming, in prep.]{salp92,boden13}.}
\label{fig:coolingcurves}
\end{figure}

Even for the low-age group, almost all solutions are in the mass regime where deuterium burning is important
for the evolution of the object. A dramatic illustration of this lies in the solutions found at lower entropies.
Whereas the models of SB12 allowed, for an age of $30_{-10}^{+20}$~Myr,
initial entropies down to only 9.5~$\Sunits$ to 1~$\sigma$ in luminosity (see Fig.~\ref{Fig:wsSB12lumteff}),
the models of MC13 find that $\kappa$~And~b could have formed with an entropy as low as $\approx8.8~\Sunits$, correcting downward from Fig.~\ref{fig:MCMC} for the entropy offset of $0.45~\Sunits$
between the two models\footnote{The implementation of the equation of state of \cite{1995ApJS...99..713S} is slightly different in the two cases, with SB12 using a simpler version of the \cite{1995ApJS...99..713S}
code that does not include a contribution from the proton spin in the partition function. See \citet{mc13} for details.}. The low-$\Si$ solutions are possible only if the models include a rise in the
object's luminosity due to deuterium burning \citep[][Marleau \& Cumming in prep.]{2012A&A...547A.105M, boden13}.
This is illustrated in Fig.~\ref{fig:coolingcurves}, which shows cooling tracks for
 different $(M,\Si)$ combinations which all reach $\log L/\LSun=-3.76$ at 30~Myr.
The lowest initial-entropy solutions undergo a `flash', where the entropy in the object
increases on a short timescale.
The combinaison of the measured $\Teff$ and luminosity (and therefore the object radius) can not help to discriminate these different possible cooling curves.  Indeed,  the high initial-entropy (``hot-start") and the flashing cooling curves have a difference of some 25~K or 0.05~$\mathrm{R_J}$ in predicted $\Teff$ and radius only.

The exact low-entropy solutions can depend on the details of deuterium burning,
for instance on the initial deuterium content or metallicity of the object \citep{spiegel11, 2012A&A...547A.105M}, but
the main conclusion is a robust one: \emph{the combustion of deuterium may play a significant role
in the cooling history of $\kap$~And~b,
irrespective of the precise age of the system.} We warn that while the lowest $\Si$ are comparable to the extrapolation of the coldest starts
\citep{marl07} to higher masses, what this implies about the formation mechanism is not clear given the major uncertainties about their outcome
(see also the discussion of possible formation processes in Section~\ref{subsec:form}).

\begin{table}[t]
\begin{minipage}[ht]{\columnwidth}
\caption{Properties of the $\kappa$ And system}
\label{tab:sum}
\centering
\renewcommand{\footnoterule}{}  
\begin{tabular}{llll}
\hline \hline 
Parameter			 &  	$\kappa$ Andromedae A    &			$\kappa$ Andromedae b	&		Ref \\
\hline 
d	(pc)										&		$51.6\pm0.5$		&		\dots		&	2		\\					
Age (Myr)										&		$30^{+120}_{-10}$	&	\dots		&		1, 3	\\					
J	(mag)									&		$4.26\pm0.04$		&	$15.86\pm0.21$		&	1			\\
H	(mag)											&	$4.31\pm0.05$	&	$14.95\pm0.13$			&	1		\\
$\mathrm{K_{s}}$ (mag)						& 	$4.32\pm0.05$	&		$14.32\pm0.09$	&	1			\\
L'		(mag)										&	$4.32\pm0.05$ & 	$13.12\pm0.1$		&	1, 3			\\
NB\_4.05	(mag)								&		$4.32\pm0.05$		&	$13.0\pm0.2$		&	 1		\\
M'		(mag)										&	$4.30\pm0.06$\tablefoottext{a}	&	$13.3\pm0.3$		&	 1			\\
$\mathrm{m_{J}}$(mag)					&		$0.70\pm0.06$		&	$12.30\pm0.22$		&	1			\\
$\mathrm{m_{H}}$	(mag)				&		$0.75\pm0.06$		&		$11.39\pm0.15$		&	1		\\
$\mathrm{m_{K_{s}}}$ (mag)						&	$0.76\pm0.06$		&	$10.75\pm0.11$		&		1		\\
$\mathrm{m_{L'}}$		(mag)				&	$\mathrm{0.76\pm0.06}$		&		$\mathrm{9.56\pm0.11}$		&		1		\\
$\mathrm{m_{NB\_4.05}}$	(mag)				&		$\mathrm{0.76\pm0.06}$	&		$9.44\pm0.23$	&		 1		\\
$\mathrm{m_{M'}}$	(mag)					&		$\mathrm{0.85\pm0.06}$	&		$\mathrm{9.75\pm0.31}$	&		1		\\
Spectral type								&		B9IVn		&	M9-L3:		&		1, 4			\\
$\mathrm{T_{eff}}$ (K)	&   	$10 900\pm300$		&	 $1900^{+100}_{-200}$   	&	 1			\\	 
					&   	$11 400\pm100$		&		&	 5 			\\
                                                  &      $10 700\pm300$        &               &    6  \\
log g		(dex)			    &		$3.78 \pm 0.08$\tablefoottext{b}      &     $4.5\pm1.0$                         &   1  \\		
                                                   &     $4.10\pm0.03$		&				&	 5		\\
                                                   &     $3.87 \pm 0.13$     &                                  &   6  \\
M/H			(dex)								&		$-0.36 \pm 0.09$		&		\dots	&		1, 5	\\
													&		$-0.32\pm 0.15$     &                    &     6   \\	
$\mathrm{log_{10}(L/L_{\odot}}$)		&		$1.83\pm0.04$\tablefoottext{c}					&		$\mathrm{-3.76\pm0.06}$ &  1 \\
Mass		($\mathrm{M_{\odot}}$)					&		$2.6\pm0.2$\tablefoottext{d}		&	 $0.013^{+0.022}_{-0.001}$ \tablefoottext{e}		&		1	\\
Mass		($\mathrm{M_{\odot}}$)									&				&	 $\geq 0.011$ \tablefoottext{f}		&		1	\\
\hline
\end{tabular}
\end{minipage}
\tablefoot{[1] - this work,  [2] - \cite{2007A&A...474..653V}, [3] \cite{2013ApJ...763L..32C}, [4] \cite{1994AJ....107.1556G}, [5] \cite{2005AJ....129.1642F}, [6] \cite{2011A&A...525A..71W}.}
\tablefoottext{a}{Estimated  from the mean K'-M' colors of B9 stars.} \\
\tablefoottext{b}{The polar surface gravity. Estimated by correcting the measured log(g) for the rapid rotation of the star using the method of \cite{2006ApJ...648..591H}.}\\
\tablefoottext{c}{Found ajusting BT-Settl spectra with $\mathrm{T_{eff}=}$10 600, 10 800, and 11 200 K, and log g=3.5 on the 0.365-22.1 $\mu$m flux distribution of the source.}\\
\tablefoottext{d}{Estimated comparing the measured $\mathrm{T_{eff}}$ and bolometric luminosity of $\kappa$ And A  to evolutionary tracks of \cite{2012A&A...537A.146E}, with and without rotation, for ages of 20 to 150 Myr.}\\
\tablefoottext{e}{Using COND  ``hot-start" evolutionary models for the most concervative age range \citep{2003A&A...402..701B}.\\} 
\tablefoottext{f}{Using the ``warm-start" models.\\} 
\end{table}

%

\section{Discussion}
\label{section:discussion}
\label{subsec:form}
The characterization of $\kappa$ And further illustrate the challenge of determining accurate masses for companions due to uncertainties on the age-dating methods and evolutionary tracks.

The de-projected (and projected) separation of $\kappa$ And b \citep[$61^{+50}_{-20}$ AU,][]{2013ApJ...763L..32C} is compatible with the size of primordial \citep[e.g.][ and ref therein]{1997ApJ...490..792M, 2009A&A...508..707P, 2013A&A...549A..92G} and debris disks \citep[][and ref therein]{2013MNRAS.428.1263B} surrounding stars in the same mass range as $\kappa$ And A, some of which show structures suggesting a clear signpost for planets \citep[e.g. HR4796,][]{1999ApJ...513L.127S, 2012A&A...546A..38L}. The companion location also fits well  with the extent and the location of cavities/gaps/spirals in young structured (or transition) disks \citep{2011ApJ...732...42A, 2013ApJ...762...48G, 2013ApJ...766L...2Q} discovered around Herbig Ae stars. $\kappa$ And b's separation is close to that of the candidate substellar embryo \citep{2013ApJ...766L...1Q} around the 2.4 $\mathrm{M_{\odot}}$ star HD 100546, which  mass might extend inside the ``brown-dwarf" regime. It is in addition intermediate between the two outermost planets orbiting HR 8799, and nearly identical to the probable exoplanet HD 95086b \citep{2013arXiv1305.7428R}, all orbiting intermediate-mass stars. Therefore, despite the large uncertainties on the mass of $\kappa$ And b and the lack of disk excess emission around $\kappa$ And A (Section \ref{section:systage}), we should still consider that the companion could have formed within a disk.

The growing population of  massive gaseous companions disovered on short-period orbits around massive stars \citep[e.g. ][ and ref therein]{2006A&A...452..709G, 2008A&A...491..889D, 2010ApJ...717..348H, 2011ApJ...728...32L, 2011A&A...533A..83B}, the discovery of unusually dense substellar companions \citep[e.g. CoRoT-20b and HAT-P-20b][]{2011ApJ...742..116B, 2012A&A...538A.145D}, and recent simulations \citep{2009A&A...501.1161M, 2013arXiv1302.1160B}, suggest that the core-accretion mechanism might still work well inside the brown-dwarf mass regime\footnote{We note that disk-instability associated to ``tidal-downsizing" might also explain the observations \citep{2010Icar..207..509B, 2011ApJ...735...30B,2010MNRAS.408.2381N, 2010MNRAS.408L..36N, 2011MNRAS.413.1462N}. But the models hasn't been extensively tested against observations yet \citep{2013arXiv1304.4978F}.}. \cite{2008ApJ...673..502K} shows that 10 $\mathrm{M_{\oplus}}$ cores can form in less than 1 Myr from $\sim$3 to $\sim$23 AU around 2.4-2.8 $M_{\odot}$ stars. \cite{2011ApJ...727...86R} also propose that core-accretion could operate out to 40-50 AU. A formation closer to the snow line associated to dynamical scattering or outward migration could explain $\kappa$ And b \citep{2008A&A...485..877P, 2008MNRAS.387.1063P, 2009ApJ...705L.148C, 2010MNRAS.401.1950P, 2011IAUS..276..225C, 2012MNRAS.421..780L}. Dynamical scattering would  nevertheless require additional, but yet undetected massive companions. Therefore, it is less suitable if the system is significantly older than 30 Myr and/or if the initial entropy of $\kappa$ And b is close to ``cold-start" conditions. The proposed modifications on the way solids are accreted to form a core \citep{2010A&A...520A..43O,2012A&A...544A..32L, 2012A&A...546A..18M} might still facilitate the formation of $\kappa$ And b by nucleated instability closer to its present separation.

   \begin{figure}
   \centering
   \includegraphics[width=\columnwidth]{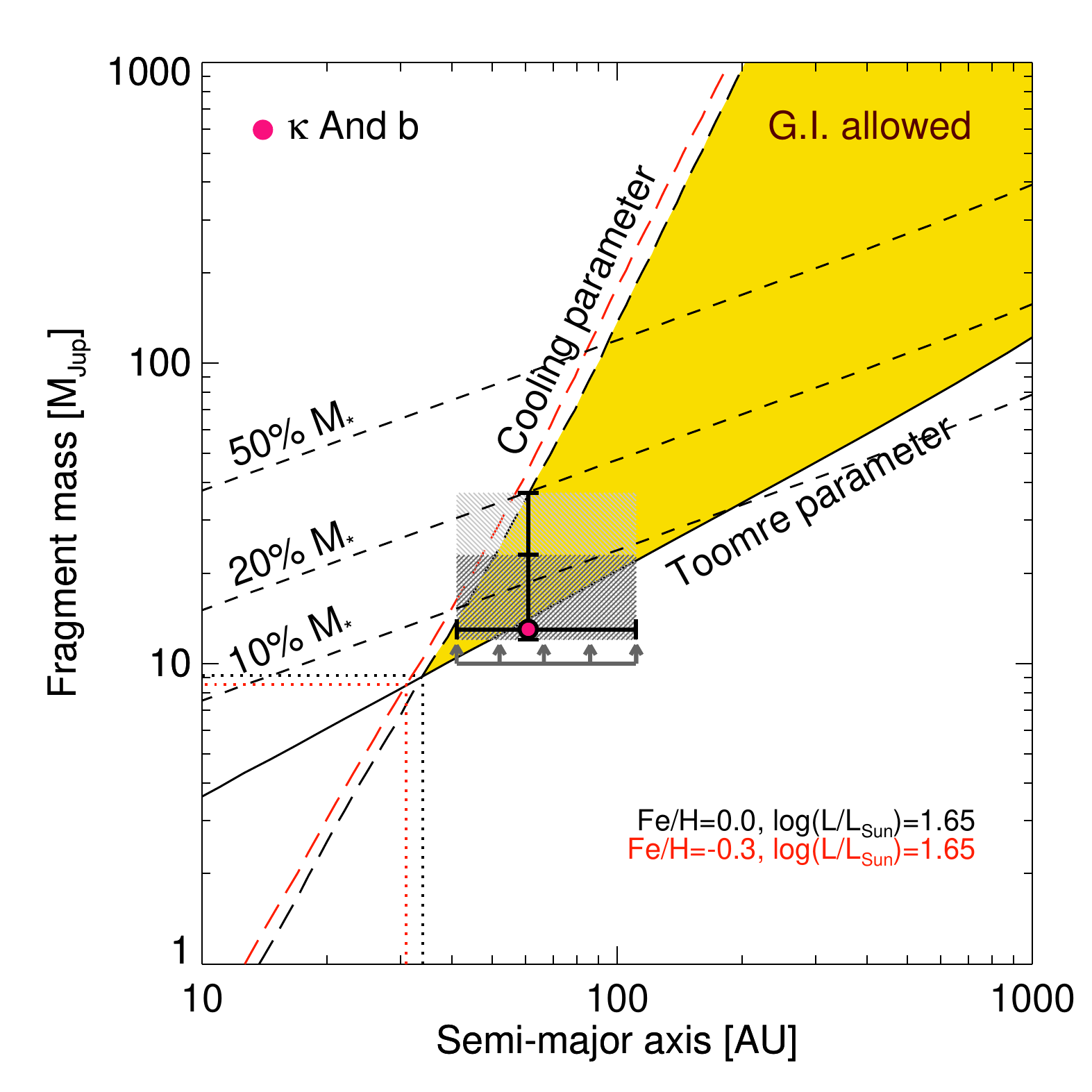}
      \caption{Predicted masses of fragments caused by disk instabilities as a function of semi-major axis for the massive star $\kappa$ And. Formation of planets by disk instability is possible when the Toomre criterium is exceeded (solid lines) and when the disk can cool sufficiently rapidly (long dashed lines). The allowed formation space (in yellow) lies in-between these two criteria. Companions can not be formed in situ at separations shorter than the ones indicated by the dotted lines. We superimpose the semi-major axis and mass estimates of $\kappa$ And b (magenta dot) estimated from ``hot-start" models for Columba ages ($\mathrm{30^{+20}_{-10}}$ Myr; dark grey shaded area) and considering an upper-limit corresponding to the highest age for the Pleiades cluster ($\mathrm{30^{+120}_{-10}}$ Myr; light grey shaded area). We also report the lower limit on the companion mass inferred from ``warm-start" models (grey  arrows, see Section \ref{subsub:warmstart}). We show the corresponding primordial disk-mass reservoir (10, 20, 50\%) at a given separation needed to form  given clumps/companions. The Toomre parameter is independent of stellar metallicity.} 
         \label{Fig:GI}
   \end{figure}

We also applied a dedicated formation model to explore the possibility that $\kappa$ And b could originate from a disk-instability \citep[Klahr et al. in prep; see also ][ for a description of the models]{2011ApJ...736...89J, 2013A&A...553A..60R}. These models can predict the range of semi-major axes and companion masses originating from clumps that  1/ formed in Toomre-unstable disks \citep{1981seng.proc..111T} and, 2/  cooled down more rapidly than the local Keplerian timescale. The models require as input the initial luminosity of the star and the system metallicity. The initial luminosity was estimated by inputting the temperature and luminosity  of the star  (see Table \ref{tab:sum}) into evolutionary models of \cite{2012A&A...537A.146E}, while considering system ages of 20 and 150 Myr.  We generated disk models with solar and under-solar abundances in order to reflect the possible metallicities of the star  \citep[Note that the star probably has a solar metallicity, see][]{2013ApJ...763L..32C}. The two metallicities do not change significantly the predictions (see Figure \ref{Fig:GI}). In contrast with  $\beta$ Pictoris b \citep[][]{2013arXiv1302.1160B, 2013arXiv1306.0610C}, or HR 8799bcde \citep{2011ApJ...736...89J}, $\kappa$ And b has an estimated mass and semi-major axis (and projected separations) compatible with a formation by disk-instability close to its present location.  The masses of $\kappa$ And b fall at the edge of the allowed range set by the Toomre criterium if the system age is 20 Myr. Adopting the lower limit on the estimate of the initial luminosity ($L/L_{\odot}=$1.55) shifts the contrain associated to the Toomre criterium to lower mass values, therefore solving the issue. This also highlight the fact that these models predictions remain highly dependent of the correct determination of these initial luminosities. Recent alternative disk-instability models support the idea that fragments tend to migrate inward on extremely short timescales \citep[e.g. ][and ref therein]{2011ApJ...737L..42M, 2011MNRAS.416.1971B, 2012ApJ...746..110Z, 2013A&A...552A.129V}, thus arguing against an in-situ formation by disk-instability for $\kappa$ And b. Nevertheless,  our model also show the companion -- if massive enougth --  might still be formed up to 220 AU. Also, migration processes  are still far from being understood \citep{2009A&A...501.1161M}. 

We applied the same models to the current population of directly-imaged brown-dwarfs companions orbiting young 2.2-2.5 $M_{\odot}$ stars (HR 7329 B, HD 1160B, HIP 78530B; see Appendix \ref{AppendixC}) in an attempt to understand how  $\kappa$ And b relates to these objects.  HR 7329 B has estimated separation and masses (using ``hot-start" models) compatible with a formation by  disk-instability. This companion and $\kappa$ And b fall in a similar area in the diagrams (Figures \ref{Fig:GI} and \ref{Fig:GIHR7329}) where the primordial disk  needed to form clumps with the estimated masses of the companions is  relatively light ($\sim$10\% M$_{*}$). Such disk masses are compatible with the observations of primordial disks  \citep[e.g.][]{1990AJ.....99..924B, 2013A&A...549A..67G}. HR 7329 B is the only companion of the sample known to orbit a star with a debris disk  \citep[see][ for HD 1160 and HIP 78530]{2006ApJ...653..675S, 2012ApJ...756..133C}. \cite{2009A&A...493..299S}  derive an outer radius of 24 AU, which might result from truncation/disruption \citep{2012ApJ...745..147R}.  HD 1160 B is part of a multiple system with a  0.22 $\mathrm{M_{Sun}}$ companion HD 1160 C imaged  at a projected separation of 533 AU (2.2") from the star, therefore possibly placed on a concentric orbit. Models indicate that HD 1160 B would need a more massive ($\sim$0.2 $\mathrm{M_{star}}$) disk to form than $\kappa$ And b.  Nevertheless, the comparison of HD 1160 B properties to the disk-instability models is more subject to caution since the formation of companions in such high-multiplicity systems is probably far more complex.  To conclude, the model predict that the extremely wide companion HIP 78530 B \cite{2011ApJ...730...42L} can not have formed in-situ, contrary to the three other studied systems. Therefore,  \textit{these comparisons would now deserve being repeated with other models, and on a larger sample of objects.} 

\section{Conclusions}
We  present the first deep-imaging observations of the $\kappa$ And system at 4.05 (NB\_4.05) and 4.78 $\mu$m (M$'$). We retrieved the companion at these wavelengths and  estimate $\mathrm{NB\_{4.05}=13.0 \pm 0.2}$ and $\mathrm{M'=13.3 \pm 0.3}$ mag. We also obtained new photometry of $\kappa$ And A in the J, H, and $\mathrm{K_{s}}$ bands, which we use to re-evaluate the 1.1-2.5 $\mu$m photometry of the companion. The  resultant photometry indicates that the companion is a late-M or early-L dwarf with a bolometric luminosity of $\mathrm{Log_{10}(L/L_{\odot})=-3.76\pm0.06}$. We obtained a low-resolution set of observations accross the 1-5 $\mu$m spectral energy distribution of $\kappa$ And b and compared  it to predictions from atmospheric models exploring the limiting and intermediate cases of dust-formation. All models converge toward a $\mathrm{T_{eff}}=1900^{+100}_{-200}$ K for $\kappa$ And b. Models with dust in the photosphere of the object  better reproduces $\kappa$ And b spectral energy distribution.  The models do not enable to constrain the companion's surface gravity.

The luminosity and temperature were then used as input of evolutionary models accounting for a wide range of initial conditions. We re-estimated and considered for that purpose a more conservative age range than in the discovery paper ($30^{+120}_{-10}$ Myr). ``Hot-start" models constrain the mass of $\kappa$ And b to $\mathrm{12-39\: M_{Jup}}$.  Conversely, ``Warm-start" models computed for initial entropies from 8 to 13 $\mathrm{k_{B}/baryon}$ provide a lower limit of 11 M$_{Jup}$.  Therefore a substantial  fraction of the allowed mass range of $\kappa$ And b is in the typical brown dwarf regime ($\gtrsim$ 13.6 $\mathrm{M_{Jup}}$). The latest ``Warm-start" models  reveal in addition that $\kappa$ Andromedae b could be undergoing a deuterium ‘flash’. This  ‘flash’ is expected to play a significant role in the cooling history of the companion. It poses a serious challenge to the companion mass determination, irrespective of the uncertainties associated to the system age.   

The formation models we used indicate that, for a large fraction of plausible masses, and given the separation of the companion, $\kappa$ And b might have formed close to its present location by disk-instability. We apply the same models to the current sample of low-mass companions on wide orbits ($\gg$ 10 AU) around stars in the same masse range. These models suggests that some, but not all, of these objects could have also formed by disk-instability.

\cite{2013Sci...339.1398K} has recently attempted a determination of the C/O ratio in the atmosphere of the young wide-orbit gas giant planet HR 8799c. Applying the same method to $\kappa$ And b could help to clarify the companion's formation scenario in the near future. Additional  measurements (radial velocity, non-redundant masking) would also be of value to set constraints on possible dynamical perturbers at smaller orbital separations, which could account for the current location of $\kappa$ And b. 
 
\bibliographystyle{aa}

\begin{acknowledgements}
We are very grateful to our anonymous referee for reviewing this article. We thank the LMIRCam instrument team for operating the instrument during our observations. The authors recognize and acknowledge the significant cultural role and reverence that the summit of Mauna Kea has always had within the indigenous Hawaiian community. We thank Ginny McSwain for her complementary analysis of the high resolution spectrum of $\kappa$ And A. We are grateful to Christiane Helling, Soeren Witte, and Peter Hauschildt for developing and providing the DRIFT-PHOENIX models. We thank Anne-Marie Lagrange for checking for past SOPHIE observations of $\kappa$ And A and Julien Rameau for providing the detection limit on HR 7329B. We also thank Johan Olofsson for checking the spectral energy distribution of the star. This research was conducted in part using the MIMIR instrument, jointly developed at Boston University and Lowell Observatory and supported by NASA, NSF, and the W.M. Keck Foundation. J. Carson, J. Wisniewski, and C. Grady were supported by NSF awards 1009314, 1009203, and 1008440. J. Kwon is supported by the JSPS Research Fellowships for Young Scientists (PD: 24·110). The research leading to these results has received funding from the  French ``Agence Nationale de la Recherche'' through project grant ANR10-BLANC0504-01, the ``Programme  National de Physique Stellaire'' (PNPS) of CNRS (INSU), and the European Research Council under the European Community's Seventh  Framework Program (FP7/2007-2013 Grant Agreement no. 247060). It was also conducted within the Lyon Institute of Origins under  grant ANR-10-LABX-66. 
 \end{acknowledgements}

\Online

 \begin{appendix} 
  \section{$\kappa$ And A color-magnitude diagram samples}
  \label{AppendixA}
  
We describe here each of the empirical sub-samples that were used to construct Figure~\ref{Fig:CMDJosh}:
  
\begin{itemize}
 \item{{\bf Scorpious-Centaurus} - For this sample, we used the list of \citet{2012ApJ...756..133C}.  We removed known spectroscopic, eclipsing, and sub-arcsecond resolved binaries and Herbig AeBe stars to minimize photometric scatter in the CMD.  Photometry and distances were collected from the Hipparcos catalog \citep{1997A&A...323L..49P, 2007A&A...474..653V}.  Photometry was corrected using the individual extinctions listed in \citet{2012ApJ...756..133C}.} 
 \item{{\bf IC 2391} - This sample is comprised of early-type stars selected from the membership list of \citet{1969PASP...81..629P}.  Distances and photometry are from Hipparcos.  The photometry was corrected using an average cluster reddening of 0.01 \citep{1996ApJS..106..489P}} 
\item{{\bf The Pleiades}- We drew early-type members from the list of \citet{2007ApJS..172..663S} and used individual distances and photometry from Hipparcos.  Individual reddening values from \citet{1986ApJ...309..311B} were used to correct the photometry.}
\item{{\bf Ursa Majoris moving group} - We chose A-type stars proposed to be members of the group nucleus by \citet{2003AJ....125.1980K}.  Photometry calculated in this reference was also adopted.  We applied no correction for reddening.}
\item{{\bf Young Moving Groups} - We compiled proposed A and B-type members of the AB Doradus, Tucana/Horologium, Columba, and $\beta$ Pictoris young kinematics groups from \citet[][and references therein]{2013ApJ...762...88M}.  For these stars, when available, we compiled the mean photometry from the catalogs of Mermilliod \citep[][including $\kappa$ And A]{1994cmud.book.....M}.  When photometry was not available in these catalogs it was taken from Hipparcos.  In the construction of the CMD, we used Hipparcos distances \citep{2007A&A...474..653V} and applied no reddening correction.}  
\end{itemize}

  \section{New determination of the atmospheric parameters of $\kappa$ And A}
 \label{AppendixB}

\begin{figure}
   \centering
   \includegraphics[width=\columnwidth]{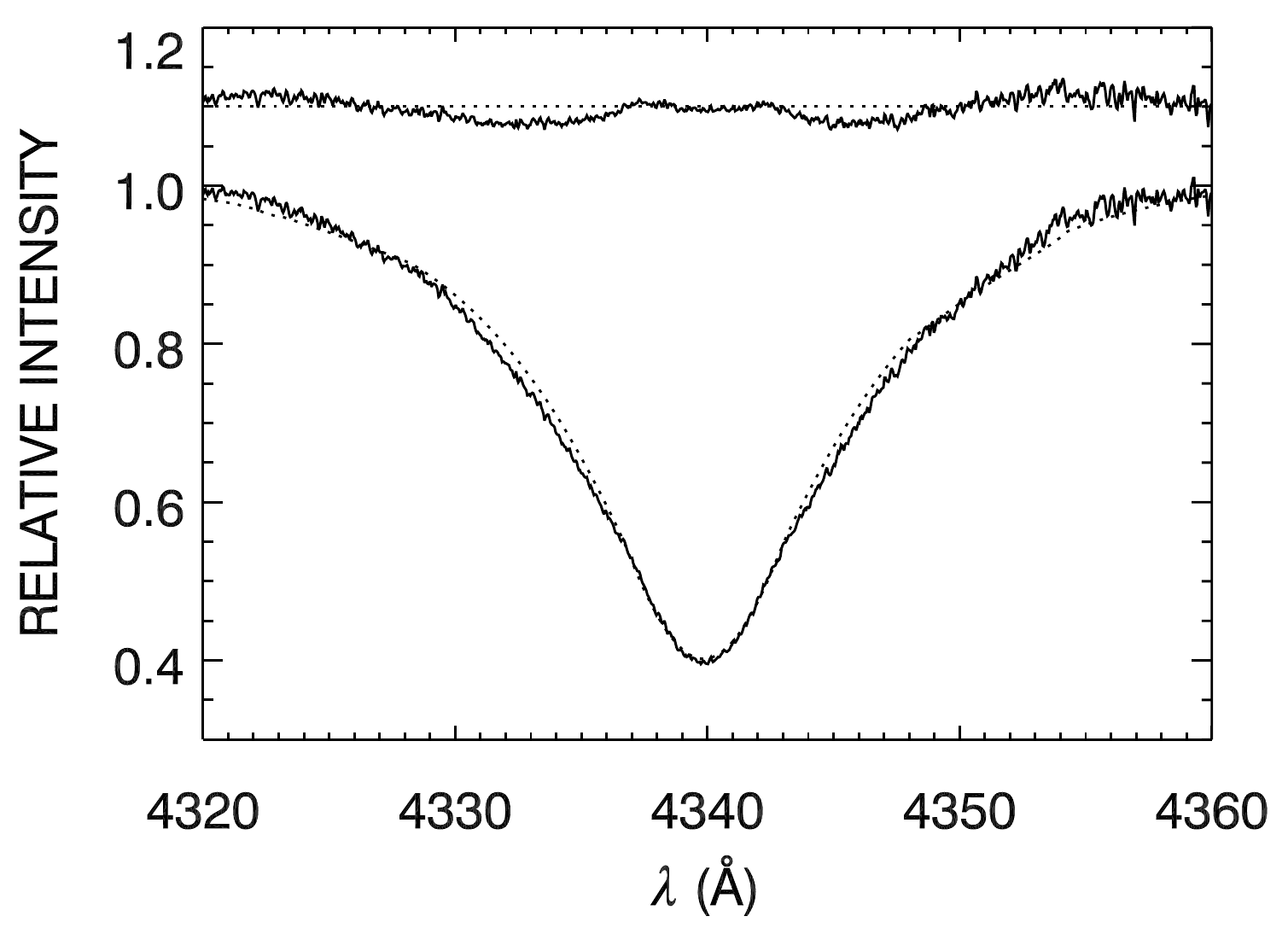}
   \includegraphics[width=\columnwidth]{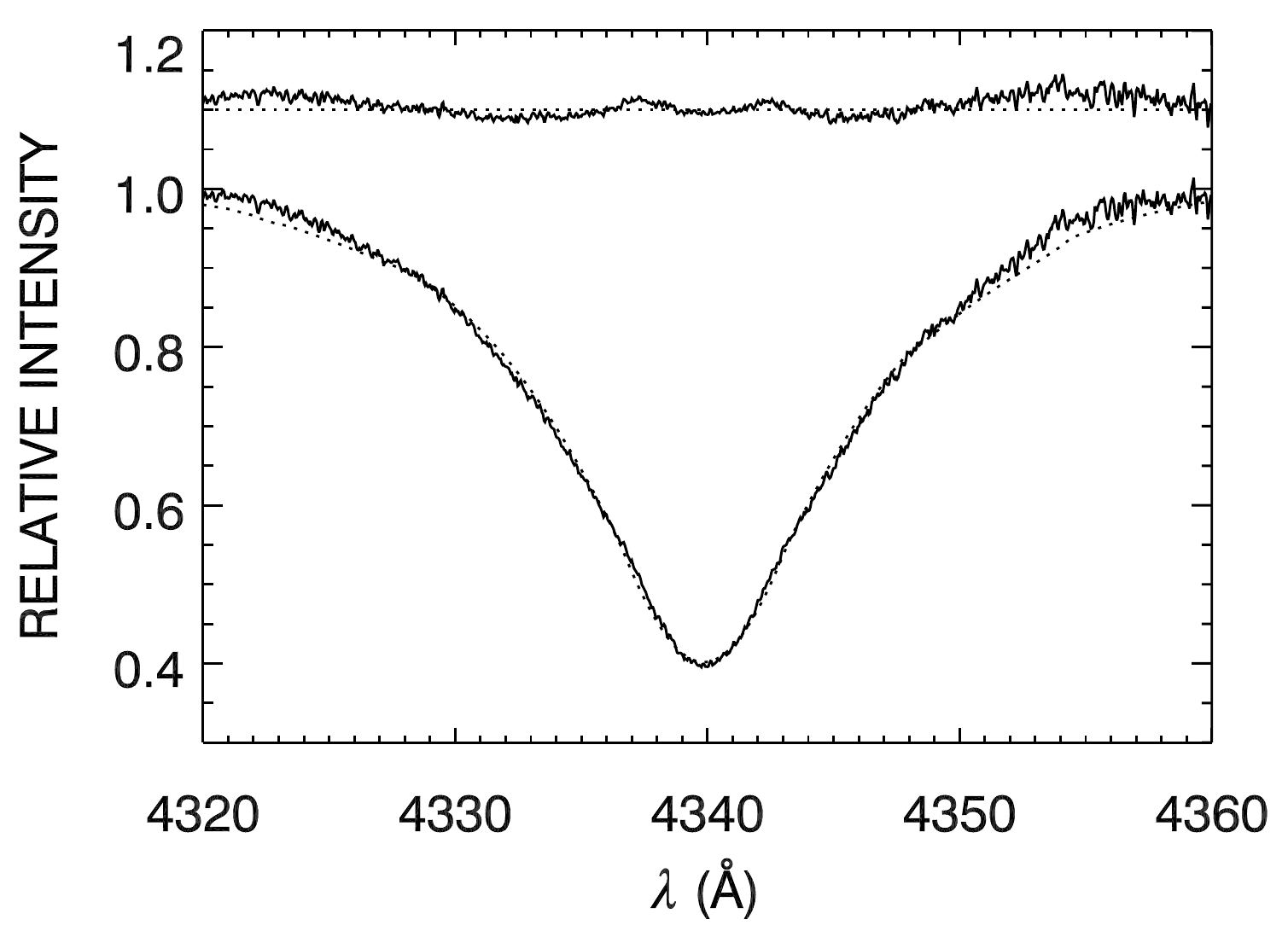}
      \caption{Hydrogen $\mathrm{H_{\gamma}}$ line of $\kappa$ And A (solid lines) fitted by atmospheric models (dotted lines). Residuals from the fit are shown in the upper part of the figures. Top: Solution where the line wings and the core are well fitted, but where the fit degrades in the line breadth. Bottom: Solution for which the line breadth is better fitted. These solutions were combined to derive the new  estimate of the temperature and the surface gravity of the star.} 
         \label{Fig:atmoKapAnd}
   \end{figure}

Our APO-ARCES spectrum of $\kappa$ And A resembles those of other of luminosity class IV and V B-type stars. Our spectral analysis followed the procedure developed by \cite{2012AJ....144..158M} which uses model spectral template fitting to estimate the atmospheric parameters of the star.  Their method uses four lines, H$\gamma$ $\lambda$4340~\AA, He I $\lambda$4387~\AA, $\lambda$4471~\AA, $\lambda$4713~\AA, and Mg II $\lambda$4481~\AA~to determine $v$sin$i$, $\mathrm{T_{eff}}$, and log g (see Figure \ref{Fig:atmoKapAnd}).  Since the lines that are sensitive to $v$sin$i$ (He I and Mg II) were weak in our spectrum, we maintained a fixed $v$sin$i$ \citep[$v$sin$i$ = 190 km s$^{-1}$,][]{2000AcA....50..509G, 2005AJ....129.1642F} with a conservative uncertainty (20 km s$^{-1}$) during the analysis of the H$\gamma$ line, which is sensitive to temperature and gravity.  This analysis found two solutions: one where the model spectrum was a better fit to the line wings and core, but was slightly off in the line width, and another where the best fit model better reproduces the line width but gave a poorer fit to the wings.  We combined the results from these two solutions to estimate $\mathrm{T_{eff} = 10900 \pm 300\:K}$, $\mathrm{log\:g = 3.50 \pm 0.08\:dex}$ where the larger uncertainties reflect the combined results.  As described in Section~\ref{section:analresults}, we applied a correction to the measured surface gravity for the rapid rotation of the star using the method of \citet{2006ApJ...648..591H}.  This method uses models to estimate the surface gravity at the pole of the star which should remain relatively unaffected by rapid rotation an give a better indication of the stars true evolutionary state.  We find $\mathrm{log\:g_{pol} = 3.78 \pm 0.08\:dex}$ for $\kappa$ And A. Our estimated $\mathrm{T_{eff}}$ is generally in agreement with those determined using other methods in the literature \citep[e.g.][]{2011A&A...525A..71W}. However, the surface gravity is lower than previous estimates, which have not been corrected for the star's rotation (see Table~\ref{tab:sum}). The spread in estimated surface gravities illustrates the inherent challenges in accurate atmospheric parameter determination for early-type stars.

 \section{Disk-instability models for HR 7329B, HD 1160B, and HIP 78530B.}
 \label{AppendixC}
We applied the disk-instability models of H. Klahr to the rare brown dwarfs companions identified around young  (age $\leq$ 100 Myr) early-A/late-B type stars. The models require as input the metallicity and initial luminosity of the star (the latter roughly scales with the stellar mass). We retrieve the zero-age main sequence luminosity of the stars  by inputting the present effective temperature or mass estimates of the stars and the known age of the system as inputs of evolutionary tracks \cite{2012A&A...537A.146E}. The properties of the systems are summarized in Table \ref{tab:appA}. We considered a solar metallicity for HR 7329B as the present measurements are roughly solar \citep[a variation of +0.17 dex does not affect the cooling time-scale significantly;][]{2008A&A...490..297S}. We also find a small sensitivity of the modelsimulation of predictions to the choice of the metallicity for HD 1160 \citep[Fe/H=0.0 and -0.3 considered here; see section 3.4 of ][]{2012ApJ...750...53N}.  Finally, we assumed HIP 78530B has a roughly solar metallicity. This is likely to be the case given recent measurements on lower mass stars of the associations \citep{2009A&A...501..965V}.

\begin{figure*}
   \centering
   \begin{tabular}{ccc}
   \includegraphics[width=5.8cm]{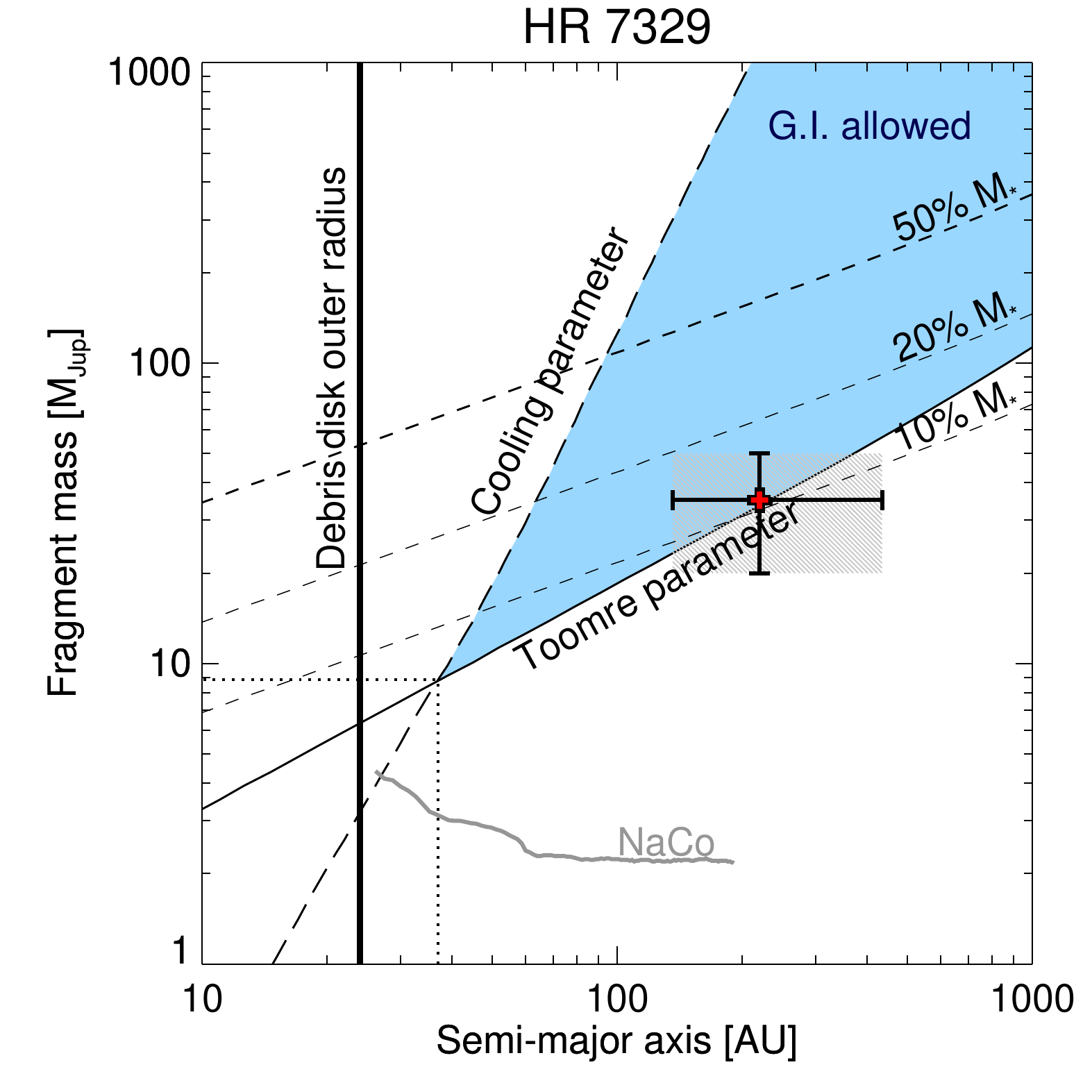} &
   \includegraphics[width=5.8cm]{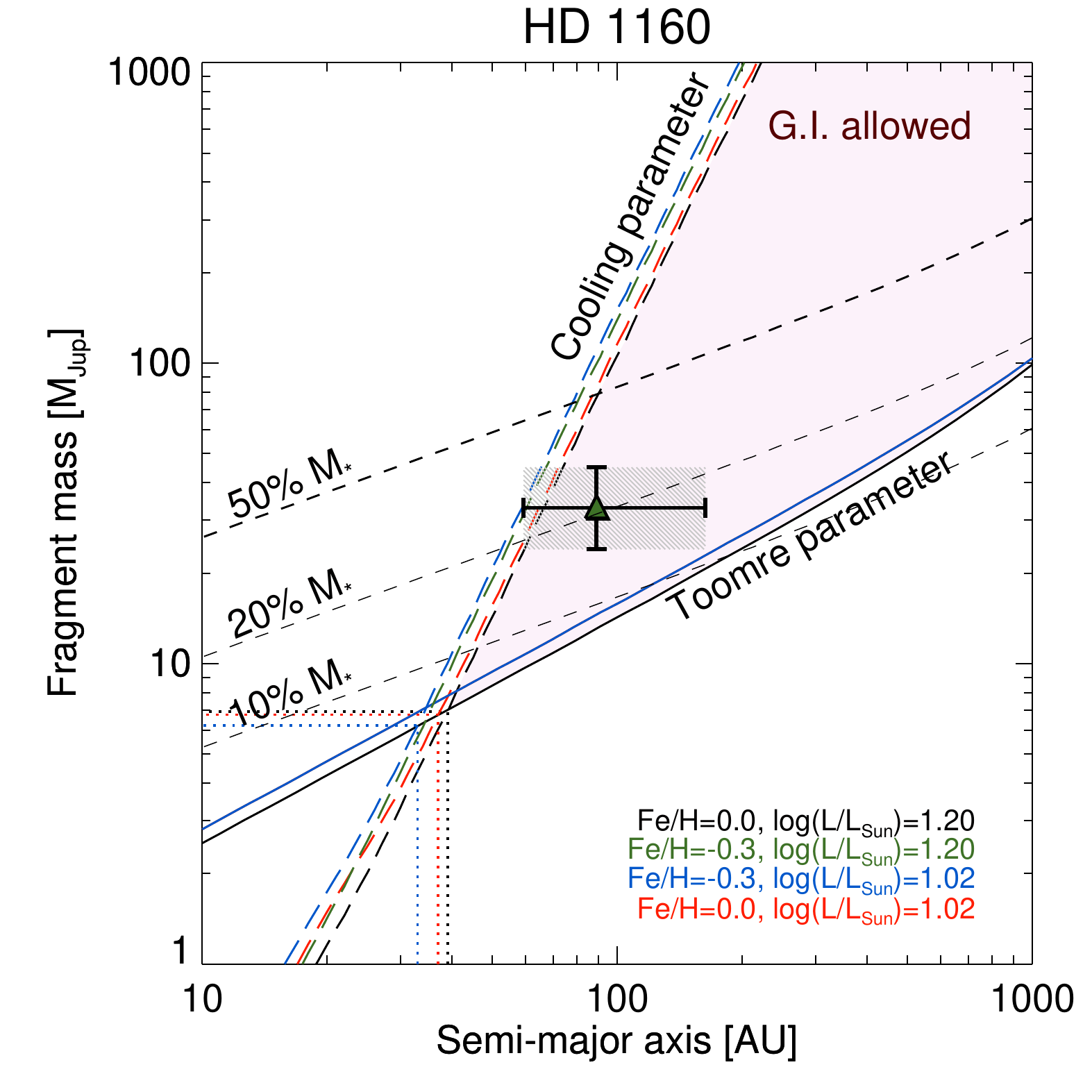} &
   \includegraphics[width=5.8cm]{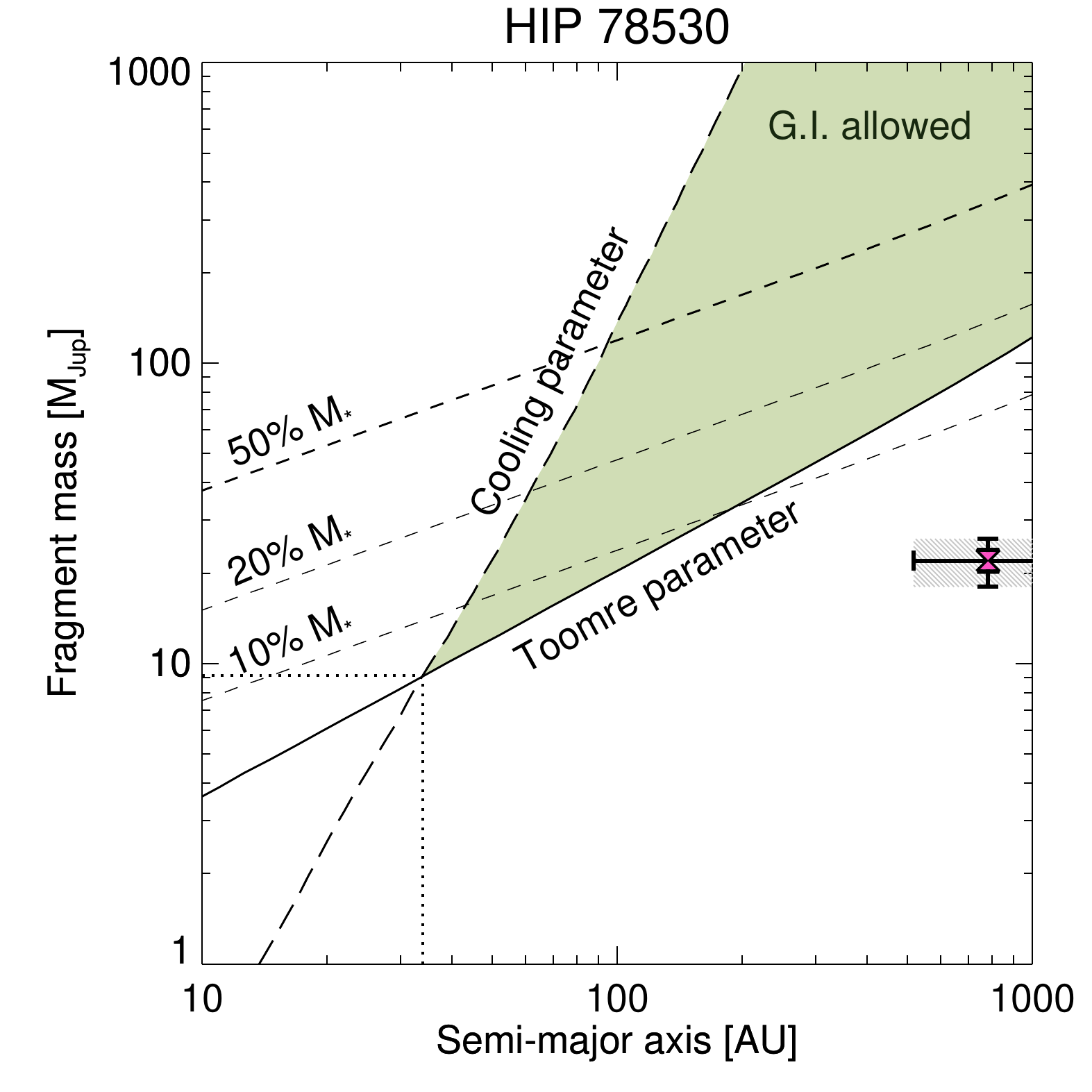}
   \end{tabular}
 
      \caption{Same as Figure \ref{Fig:GI}, but for the brown-dwarf companions HR 7329B (left, red cross), HD 1160B (middle, green triangle), and HIP 78530B (right, pink hourglass). We added the NaCo detection limit obtained by \cite{2013A&A...553A..60R} for HR7329 B. Model predictions do not extend beyond 1000 AU.} 
         \label{Fig:GIHR7329}
   \end{figure*}

We deprojected the observed separation of  HIP 78530B,  HD 1160B and C following the values from the Monte Carlo simulation of \cite{2009ApJ...697..824A}.  We use the values reported in \cite{2011MNRAS.416.1430N} for HR 7329 B. The properties of the three companions are compared to models predictions in Figure \ref{Fig:GIHR7329}.

 \begin{table*}[t]
\begin{minipage}[ht]{\textwidth}
\caption{Young 2.2-2.5 $\mathrm{M_{\odot}}$ stars with brown-dwarf (or low-mass) companions on wide orbits.}
\label{tab:appA}
\centering
\renewcommand{\footnoterule}{}  
\begin{tabular}{lllllllllll}
\hline \hline 
Name         &   d   &   age   &  FeH - star  &    Spectral    &    Spectral  &   $\mathrm{M_{A}}$         &    $\mathrm{M_{B}}$\tablefoottext{a}  &  $\rho$   &   $a$\tablefoottext{b}    &   Reference  \\
					&  (pc)    &   (Myr)   &    (dex)  &     type A                &		 type B 						&	($\mathrm{M_{\odot}}$)  & 	($\mathrm{M_{Jup}}$)    &   (")   &   (AU)    &    \\   
\hline 
HR 7329B    &   $47.7\pm 1.5$   &    $12^{+8}_{-4}$   & 0.17 & A0V   &   M7-8   &   $2.2\pm0.1$    &    $35\pm15$  &   4.2  &   $220^{+214}_{-84}$   &   1, 2, 3  \\
HD 1160B    &  $103\pm 5$       &  $50^{+50}_{-40}$  & \dots & A0V    &  \dots   &    $\sim$2.2   &   $37\pm12$   &   0.8  &   $89.1^{+73.9}_{-29.7}$  &   4, 5, 6 \\
HIP 78530B   &   $156.7\pm13.0$  &   3-11   & \dots &  B9V   &   M$8\pm1$   &  $\sim2.5$   &  $23\pm3$   &   4.5  &  $781^{+649}_{-264}$  &  6, 7  \\
\hline
\end{tabular}
\end{minipage}
\tablefoot{[1] - \cite{2011MNRAS.416.1430N}, [2] - \cite{2011MNRAS.410..190T}, [3] - \cite{2008A&A...490..297S}, [4] - \cite{2012ApJ...750...53N}, [5] - \cite{2007A&A...474..653V}, [6] - This work, [7] - \cite{2011ApJ...730...42L}.}
\tablefoottext{a}{Estimated from ``hot-start" models.} \\
\tablefoottext{b}{De-projected semi-major axis.} \\
\end{table*}

 \end{appendix}

\bibliography{KapAnd_charac}

\end{document}